\documentstyle{article}\begin{document}
\title{Functional formulation of  quantum theory of a scalar field in a metric
with Lorentzian and Euclidean signatures}

\author{ Z. Haba\\
Institute of Theoretical Physics, University of Wroclaw,\\ 50-204
Wroclaw, Plac Maxa Borna 9, Poland\\
email:zbigniew.haba at uwr.edu.pl} \maketitle
\begin{abstract}We study the Schr\"odinger equation
in quantum field theory (QFT) in its functional formulation. In
this approach  quantum correlation functions can be expressed as
classical expectation values over (complex) stochastic processes.
We obtain a stochastic representation of the Schr\"odinger time
evolution on Wentzel-Kramers-Brillouin (WKB) states by means of
the Wiener integral. We discuss QFT in a flat expanding metric and
in de Sitter space-time. We calculate the evolution kernel in an
expanding flat metric in the real time formulation. We discuss a
field interaction in pseudoRiemannian and Riemannian metrics
showing that an  inversion of the signature leads to some
substantial simplifications of the singularity problems  in QFT.
\end{abstract}

\section{Introduction}
We discuss the functional integral approach to quantum field
theory (QFT) in a complete analogy to the Schr\"odinger picture in
quantum mechanics \cite{jackiw,jack}. In contradistinction to the
Heisenberg picture, we insist on states and their time evolution.
The functional approach to QFT is a realization of the
conventional Schr\"odinger picture of quantum mechanics in the
space of functions of an infinite number of variables. Its
efficiency has been demonstrated in refs.~\cite{jackiw2,jackiw3},
in applications to quantum fields in de Sitter space. In this
paper, we discuss the functional approach to QFT   on general
hyperbolic manifolds.  In the functional formulation, we can
develop the path integral methods for a calculation of expectation
values earlier exploited at imaginary time as a rigorous tool in
quantum mechanics \cite{simonbook}. We can extend these methods to
real time by means of complex-valued stochastic processes.  The
extension to real time requires a complex extension of the
configuration (field) space. For fields on a manifold, such an
extension means a generalization of the path integral to complex
manifolds (for some recent attempts on such generalizations, see
\cite{witten,gibbons,hartle,turok,visser,sorkin,segal,lehners,halliwel}).
The standard approach to field quantization on a manifold
\cite{davis,ashtekar} implicitly assumes a non-unique ground
state. The field can be expanded in creation--annihilation
operators. These operators are defined in the Fock space which, in
functional representation, is a Gaussian normalized state. The
functional representation of quantum fields does not require the
Fock space. We can define the Feynman path integral by means of
the Wiener integral, which provides a particular Gaussian
functional measure for a field configuration space  (this could
also be considered as a particular Fock space). Such a formulation
does not require that there is a Fock space as a ground state  for
quantum fields. In this paper, the field is determined by a  WKB
Gaussian wave function as a solution of the Schr\"odinger equation
for the canonical Hamiltonian defined by canonical quantization.
We do not assume that the particular WKB wave function is
normalizable. Nevertheless, we can define correlation functions in
some other normalizable states with WKB phase factors. These
correlation functions determine the quantum fields. The
computation of correlations can be reduced to a calculation of
expectation values over stochastic processes. For this reason, the
functional formulation can be useful for numerical calculations.
In earlier papers~\cite{habajp,habajmp}, we have derived a
representation of the Feynman integral by means of the Brownian
motion in quantum mechanics, following an approach of
mathematicians Cameron and Doss \cite{doss1,doss2}. For quantum
field theory, the paths must be transformed into the paths of a
quantum oscillator as a quantum field is a collection of
oscillators. For a free field in Minkowski space-time, we have
discussed this approach in \cite{habaepj}, introducing a random
field satisfying a stochastic equation defined by the ground
state. For a free field on a general Riemannian manifold, a random
field determined by a Gaussian state defined on the Riemannian
manifold is needed. It can be seen that, in order to obtain an
interacting field, a non-linear stochastic equation is required.
The presence of noise  in semiclasical quantum gravity
correlation functions  has been discovered in
refs.~\cite{hu1,hu2}. Our approach relies on the construction of a
Gaussian random field.
 Instead of looking for a non-linear stochastic equation of the random field, we
introduce an interaction via the Feynman--Kac formula. There is no
problem of such a construction on the perturbative level.
Perturbation theory reduced to a polynomial expansion in quantum
fields will be equivalent to the standard QFT (hence, the standard
renormalization of polynomials applies). However, there arises a
difficult task to show that the corresponding Feynman--Kac factor
has a finite expectation value with respect to the Wiener measure.
Then, the functional integral over the solutions defining the
correlation functions must be finite. We show that a change in the
signature of the spatial metric (the time remains real) can
substantially facilitate the proof of integrability. The problem
has been studied in constructive quantum field theory in the
Euclidean framework. In the Feynman integral formulation (real
time), we deal with oscillatory integrals. In such a case, some
stability problems can be avoided. We assume that time is
well-defined in an evolution of the quantum scalar field on a
classical gravitational background. The appearance of time and the
Schr\"odinger evolution in a semiclassical framework for quantum
gravity have been discussed in
\cite{kiefer,unruh,brown,kuchar,kiefer0}. We repeat (in a somewhat
modified way) a derivation of the Schr\"odinger equation from the
Wheeler--DeWitt equation~\cite{dw} in Appendix H. We
hope that the study of the Schr\"odinger equation in an external
metric can help to solve the Wheeler--DeWitt constraint in quantum
gravity.

The plan of this paper is the following. First, we explain the
mathematically rigorous probabilistic method
\cite{albeverio,daletski} of solving the imaginary time
Schr\"odinger equation for a quantum (Euclidean) scalar field  as
a perturbation of the ground state solution .
  Then, in
Section \ref{sec3}, the method is extended by an analytic
continuation to apply to solutions for the real-time Schr\"odinger
equation. We discuss the functional Schr\"odinger equation for a
scalar field on a globally hyperbolic manifold
\cite{ellis-hawking,hyperbolic}. We consider general complex
metrics which may arise as saddle points in an average over
metrics in quantum gravity.  In Section \ref{sec4}, we show that a
free scalar quantum field has the Schr\"odinger wave function
solution, which is a pure WKB Gaussian phase if the initial state
is a pure Gaussian phase. Using this Gaussian solution, we
construct a random wave field (as a solution of a stochastic
equation),  which allows for the construction of a general
solution of the Schr\"odinger equation as an expectation value
over the Brownian motion. The solution can be calculated in a
simple way if the initial state is of the form of the Gaussian
factor times a mild perturbation. It can have an explicit form if
the manifold has a large group of symmetry. In Section \ref{sec5},
we discuss the scalar field on de Sitter space-time and on its
Euclidean continuation (the sphere $S^{4}$). Stochastic equations
for the de Sitter field and fields in an expanding (homogeneous)
universe are discussed in Section \ref{sec6}. The large symmetry
group  allows for an expansion in terms of eigenfunctions of the
algebra of this group. In this section, some field correlations
are calculated. The explicit Gaussian solutions of the
Schr\"odinger equation are not available for general initial
conditions. In Section \ref{sec6}, we derive an asymptotic formula
for the solution at a large angular momentum. An analogous formula
at large momenta for  the homogeneous expanding metric is obtained
in Appendix B. The asymptotic formulas allow for an
approximate calculation of correlation functions of stochastic
fields. In Section \ref{sec7}, the free field time evolution is
expressed by an evolution kernel. Then, a computation of
multi-time correlation functions can be reduced to Gaussian
integrals with the evolution kernels. The method can be extended
to interacting fields by means of the Feynman--Kac formula. The
correlation functions determine QFT (Wightman's reconstruction).
In \mbox{Sections \ref{sec8}--\ref{sec10}}, the power-law
evolution of massless scalar fields is discussed. The most
detailed results concern the free field  in a radiation background
(\mbox{Section \ref{sec10}}) with the spatial metric $g_{jk}\simeq
\delta_{jk}t$, which changes signature at $t=0$. In
Sections~\ref{sec11} and \ref{sec12}, we discuss  polynomial and
trigonometric interactions. We show, in Section~\ref{sec11}, that
for a finite mode approximation the expectation value of the
Feynman--Kac factor can be finite in the Wiener integral
formulation of the Feynman integral when we average the potentials
over the random fields constructed from the Gaussian WKB wave
functions in the earlier sections. In Section~\ref{sec12}, it is
shown that an inversion of the spatial signature allows to show
that the expectation value of the Feynman--Kac factor is finite
for an infinite number of modes. The inversion of the signature
may be considered either as a technical tool or as a quantum
effect of an average of the functional integral over quantum
metrics. Some supplementary material is located in the Appendices
A-H. Appendices C-E describe for the
convenience of the reader some simplified  models (discussed
earlier) of the ones studied in the main part.
Appendices A, B, F
and G contain some additions to the results in the
main text, which illustrate the method of stochastic equations
(this is just a transformation of sample paths). We have studied
the Brownian motion formulation of quantum physics for some time.
The motivation was twofold: 1. to formulate the Feynman integral
as a rigorous mathematical tool, and 2.~to make quantum
correlations susceptible to standard simulations by means of
random variables. We obtained the stochastic formulation of the
Feynman integral first in quantum
mechanics~\cite{habajp,habajmp,hababook}. The stochastic free
field is introduced in~\cite{habaepj,habaspringer,habarxiv}. The
inversion of the metric in a radiation background is discussed in
\cite{habaeuro}. In this paper, we present a comprehensive
approach to the stochastic representation of QFT (beyond the
homogeneous expanding metric), including the previously studied
models as special cases. The choice of the stochastic field is not
unique. It depends on the selection of the Gaussian WKB state. We
discuss, in detail, the form of the field which enables a
construction of a well-defined (integrable) Feynman--Kac factor
for the field theory with an interaction.

\section{The Imaginary Time Schr\"odinger Equation}\label{sec2}
First, we discuss  the standard canonical field theory in a
mathematically rigorous  imaginary time formulation
\cite{albeverio}. The Hamiltonian is defined as
\begin{equation}{\cal H}=\frac{1}{2}\int d{\bf x}\Big(
\Pi^{2}+(\nu \Phi)^{2}\Big)+\int d{\bf x}V(\Phi)={\cal H}_{0}+V,
\end{equation}where
\begin{equation}
 \nu=\sqrt{-\triangle+M^{2}}.
 \end{equation}
The canonical momentum $\Pi$ satisfies the commutation relations
with the field $\Phi$
\begin{equation}
[\Phi({\bf x}),\Pi({\bf y})]=i\hbar \delta({\bf x}-{\bf y}).
\end{equation}
Let $\psi_{t}^{g}$ be a  solution (usually the ground state) of
the imaginary time Schr\"odinger equation
\begin{equation}-\hbar\partial_{t}\psi={\cal H}\psi.
\end{equation}

Let us consider the general solution of the Schr\"odinger Equation
(4)
 with the initial condition $\psi=\psi_{0}^{g}\chi$
\begin{equation}
\psi_{t}=\psi_{t}^{g}\chi_{t}.
\end{equation}
Then, $\chi_{t}$ satisfies the equation
\begin{equation} -\hbar\partial_{t}\chi=\int d{\bf
x}\Big(\frac{1}{2} \Pi^{2}-(\Pi\ln\psi_{t}^{g})\Pi\Big)\chi
\end{equation}
with
\begin{equation}
\Pi({\bf x})=-i\hbar\frac{\delta}{\delta \Phi({\bf x})}.
\end{equation}
$\psi_{t}^{g}$ may be an arbitrary solution of the Schr\"odinger
Equation (4)  (with an arbitrary initial condition). The
efficiency of the representation ((5) and (6)) depends on the
choice of $\psi_{t}^{g}$ and the assumption that initial states
are under consideration (these will be the WKB states). It can be
seen that, in Equation (6), the kinetic term $(\nu\Phi)^{2}$, as
well as the potential $V$, are absent. Equation (6) is a diffusion
equation in infinite dimensional spaces \cite{daletski}. The
solution of Equation (6) can be expressed as \cite{freidlin}
\begin{equation}
\chi_{t}(\Phi)=E\Big[\chi\Big(\Phi_{t}(\Phi)\Big)\Big],
\end{equation}
where $\Phi_{t}(\Phi)$ is the solution of the stochastic equation
($t\geq s\geq 0$)
\begin{equation}
d\Phi_{s}({\bf x})=\hbar\frac{\delta}{\delta \Phi_{s}({\bf
x})}\ln\psi_{t-s}^{g}dt+\sqrt{\hbar}dW_{s}({\bf x})
\end{equation}
with the initial condition $\Phi$.
 $E[\ldots]$ denotes an expectation value with respect to the Wiener
process (Brownian motion) with the mean zero and the covariance
($t\geq 0$, $s\geq 0$)
\begin{equation}
E\Big[W_{t}({\bf x})W_{s}({\bf y})\Big]=min(t,s)\delta({\bf
x}-{\bf y}).
\end{equation}
The correlation functions of the quantum Euclidean  field in the
state $\psi_{0}^{g}$ can be expressed by the correlation functions
of the stochastic process $\Phi_{t}$. If we could find a
particular solution of the Schr\"odinger equation and solve the
non-linear stochastic differential Equation (9), then the problem
of solving QFT and  calculating the field correlation functions
could be reduced to
a calculation of expectation values with respect to the Wiener process. We do not know any solution of the Schr\"odinger equation for  scalar field theory (for a potential $V(\Phi)$ which is not quadratic). However, in field theories with large symmetry, this could be possible (let us mention Chern--Simons 
 states in gauge theories \cite{witten} and Kodama states in gravity \cite{kodama}). The imaginary time in this section has a rigorous mathematical formulation
for super-renormalizable field theories in two dimensions
\cite{albeverio}. In general, for arbitrary states $\psi_{t}^{g}$,
higher dimensions and real time, we expect difficulties with a
derivation of solutions of Equation (9) and their renormalization.
 We can manage in this paper linear stochastic equations (corresponding to Gaussian $\psi_{t}^{g}$) for free field theory.
Then, the interaction is introduced as usual by means of the
Feynman--Kac~formula.

 Let us consider the simplest
example: the free field. Then, the ground state is
\begin{equation}
\psi^{g}=Z^{-1}\exp(-\frac{1}{2\hbar}\Phi\nu\Phi),
\end{equation}
where $Z$ is the state normalization.
 The stochastic Equation (9)
reads
\begin{equation}
d\Phi_{t}=-\nu\Phi_{t} dt+\sqrt{\hbar}dW_{t}.
\end{equation}
The solution is (with the initial condition $\Phi$ at $t_{0}$)
\begin{equation}
\Phi_{t}=\exp(-\nu(t-t_{0}))\Phi+\sqrt{\hbar}\int_{t_{0}}^{t}
\exp(-\nu(t-s))dW_{s}.\end{equation} We calculate
\begin{equation}\begin{array}{l} \int d\Phi \psi_{g}^{2}
E\Big[\exp(\int dt d{\bf x}f_{t}({\bf x})\Phi_{t}({\bf x}))\Big]
\cr=\exp\Big(\frac{1}{2}\int dtdt^{\prime}
\Big(f_{t},(2\nu)^{-1}\exp(-\nu \vert
t-t^{\prime}\vert)f_{t^{\prime}}\Big)\Big)
\end{array}\end{equation}
On the rhs of Equation (14), we have the generating functional for
the correlation functions of the quantum Euclidean free field.

 As a
time-dependent solution of Equation (4) (with $V=0$), we may
consider
\begin{equation}
\psi_{t}^{g}=A(t)\exp(i\frac{1}{2\hbar}(\Phi\Gamma_{t}\Phi)).
\end{equation}
In Equation (15), $\Gamma$ is an operator with an integral kernel
$\Gamma({\bf x},{\bf y})$.
 We can derive an equation for this operator, demanding that (15) is the solution of the
 Schr\"odinger Equation (4).

Then, $\psi_{t}^{g}$ is the solution of the  free imaginary time
Schr\"odinger Equation (4) ($V=0$) if
\begin{equation}
	i\partial_{t}\Gamma+\Gamma^{2}+\nu^{2}=0.
\end{equation}
Equation (16) is equivalent to
\begin{equation}
	(\partial_{t}^{2}-\nu^{2})u=0
\end{equation}
if
\begin{equation}
	-i\Gamma=u^{-1}\frac{d}{dt}u.
\end{equation} 
The general solution of Equation (17) is
\begin{equation}
u=C_{1}\sinh(\nu t)+C_{2}\cosh(\nu t)
  \end{equation}
If $C_{1}=C_{2}$, then, from Equations (15),(18) and (19)
 we obtain the ground state solution (11).
 We discuss the case $C_{1}=0$ in Appendix A.
 It defines another field $\Phi_{t}$ whose correlation functions
 are equal to the ones of the standard free Euclidean field, but in
 another time-dependent state (15). We obtain another realization of
 the solution (5) of the Schr\"odinger equation for the free field.

With the potential $V$ in Equation (1), the solution of the
Schr\"odinger Equation (4) reads \cite{simonbook,freidlin} (the
Feynman--Kac formula
 requires $V(\Phi)$ to be bounded from below)\begin{equation} \chi_{t}(\Phi)= E\Big[
\exp\Big(-\frac{1}{\hbar}\int_{0}^{t}V(\Phi_{s})ds\Big)\chi(\Phi_{t}(\Phi))\Big].
\end{equation}
At the end of this section, we wish to point out some problems
with the definition of the Feynman--Kac integral (20) in Euclidean
field theory (even if $V(\Phi)$ is bounded from below). We
consider in subsequent sections exponential potentials
\begin{displaymath}
V(\Phi)=\lambda\int d{\bf x}\exp(\alpha\Phi({\bf x})),
\end{displaymath}
where $B\subset R^{d}$ is a bounded region in $R^{d}$. It can be
seen that higher orders of the perturbation expansion in $\lambda$
of the normal ordered exponentials  in the Feynman--Kac Formula
(20) are divergent if the dimension of the space-time is $d\geq
3$, e.g., in the second order ($:-:$    denotes the normal
    ordering), we obtain
\begin{displaymath}\begin{array}{l} \lambda^{2}\int_{0}^{t} ds \int_{0}^{t}d\tau \int_{B}d{\bf
x}\int_{B}d{\bf y}E\Big[:\exp(\alpha\Phi_{s}({\bf x})):
:\exp(\alpha\Phi_{\tau}({\bf y})):\Big]\cr=\lambda
^{2}\int_{0}^{t} ds \int_{0}^{t}d\tau \int_{B}d{\bf
x}\int_{B}d{\bf y}\exp\Big(\alpha^{2}E\Big[\Phi_{s}({\bf
x})\Phi_{\tau}({\bf y})\Big]\Big).\end{array}
\end{displaymath}
The two-point function is positive and at short distances
\begin{displaymath}
E\Big[\Phi_{s}({\bf x})\Phi_{\tau}({\bf y})\Big]\simeq \Big(
\Big((s-\tau)^{2}+({\bf x}-{\bf y})^{2}\Big)^{1-\frac{d}{2}}.
\end{displaymath}
The $\lambda^{2}$ term of the perturbation series of Equation (20)
is infinite if $\alpha$ is real and $d\geq 3$ (this follows from
$\exp (x)\geq 1+\frac{x^{4}}{4!}$ for $x\geq 0$).

If $\alpha=i\beta$ is purely imaginary, then
\begin{displaymath}\begin{array}{l} \lambda^{2}t^{2}\vert B\vert^{2}\geq\lambda^{2}\int_{0}^{t} ds \int_{0}^{t}d\tau \int_{B}d{\bf
x}\int_{B}d{\bf y}E\Big[:\exp(i\beta\Phi_{s}({\bf x})):
:\exp(i\beta\Phi_{\tau}({\bf y})):\Big]\cr= \lambda
^{2}\int_{0}^{t} ds \int_{0}^{t}d\tau \int_{B}d{\bf
x}\int_{B}d{\bf y}\exp\Big(-\beta^{2}E\Big[\Phi_{s}({\bf
x})\Phi_{\tau}({\bf y})\Big]\Big) \cr\geq\lambda ^{2} t^{2}\vert
B\vert^{2}\exp\Big(-\beta^{2}t^{-2}\vert B\vert^{-2}\int_{0}^{t}
ds \int_{0}^{t}d\tau \int_{B}d{\bf x}\int_{B}d{\bf
y}E\Big[\Phi_{s}({\bf x})\Phi_{\tau}({\bf y})\Big]\Big)
\end{array}
\end{displaymath}
from Jensen inequality. We obtain an upper bound and a non-zero
lower bound if $d<6$. However, if we form real potentials as,
e.g., $\lambda \cos\beta\Phi$, then at the order $\lambda^{2}$,
there will be terms without an upper bound because of the
multiplication of the terms $\exp(i\beta\Phi({\bf x}))$ with
$\exp(-i\beta\Phi({\bf y}))$.

\section{Real Time: The Free Field on a Manifold}\label{sec3}

We consider a globally hyperbolic manifold \cite{hyperbolic} with
a choice of coordinates such that  the metric is of the form
\begin{equation} ds^{2}=g_{00}dx^{0}dx^{0}-g_{jk}dx^{j}dx^{k}.
\end{equation}

 The Lagrangian of the free field is \cite{fulling}
\begin{equation}
L=\frac{1}{2}\sqrt{-g}g^{\mu\nu}\partial_{\mu}\phi\partial_{\nu}\phi-\frac{M^{2}}{2}\sqrt{-g}\phi^{2},
\end{equation} where $g=\det(g_{\mu\nu})$.
The canonical momentum is
\begin{equation}
 \Pi=g^{00}\sqrt{-g}\partial_{0}\phi.
\end{equation}
From the Lagrangian (22), we drive the canonical  Hamiltonian
\begin{equation}\begin{array}{l} {\cal H}(t)\equiv \int d{\bf x}{\cal H}(g,{\bf x})=\int d{\bf x}\Big(\Pi\partial_{0}\phi
-L\Big) \cr=\int d{\bf
x}\Big(\frac{1}{2}g_{00}\frac{1}{\sqrt{-g}}\Pi^{2}
+\frac{1}{2}\sqrt{-g}g^{jk}\partial_{j}\phi\partial_{k}\phi+\frac{M^{2}}{2}\sqrt{-g}\phi^{2}\Big).
\end{array}\end{equation}

In subsequent sections, we shall discuss the Lagrangian (22) and
the Hamiltonian (24) for the metric tensors $g_{\mu\nu}$, which
can arise as saddle points in the Feynman path integral. Such
metrics satisfy Einstein equations, but they do not fulfill the
requirement $-g>0$. We must choose the square root  $\sqrt{-g}$ in
such a way (see  \cite{witten,gibbons,turok,sorkin,visser,segal})
that  the path integral and the Schr\"odinger equation
\begin{equation}
i\hbar\partial_{t}\psi_{t}={\cal H}(t)\psi_{t}
\end{equation}
are well-defined. The solution of Equation (25) must define an
operator which is a contraction in a Hilbert space (otherwise
$\psi_{t}$ may have an infinite norm). We assume (as in
\cite{gibbons}) that, for $t\geq 0$, the metric is real and
Lorentzian. In the past, the stationary points of the Lagrangian
with matter satisfying some positive energy conditions
\cite{ellis-hawking} will necessarily exhibit a Big Bang
singularity (then, e.g., $\frac{1}{\sqrt{-g}}$ is not defined). We
assume that the solutions of Einstein equations have a
continuation to $t<0$, but do not satisfy the requirements of the
classical general relativity (they may be Euclidean or even
complex). So, for positive time, we shall have the Schr\"odinger
equation, whereas for negative time with the Hamiltonian (24), a
diffusion-type Equation (a contraction \cite{yoshida}). The
necessary condition for a contraction  at $t<0$ is that the
infinite dimensional diffusion generator has the imaginary part of
the second order differential operator, which is a positive
operator. From the Hamiltonian (24), we can see that this will be
the case if

$g_{00}\frac{1}{\sqrt{-g}}=R+iI$ where $I\geq 0$,

where $R$ and $I$ are the real and imaginary parts of a complex
function. In such a  case, we write ${\cal H}=i{\cal H}_{E}$  and
write the Schr\"odinger Equation (25) as a diffusion equation
\vspace{-6pt}
\begin{equation}
\hbar\partial_{t}\psi_{t}={\cal H}_{E}(t)\psi_{t}
\end{equation}
In our models, we choose $\sqrt{-g}=-i\sqrt{\vert g\vert}$. Then,
${\cal H}_{E}$ will be (on a formal level) a Hermitian positive
operator. Hence, Equation (26) defines a contraction (diffusion)
for $t<0$. For the inverted metric, the mass term has an opposite
sign to the kinetic term. So, we invert its sign as
$M^{2}\rightarrow -\mu^{2}$.

 In Section~\ref{sec10}, we discuss
a solution for a homogeneous radiation metric $a^{2}\simeq t$
which leads to an infinite energy density when $t\rightarrow 0$,
but it is also a solution of Einstein equations at $t<0$ with
$a^{2}<0$. Examples which have a continuation from the real time
in  de Sitter space to the four-sphere  at the imaginary time have
been discussed in \cite{gibbons,hartle,halliwel}.

Let us still mention another interpretation of the
Hamiltonian (24) for an inverted~metric
\begin{equation}\begin{array}{l} {\cal H}(t)=\int d{\bf x}\Big(\Pi\partial_{0}\phi
-L\Big) \cr=\int d{\bf
x}\Big(\frac{1}{2}g_{00}\frac{1}{\sqrt{\vert g\vert}}\Pi^{2}
+\frac{1}{2}\sqrt{\vert
g\vert}g^{jk}\partial_{j}\phi\partial_{k}\phi+\frac{M^{2}}{2}\sqrt{\vert
g\vert}\phi^{2}\Big).
\end{array}\end{equation}
Such a modification of the Hamiltonian (24)  transforms the
Schr\"odinger evolution with the Lorentzian metric into the one
with an Euclidean metric. It is still unitary. The inversion of
the spatial  metric in Equation (27) (so that $g^{jk}$ is
negatively definite) at $t=0$ is an analog of a transformation of
the oscillator for $t\geq 0$ into an upside-down oscillator for
$t< 0$ \cite{guth,barton,habarxiv}. The Hamiltonian (27) results
from the replacement $\sqrt{-g}\rightarrow \sqrt{\vert g\vert}$ in
the Lagrangian (22) as a possible candidate for quantum gravity.

 Inserting  the WKB wave function (15) in Equation (25), we obtain an
equation for the operator $\Gamma$ and the normalization
coefficient $A$
\begin{equation}
\partial_{t}\Gamma_{t}+\Gamma_{t}{\cal J}\Gamma_{t}+{\cal M}=0,
\end{equation}
\begin{equation}
\partial_{t}\ln A=-\frac{1}{2}\int d{\bf x}\frac{g_{00}}{\sqrt{-g}}\Gamma_{t}({\bf x},{\bf
x}),
\end{equation}
where
\begin{equation}
{\cal J}({\bf y},{\bf
y}^{\prime})=\frac{g_{00}}{\sqrt{-g}}\delta({\bf y},{\bf
y}^{\prime})
\end{equation}
and ${\cal M}$ is the differential operator
\begin{equation}
{\cal M}={\cal
J}^{-1}K^{2}=M^{2}\sqrt{-g}-\partial_{j}\sqrt{-g}g^{jk}\partial_{k}.
\end{equation}where the operator \cite{fulling}
\begin{displaymath}
K^{2}=-g_{00}\frac{1}{\sqrt{-g}}\partial_{j}g^{jk}\sqrt{-g}\partial_{k}+g_{00}M^{2}
\end{displaymath}
is self-adjoint in $L^{2}(d\mu)$ with respect to the measure
\begin{displaymath}
d\mu=g^{00}\sqrt{-g}d{\bf x}.
\end{displaymath}

Let us define the operator
\begin{equation}
{\cal G}_{t}=\exp(\int^{t}{\cal J}_{s}\Gamma_{s}ds).
\end{equation}
$\Gamma_{t}$ can be expressed by ${\cal G}_{t}$

\begin{equation}
\Gamma_{t}={\cal J}_{t}^{-1}\partial_{t}{\cal G}{\cal
G}^{-1}={\cal J}_{t}^{-1}\partial_{t}\ln {\cal G}.
\end{equation}
${\cal G}_{t}$ satisfies a linear operator equation
\begin{equation}
\partial_{t}^{2}{\cal G}=\partial_{t}{\cal J}{\cal J}^{-1}\partial_{t}{\cal G}-{\cal J}{\cal M}{\cal
G}.
\end{equation}
Let us note that $\sqrt{-g}$ cancels in ${\cal J}{\cal M}$ in
Equation (34). Hence, this equation does not depend on the
interpretation of $\sqrt{-g}$ for an inverted metric. It can be
shown that the operator Equation (34) coincides with the wave
equation corresponding to the Lagrangian~(22). Then, as discussed
in Section~\ref{sec2}, $\Gamma\Phi$ is the drift in the stochastic
Equation (9). For the general metric (21), it is not simple to
find a solution $\Gamma$. In subsequent sections, we find explicit
solutions assuming the invariance either with respect to the de
Sitter group or under translations (homogeneous spatially flat
manifold). Some information about $\Gamma$ is needed in order to
determine whether $\psi_{t}^{g}$ is square integrable. We need to
know $\Gamma$  for a construction of an interaction in such a way
that the Feynman--Kac factor has a finite expectation value.

 We express  the Feynman path integral solution of the
   Schr\"odinger Equation (25)
($V=0$)
 with the initial condition
     \begin{equation}
     \psi^{g}_{0}=\exp(\frac{i}{\hbar}S_{0})
   \end{equation}
   in the form
 \begin{equation} \begin{array}{l}
   \psi_{t}^{g}(\Phi)=\int d\Phi(.)\exp\Big(\frac{i}{2\hbar}\int d\tau d{\bf
   x}\Big(\sqrt{-
   g}g^{\mu\nu}\partial_{\mu}\Phi\partial_{\nu}\Phi-M^{2}\sqrt{ -g}\Phi^{2}\Big)
    \exp(\frac{i}{\hbar}S_{0}(\Phi_{t}(\Phi)).
   \end{array}\end{equation}
Formula (36) is well-established with real values of
   $\sqrt{-g}$ as a solution of Equation~(25)~\cite{freidlin}. However, a formal
   derivation of Equation (36) does not use any assumption on the signature
   of the metric, as long as the exponential (36) is bounded as a function of $\Phi$.
   As in the Hamiltonian Equation (25), we assume that, for $t\geq 0$, the metric
   is real and Lorentzian.
   For $t<0$, we admit complex $g_{\mu\nu}$ and complex $\sqrt{-g}$.
   If we require that the quadratic factor $(\partial_{0}\Phi)^{2}$ in the Feynman integral (36) does not grow
   for  $t<0$, then the conditions upon the metric will be
 
  $ \sqrt{-g}g^{00}=\tilde{R}+i\tilde{I}$ 
where $\tilde{I}\leq 0$.

If the spatial part in the action in the exponential (36) is to be
positive definite for $t<0$, then

$\sqrt{-g}g^{jl}=R^{jl}+iI^{jl}$,

where $I^{jl}$ is a positive definite matrix. If
$\sqrt{-g}=-i\sqrt{\vert g\vert}$, then the real part of the
integral of the quadratic term $(\partial_{0}\Phi)^{2}$ in
Equation (36) is negative, and if $g^{jl}$ is negatively definite
(inverted metric), then the real part of the spatial quadratic
term in (36) is also negative. The mass term has an opposite sign;
hence, if it is to be negative, we must change $M^{2}\rightarrow
-\mu^{2}$ for $t<0$. The conditions for the path integral coincide
with the ones derived from the Hamiltonian (below Equation (25)).
We shall still discuss the Schr\"odinger equations and path
integrals in detail in specific models in subsequent sections.
 The requirements for the complex metric have been discussed in
\cite{segal,witten,lehners}. They look different then the ones
required for the Hamiltonian at the beginning of this section. We
shall discuss these conditions in Sections 9 and
10 when we discuss a change in signature.

   We assume that $S_{0}$ is a quadratic form in $\Phi$
   \begin{equation}
   S_{0}(\Phi)=(\Phi,\Gamma_{0}\Phi),
   \end{equation}
 where $(,)$ denotes  the scalar product in $L^{2}(d{\bf x})$.
   We solve the Schr\"odinger Equation (25) by means of the stationary phase method.
   We expand the Feynman integral (36) around the stationary point
   $\phi_{s}^{c}(\Phi)$. The stationary point is obtained as a solution $\phi_{s}^{c}(\Phi)$
   of the Cauchy problem
    with the initial field value $\Phi$,
and the final boundary condition on the time derivative
\cite{truman}
\begin{equation}
\frac{d\phi_{t}^{c}}{dt}=-\frac{\delta
S_{0}(\phi)}{\delta\phi}(\phi_{t}^{c}).
\end{equation}
The solution $\phi_{t}^{c}(\Phi)$
   is linear in $\Phi$. We write  \begin{equation}
   \Phi_{s}=\phi_{s}^{c}(\Phi)+\sqrt{\hbar}\phi^{q}_{s}.
   \end{equation}
 Then,
   \begin{equation}
   \psi_{t}^{g}(\Phi)=A_{t}\exp(\frac{i}{\hbar}S_{t}(\phi_{c}(\Phi))
   \equiv A_{t}\exp(\frac{i}{2\hbar}(\Phi,\Gamma_{t}\Phi)).\end{equation}
  The classical action $S_{t}$
    is a bilinear form in $\Phi$ defined by a real kernel $\Gamma_{t}({\bf x},{\bf y})$
    (if $S_{0}$ is a real function),
    $A_{t}$ is expressed by the determinant of an operator
   defined by the quadratic form in $\phi^{q}$.
    The determinant depends only on time and the
    metric. It follows that if $S_{0}$ is real, then $S_{t}$ is a real bilinear form (if $S_{0}$ is complex,
then $S_{t} $ is also complex). Hence, $\psi_{t}^{g}$, as a
function of $\Phi$, is a pure phase factor.

\section{Gaussian Solution of the Schr\"odinger Equation and Linear Stochastic Equations}\label{sec4}

We approach the quantum field theory (QFT) in Minkowski space-time
by means of stochastic equations
\cite{habaspringer,habaepj,hababook}. The stochastic equations
determine the solution of the Schr\"odinger equation as in
Equation (8).
 We  continue the imaginary time  in Equations~(8) and (9) to the real
time \cite{habaepj}
  \begin{equation}
 d\Phi_{s}({\bf x})=i\hbar\frac{\delta}{\delta \Phi_{s}({\bf
 x})}\ln\psi_{t-s}^{g}dt+\sqrt{i\hbar}dW_{s}({\bf x}),
 \end{equation}
 where $\sqrt{i}=\exp(i\frac{\pi}{4})=\frac{1}{\sqrt{2}}(1+i)$. Let $\hat{\Phi}$ be the relativistic quantum free field. Then,
the generating functional of the time-ordered correlation
functions of $\hat{\Phi}$ in the vacuum $\psi^{g}$ (11) can be
expressed \cite{habaepj} by the solution $\Phi_{t}(\Phi)$ of the
stochastic Equation (41)
  \begin{equation}\begin{array}{l}(\psi_{g},T\Big(\exp\int dtd{\bf x} \hat{\Phi}_{t}({\bf x})f_{t}({\bf x})\Big)
  \psi_{g})
  =\int d\Phi \psi_{g}^{2}(\Phi)E\Big[\exp(\int dtd{\bf x}  f_{t}({\bf x})\Phi_{t}(\Phi,{\bf x}))\Big] \cr =
\exp\Big(\frac{1}{2}\int dtdt^{\prime}
 \Big(f_{t},(2\omega)^{-1}\exp(-i\omega \vert
 t-t^{\prime}\vert)f_{t^{\prime}}\Big)\Big),
 \end{array}\end{equation}
 where
 $T(\ldots)$ is the time-ordered product and $(f,g)$ denotes the scalar product in $L^{2}(d{\bf x})$ .

 When the initial condition is $\psi=\psi_{0}^{g}\chi$, and $\psi_{t}^{g}$ is the solution
  of the Schr\"odinger Equation (25), then $\chi$ solves the
equation
\begin{equation} i\hbar\partial_{t}\chi_{t}=\int d{\bf
x}\Big( \frac{1}{2}\Pi^{2}+(\Pi\ln\psi_{t}^{g})\Pi\Big)\chi_{t}.
\end{equation}
  If the solution $\psi_{t}^{g}$ of the Schr\"odinger equation
  for quantum fields defined  on the Minkowski space-time
  is of the form (40) then Equation (41) reads \cite{freidlin}
  \begin{equation}
  d\Phi_{s}=-\Gamma_{t-s}\Phi_{s}ds+\sqrt{i\hbar}dW_{s}.
\end{equation}
Let $\Phi_{s}(\Phi)$ be the solution of Equation (44) with the
initial condition $\Phi$. If $\chi$ is a holomorphic function,
then the solution of the Schr\"odinger Equation (25) is
\begin{equation}
\psi_{t}=\psi^{g}_{t}E\Big[\chi\Big(\Phi_{t}(\Phi)\Big)\Big].
\end{equation}
With the interaction $V_{t}$, the Feynman formula reads
\begin{equation}
\psi_{t}=\psi^{g}_{t}E\Big[\exp\Big(-\frac{i}{\hbar}\int_{0}^{t}V_{t-s}(\Phi_{s})ds\Big)
\chi\Big(\Phi_{t}(\Phi)\Big)\Big].
\end{equation}

The solution of the stochastic equation determines the free field
correlation functions ($V=0$)
\begin{equation}\begin{array}{l}(\psi_{0}^{g},F_{1}(\Phi_{t})F_{2}(\Phi)\psi_{0}^{g})
=\int d\Phi
\vert\psi_{t}^{g}(\Phi)\vert^{2}F_{1}(\Phi)E\Big[F_{2}\Big(\Phi_{t}(\Phi)\Big)\Big].
\end{array}\end{equation}
For more general states $\psi^{g}\chi$ and $V\neq 0$, we have

\begin{equation}\begin{array}{l}(\psi_{0}^{g}\chi,F_{1}(\Phi_{t})F_{2}(\Phi)\psi_{0}^{g}\chi)
=\int d\Phi \vert\psi_{t}^{g}(\Phi)\vert^{2}F_{1}(\Phi)\cr
E\Big[\chi\Big(\Phi_{t}(\Phi)\Big)\exp(-\frac{i}{\hbar}\int_{0}^{t}V_{t-s}(\Phi_{s})ds)
\Big]^{*}\cr\times
E\Big[F_{2}\Big(\Phi_{t}(\Phi)\Big)\chi\Big(\Phi_{t}(\Phi)\Big)
\exp(-\frac{i}{\hbar}\int_{0}^{t}V_{t-s}(\Phi_{s})ds)\Big].
\end{array}\end{equation}
We generalize these correlation functions in Section~\ref{sec7} to
multitime correlation functions after a derivation of a formula
for the evolution propagator. In principle,
  QFT (with the Hilbert space and quantum fields) can be determined
by the correlation functions  (Wightman construction).

\section{De Sitter Space}\label{sec5}

In this and subsequent sections, we obtain quantum field evolution
on the Lorentzian and Euclidean backgrounds. First, we discuss
particular solutions (de Sitter and the sphere $S^{4}$) of
Einstein equations with a cosmological constant (obtained in
\cite{gibbons,hartle}). De Sitter space-time can describe an early
inflationary stage of the universe, as well as the final stage of
an acceleration driven by dark energy. The authors
\cite{gibbons,hartle} glue together the Lorentzian solution for
positive time with the Euclidean solution ($S^{4}$) for an
imaginary time.
 We
consider several coordinate systems on de Sitter space
\cite{spradlin,gibbons} which can be considered as a submanifold
of the complex quadric
\begin{equation}
z_{1}^{2}+z_{2}^{2}+z_{3}^{3}+z_{4}^{2}+z_{5}^{2}=\frac{1}{H^{2}},
\end{equation}
where $H$ has the meaning of the Hubble constant. We first
consider a real form of the quadric (49), defining the de Sitter
space
\begin{equation}
x_{1}^{2}+x_{2}^{2}+x_{3}^{3}+x_{4}^{2}-x_{5}^{2}=\frac{1}{H^{2}}.
\end{equation}
In the coordinates $(t,\omega)$, where $\omega\in S^{3}$, the
metric on the hyper-sphere (50) is
\begin{equation}
ds^{2}=dt^{2}-\frac{1}{H^{2}}\cosh^{2}(Ht)d\omega^{2},
\end{equation}
where $d\omega^{2}$ is the metric on $S^{3}$.

In conformal coordinates \begin{equation}
\cos(\tau)=\frac{1}{\cosh(Ht)},
\end{equation}
where $ 0<\tau<\frac{\pi}{2}$, we obtain the metric
\begin{equation}
ds^{2}=\frac{1}{H^{2}\cos^{2}(\tau)}(d\tau^{2}-d\omega^{2}).
\end{equation}
The Euclidean version of the manifold (53) describes the
four-dimensional sphere of radius~$\frac{1}{H}$
\begin{displaymath}
x_{1}^{2}+x_{2}^{2}+x_{3}^{3}+x_{4}^{2}+x_{5}^{2}=\frac{1}{H^{2}}
\end{displaymath}
Then, the metric is
\begin{equation}
ds^{2}=dt^{2}+\frac{1}{H^{2}}\cos^{2}(Ht)d\omega^{2}.
\end{equation}
The introduction of conformal coordinates
\begin{equation}
\cosh(\tau)=\frac{1}{\cos(Ht)},
\end{equation}
where $-\frac{\pi}{2H}<t\leq 0$ (we choose negative time in the
Euclidean domain) gives the metric
\begin{equation}
ds^{2}=\frac{1}{H^{2}\cosh^{2}(\tau)}(d\tau^{2}+d\omega^{2}).
\end{equation}
We can also introduce spatially flat coordinates describing the
expanding universe (which will be discussed in detail in
subsequent sections)
\begin{displaymath}
ds^{2}=dt^{2}-a^{2}d{\bf x}^{2}.
\end{displaymath}
The expanding flat metric in coordinates which cover the  half of
de Sitter space-time (visible by an observer at the origin) is
\begin{equation}
ds^{2}=dt^{2}-\exp(2Ht)d{\bf x}^{2}.
\end{equation}
The ``Euclidean'' version of the metric (57) (${\bf x}\rightarrow
i{\bf x}$)
\begin{equation}
ds^{2}=dt^{2}+\exp(2Ht)d{\bf x}^{2}
\end{equation}
does not represent a metric on  the sphere $S^{4}$. This is the
metric on the hyperbolic space which is the Euclidean version of
the anti-de Sitter space-time. In fact, the metric (58) can be
expressed in a form familiar from the realization of the anti-de
Sitter space as a generalized Poincare upper half-plane
\begin{displaymath}
ds^{2}=y^{-2}(H^{-2}dy^{2}+d{\bf x}^{2})
\end{displaymath} with $y=\exp(-Ht)$.
The sphere $S^{4}$,
 Euclidean anti-de Sitter space  and Euclidean continuations of  de Sitter space
 are closely related
\cite{polyakov}. We are allowed to treat the coordinates $t$ and
$\tau$ in Equations (51)--(58) in Lorentzian and Euclidean metrics
as time in the Lagrangian formalism (in the Euclidean version, the
time evolution will be a rotation of the sphere).

There remains to study the Schr\"odinger Equations (25)--(27)
resulting from the definition of ${\cal H}$ in Equations (24) and
(27). First, we look for a Gaussian solution (40) of these
equations. In the coordinates (51)--(56), we can use the $O(4)$
symmetry in order to diagonalize the equation for $\Gamma$. First,
we expand the fields in the spherical harmonics \cite{kirsten}
\begin{displaymath}
\Phi(\tau,\omega)=\sum_{lm}Y_{lm}(\omega)\Phi_{lm}(\tau),
\end{displaymath}
 where
\begin{displaymath}
\triangle_{S}Y_{lm}=-l(l+2)Y_{lm}.
\end{displaymath}
 $\triangle_{S}$ is the Laplace--Beltrami operator on $S^{3}$, $l$ is a natural number, and $m=(j,\sigma)$ where $j=0,1,\ldots,l$
and $-j\leq\sigma\leq j$, is the indexing solution of the
Laplace--Beltrami operator with the eigenvalue $-l(l+2)$
\cite{vilenkin}.

 Then,
we expand $\Gamma$ as
\begin{equation}
\Gamma_{\tau}(\omega,\omega^{\prime})=\sum_{lm}Y_{lm}(\omega^{\prime})Y^{*}_{lm}(\omega)\Gamma_{lm}(\tau).
\end{equation}
Using Formula (24), we can define the Hamiltonian in each of the
metrics (51)--(58). Subsequently, solving Equation (25), we find a
Gaussian solution (40) of the Schr\"odinger equation in each of
these coordinates.
 We obtain for the metric (53) the Hamiltonian (24)
\begin{equation}\begin{array}{l} {\cal H}= \sum_{lm}
\Big((H\cos(\tau))^{2}\Pi_{lm}^{2}+(H\cos(\tau))^{-2}l(l+2)\Phi_{lm}^{2}
+M^{2}(H\cos(\tau))^{-4}\Phi_{lm}^{2}\Big)\end{array}
\end{equation}
defining for $t\geq 0$ the Schr\"odinger Equation (25).

The Hamiltonian for the Euclidean metric (56) in the
interpretation (27) is analogous to the upside-down oscillator
\cite{barton,guth,habarxiv} (we change $M^{2}\rightarrow
-\mu^{2}$; the Schr\"odinger equation is still
$i\hbar\partial_{\tau}\psi={\cal H}\psi$)

\begin{equation}\begin{array}{l}{\cal H}= \sum_{lm}
\Big((H\cosh(\tau))^{2}\Pi_{lm}^{2}-(H\cosh(\tau))^{-2}l(l+2)\Phi_{lm}^{2}
-\mu^{2}(H\cosh(\tau))^{-4}\Phi_{lm}^{2}\Big)
\end{array}\end{equation}

We still consider the Schr\"odinger Equation (26). In the
Euclidean metric (56) with the Hamiltonian (24) and the choice
$\sqrt{-g}=-i\sqrt{\vert g\vert}$, we obtain the diffusion
Equation (26) for $\tau <0$  with \vspace{-12pt}

\begin{equation}\begin{array}{l}{\cal H}_{E}= \sum_{lm}
\Big((H\cosh(\tau))^{2}\Pi_{lm}^{2}+(H\cosh(\tau))^{-2}l(l+2)\Phi_{lm}^{2}
+\mu^{2}(H\cosh(\tau))^{-4}\Phi_{lm}^{2}\Big)
\end{array}\end{equation}

so that the Schr\"odinger equation takes the form
\begin{displaymath}
\hbar\partial_{t}\psi={\cal H}_{E}\psi.
\end{displaymath}
The operators $\Gamma$ (28) and ${\cal G}$ (32) are diagonalized
by an expansion is spherical functions. We denote a function
satisfying the wave equation (34) for ${\cal G}$ by
$u_{\tau}(\omega)$. Expanding  $u_{\tau}(\omega)$
\begin{displaymath}u(\tau,\omega)=\sum_{lm}Y_{lm}(\omega)u_{lm}(\tau)\end{displaymath}
we obtain that in the metric (53) the coefficients $u_{lm}$
satisfy the equation
\begin{equation}
\partial_{\tau}^{2}u_{lm}+2\tan(\tau)
\partial_{\tau}u_{lm}+(l(l+2)+M^{2}(H\cos(\tau))^{-2})u_{lm}=0.
\end{equation}
$\Gamma$ is related to $u$, as follows from Equation (32)
\begin{equation}
u=\exp\Big(\int^{\tau}ds (H\cos(s))^{2}\Gamma(s)ds\Big).
\end{equation}
For the Euclidean metric with $\sqrt{-g}\rightarrow \sqrt{\vert
g\vert}$, the corresponding formulas read (this equation also
follows from the Hamiltonian (61))
\begin{equation}\begin{array}{l}
\partial_{\tau}^{2}u^{E}_{lm}-2\tanh(\tau)
\partial_{\tau}u^{E}_{lm}-l(l+2)u^{E}_{lm}-\mu^{2}(H\cosh(\tau)^{-2}u^{E}_{lm}=0
\end{array}\end{equation}
Note that the definition $\sqrt{-g}\rightarrow -i\sqrt{\vert
g\vert}$  does not change the ``wave equation'' (65) for the
inverted signature. With the inverted signature, the ``wave
equation'' becomes an elliptic equation, hence it does not
describe a wave propagation anymore.

 $\Gamma^{E}$ is related to $u^{E}$ by
\begin{equation}
u^{E}=\exp\Big(i\int^{\tau}ds (H\cosh(s))^{2}\Gamma^{E}(s)ds\Big).
\end{equation}
Concerning the flat expanding metric of the general form
\begin{equation}
ds^{2}=g_{\mu\nu}dx^{\mu}dx^{\nu}\equiv dt^{2}-a^{2}(t)d{\bf
x}^{2}
\end{equation}
introduced first for de Sitter in Equation (57) (and further
discussed in models of subsequent sections) we note that Friedmann
equations governing the evolution of $a(t)$
\begin{displaymath}
(a^{-1}\frac{d}{dt}a)^{2}-\frac{1}{3}\Lambda=\frac{8\pi G}{3}\rho,
\end{displaymath}
\begin{displaymath}2a^{-1}\frac{d^{2}a}{dt^{2}}+(a^{-1}\frac{d}{dt}a)^{2}
-\Lambda=-8\pi G p
\end{displaymath}are invariant under   $a^{2}\rightarrow
-a^{2}$. Here, $\Lambda$ is the cosmological constant
($H=\sqrt{\frac{\Lambda}{3}}$), $\rho$ is the energy density and
$p$ is the pressure. This transformation  can  equivalently  be
treated as $a\rightarrow ia$. The physical interpretation forces
us to choose as a solution the metric with $a^{2}>0$ (such a
requirement may be not applicable in quantum gravity).

For the homogeneous metric (67), owing to the translation
invariance, we can decompose $\Gamma$ in Fourier components
 \begin{displaymath}
\Gamma({\bf x}-{\bf y})=(2\pi)^{-3}\int d{\bf k}\Gamma({\bf
k})\exp(i{\bf k}({\bf x}-{\bf y})).
\end{displaymath}
We consider solutions satisfying the condition $\Gamma({\bf
k})=\Gamma(-{\bf k})=\Gamma(k)$, where
$k=\vert{\bf k}\vert$.
 Then, in Fourier transform Equation (28)
in an expanding metric is
\begin{displaymath}
\partial_{t}\Gamma+a^{-3}\Gamma^{2}+a{\bf k}^{2}+M^{2}a^{3}=0.
\end{displaymath}
We Fourier transform the wave function
\begin{displaymath}
u({\bf x})=(2\pi)^{-\frac{3}{2}}\int d{\bf k}\exp(-i{\bf
kx})u_{k}.
\end{displaymath}
Then, the  wave Equation (34) is
\begin{equation}
\partial_{t}^{2}u_{k}+3a^{-1}\partial_{t}a
\partial_{t}u_{k}+a^{-2}k^{2}u_{k}+M^{2}u_{k}=0.
\end{equation}
The relation between $u$ and $\Gamma$  is determined by Equation
(32)
\begin{equation} u_{k}=\exp(\int^{t}ds a^{-3}\Gamma_{s}(k))
\end{equation}
When $a^{-2}<0$ for $t<0$, we chose $\sqrt{-g}=-i\sqrt{\vert
g\vert}=-i \vert a^{2}\vert^{\frac{3}{2}}$.  The Schr\"odinger
equation for $t>0$ is
\begin{equation}\begin{array}{l}
i\hbar\partial_{t}\psi_{t}= \frac{1}{2}\int d{\bf x}
\Big(-\hbar^{2}a^{-3}\frac{\delta^{2}}{\delta\Phi({\bf
x})^{2}}+a(\nabla\Phi)^{2}+M^{2}a^{3}\Phi^{2}\Big)\psi_{t}\end{array}
\end{equation}
whereas for negative time, when $a^{2}<0$ and $\sqrt{-g}$ is
imaginary, we have the diffusion Equation (26) (as discussed in
Section~\ref{sec10})
\begin{equation}\begin{array}{l}
\hbar\partial_{t}\psi_{t}= \frac{1}{2}\int d{\bf x}
\Big(-\hbar^{2}\vert
a^{2}\vert^{-\frac{3}{2}}\frac{\delta^{2}}{\delta\Phi({\bf
x})^{2}}+\sqrt{\vert a^{2}\vert}(\nabla\Phi)^{2}+\mu^{2}\vert
a^{2}\vert^{\frac{3}{2}}\Phi^{2}\Big)\psi_{t}\end{array}
\end{equation}
where, for negative time, we changed the notation
$M^{2}=-\mu^{2}$, suggesting that $M^{2}$ should be chosen to be
negative.

If in the Lagrangian (22) $\sqrt{-g}\rightarrow \sqrt{\vert
g\vert}$, then the Schr\"odinger equation reads
\begin{equation}\begin{array}{l}
i\hbar\partial_{t}\psi_{t}= \frac{1}{2}\int d{\bf x}
\Big(-\hbar^{2}(\vert
a^{2}\vert)^{-\frac{3}{2}}\frac{\delta^{2}}{\delta\Phi({\bf
x})^{2}}+a^{-2}(\vert
a^{2}\vert)^{\frac{3}{2}}(\nabla\Phi)^{2}-\mu^{2}(\vert
a^{2}\vert)^{\frac{3}{2}}\Phi^{2}\Big)\psi_{t}\end{array}
\end{equation}
Equation (72) is an analog of the one for an inverted oscillator
(see \cite{habarxiv}, so that the evolution is still unitary).

 We note that from Equation (68) that it follows that the Wronskian
is a constant as
\begin{displaymath}
\partial_{t}(a^{3}(u\partial_{t}u^{*}-u^{*}\partial_{t}u))=0
\end{displaymath}
For complex solutions, we choose  the normalization
\begin{equation}
u\partial_{t}u^{*}-u^{*}\partial_{t}u=-ia^{-3}.
\end{equation}
{which is fixing the constant in canonical commutation relations.
 If $a=\exp(Ht)$, as in Equation~(57) then the (complex) solution of Equation (68)
is the cylinder function
$Z_{\nu}$~\cite{gradshtein,tagirov,schom,bunch}}
\begin{equation}
u=a^{-\frac{3}{2}}Z_{\nu}(\frac{k}{H}\exp(-Ht)),
\end{equation}
where
\begin{displaymath}
\nu=\frac{3}{2}\sqrt{1-\frac{4M^{2}}{9H^{2}}}.
 \end{displaymath}
In general, as solutions of the wave equation we can take
superpositions of solutions with different $k$. The canonical
quantization is realized with the requirement \cite{tagirov} that,
in the remote past $t_{0}\rightarrow -\infty$, the solution tends
to the plane wave (in conformal coordinates). Then,
$Z_{\nu}=H^{(2)}_{\nu}$, where $H^{(2)}_{\nu}$ is the Hankel
function of the second kind \cite{gradshtein}.  If $M=0$, then we
can obtain real solutions of Equation (68)
$J_{\frac{3}{2}}(\frac{k}{H}\exp(-Ht))$ and
$Y_{\frac{3}{2}}(\frac{k}{H}\exp(-Ht))$, important for the
construction of interactions in Sections~\ref{sec11}
and~\ref{sec12}. These real solutions will give oscillatory
evolution kernels with caustic singularities.

 The solution of Equation (68) with an inverted metric ($k\rightarrow ik$ and
$M\rightarrow i\mu$ in Equation (68)) is
\begin{displaymath}
u=a^{-\frac{3}{2}}J_{\nu}(i\frac{k}{H}\exp(-Ht))=Ca^{-\frac{3}{2}}K_{\nu}(\frac{k}{H}\exp(-Ht))
\end{displaymath}with a certain constant $C$, and
\begin{displaymath}
\nu=\frac{3}{2}\sqrt{1+\frac{4\mu^{2}}{9H^{2}}}
 \end{displaymath}
It can be seen that this is a solution of the Euclidean ``wave
equation'',  corresponding to the field theory on the Euclidean
version of the anti-de Sitter space \cite{polyakov,ads}.

Summarizing, the aim of this section was to reduce the general
Equations (28)--(34) to a manageable form using the symmetry of
the manifold. In the homogeneous expanding coordinates, we can use
the Fourier transform to represent the operator $\Gamma$ as a
multiplication operator in the Fourier space. In angular
coordinates (covering the whole of de Sitter space), we can expand
the solution in terms of spherical harmonics. A change in the
spatial signature in angular coordinates transforms de Sitter
space into a sphere. The Hamiltonian and the solution of the
Schr\"odinger equation are expressed in terms of a discrete set of
variables. The quantization of these variables (as outlined in
Sections~\ref{sec2}--\ref{sec4}) is achieved by stochastic
equations in the next section.

\section{Stochastic Equations for  de Sitter Field and Fields in an Expanding Flat Metric}\label{sec6}

In this and in subsequent sections, we discuss the stochastic time
evolution for positive, negative and imaginary time. Until now,
only positive time was considered in the stochastic representation
(45), because the Brownian motion is defined for a positive time.
We can obtain a stochastic representation for a negative time,
taking the complex conjugation of Equation (25)
\begin{equation}
i\hbar\partial_{-t}\psi_{t}^{*}={\cal H}^{*}\psi_{t}^{*}\equiv
\tilde{H}(-t)\psi^{*}_{t}.\end{equation} If the Schr\"odinger
Equation (25) is to be defined for  positive and  negative time,
then the expressions for the Hamiltonian as a function of the
metric tensor must have a meaning in this range of time. This may
be not possible if Hawking--Penrose positivity conditions of the
energy-momentum are to be satisfied \cite{ellis-hawking} (then the
metric tensor may become degenerate and $\frac{1}{\sqrt{-g}}$
infinite).
  The Einstein equations
are invariant under the time reflection. However, the reflected
metric can violate the requirement of $-g>0$, as will be discussed
in Section~\ref{sec10}. In such a case, $\tilde{{\cal H}}(-t)$ for
$t>0$ in the interpretation (24) does not define a Hermitian
operator.  In fact, in Section~\ref{sec10}, ${\cal H}(-t)$ will be
anti-Hermitian. In such a case, the Schr\"odinger equation for a
positive time is transformed into a diffusion equation for a
negative time.

 In general, on a globally hyperbolic manifold, if we
find $\Gamma_{t}$ from Equation (28), then the stochastic equation
generated by the Hamiltonian (24) reads (where ${\cal J}$ is
defined in Equation (30))\begin{equation} d\Phi_{s}=-{\cal
J}(t-s)\Gamma_{t-s}\Phi_{s}ds+\sqrt{i\hbar}\sqrt{g_{00}}(-g)^{-\frac{1}{4}}(t-s)dW_{s}.
\end{equation}
For a negative time,  Equation (76) reads
\begin{equation} d\Phi_{s}={\cal
J}(t-s)\Gamma_{t-s}\Phi_{s}d(-s)+\sqrt{-i\hbar}\sqrt{g_{00}}(-g)^{-\frac{1}{4}}(t-s)dW_{-s}.
\end{equation}
With $\sqrt{-g}=-i\sqrt{\vert g\vert}$  and
$(-g)^{\frac{1}{4}}=\sqrt{-i}\vert g\vert^{\frac{1}{4}}$ in
Equation (77) for $-g<0$. Such a choice of square roots  will give
the real noise ($\sqrt{-i}$ cancels) and a positive operator
${\cal H}^{E}$ in the diffusion Equation (26).

Equation (77) can be rewritten as
\begin{equation} d\Phi_{s}={\cal
G}^{-1}\partial_{t}{\cal
G}(t-s)\Phi_{s}d(-s)+\sqrt{\hbar}\sqrt{g_{00}}\vert
g\vert^{-\frac{1}{4}}(t-s)dW_{-s}.
\end{equation}
where ${\cal G} $ is the solution of the wave Equation (34) (with
an inverted metric).
 Equation~(76) is a generalization
of Equation (44) (considered in the Minkowski space-time).

In the expanding metric (67), the Hamiltonian (24) for $t>0$ is
\begin{equation}
{\cal H}=\frac{1}{2}\int d{\bf x}
\Big(-\hbar^{2}a^{-3}\frac{\delta^{2}}{\delta\Phi({\bf
x})^{2}}+a(\nabla\Phi)^{2}+M^{2}a^{3}\Phi^{2}\Big).
\end{equation}
The stochastic Equation (76) for $t\geq 0$ reads
 \begin{equation}d\Phi_{s}=-a(t-s)^{-3}\Gamma(t-s)\Phi_{s}
 ds+a(t-s)^{-\frac{3}{2}}\sqrt{i\hbar}dW_{s}.
 \end{equation}
By a  differentiation of Equation (80), we obtain a random wave
equation
 \begin{equation}\begin{array}{l}
 (\partial_{s}^{2}-a^{-2}(t-s)\triangle+M^{2})\Phi_{s}
 +3a^{-1}(t-s)\partial_{t}a(t-s)\partial_{s}\Phi_{s}\cr
 =\Big(\frac{9}{2} a^{-1}(t-s)\partial_{t}a(t-s)
 -a^{-3}(t-s)\Gamma(t-s)\Big)\sqrt{i\hbar}\partial_{s}W\cr
 +\sqrt{i\hbar}a^{-\frac{3}{2}}(t-s)\partial^{2}_{s}W.
\end{array}\end{equation}
Equation (80) has the solution (with the initial condition $\Phi$
at $s=0$)
 \begin{equation}
 \Phi_{s}=u_{t-s}u_{t}^{-1}\Phi
 +\sqrt{i\hbar}u_{t-s}\int_{0}^{s}u_{t-\tau}^{-1}a_{t-\tau}^{-\frac{3}{2}}dW_{\tau}.
 \end{equation}
Equation (80) can be interpreted in the semi-classical
approximation if in the solution (3.20) $\Gamma$   is real (so $u$
is real). Then, in the limit $\hbar\rightarrow 0$, Equation (80)
reads

 \begin{equation}\frac{d\Phi_{s}}{ds}=-a^{-3}(t-s)\Gamma(t-s)\Phi_{s}.
 \end{equation}
Equation (83) relates the Hamilton--Jacobi limit of the
Schr\"odinger
 equation with its classical solution $\Phi_{s}$. In the standard formulation of the Hamilton--Jacobi
 theory,  if $S$ is the
 classical action, then the classical trajectory is defined by \cite{truman}\begin{displaymath}
 a^{-3}\frac{d\Phi_{s}}{ds}=\frac{\delta S}{\delta \Phi_{s}}.
\end{displaymath}

From Equation (82), the classical solution with the initial
condition $\Phi$ is
\begin{displaymath}
 \Phi_{s}=u_{t-s}u_{t}^{-1}\Phi,
 \end{displaymath}where $u_{t-s}$ is the classical solution  of the
wave equation resulting  from the Lagrangian (22).

For the negative time, the stochastic evolution is a simple
reflection if $a(t)=a(-t)$,  as appears in the effective field
theories resulting from the string theory \cite{string,string2}.
Then,  we have $\tilde{{\cal H}}(-t)= {\cal H}(t)$, and $a$ is
contracting to zero as $\vert t\vert\rightarrow 0$ and expanding
to infinity when $t\rightarrow \infty$. In such a case, the
stochastic Equation (77) takes the form \vspace{-6pt}
\begin{displaymath}
d\Phi_{s}=\Gamma(s-t)a^{-3}(t-s)\Phi_{s}d(-s)+\sqrt{-i\hbar}a^{-\frac{3}{2}}(t-s)dW_{-s},
\end{displaymath}
where $a^{-3}\Gamma=u^{-1}\partial_{t}u$. It has the solution
\begin{equation}
\Phi_{s}(\Phi)=u_{t-s}^{(-)}(u_{t}^{(-)})^{-1}-\sqrt{-i\hbar}u_{t-s}^{(-)}\int_{0}^{-s}
(u_{t+\tau}^{(-)})^{-1}a_{t+\tau}^{-\frac{3}{2}}dW_{\tau},
\end{equation}
where $u_{t}^{(-)}$ is the solution of the wave Equation (68) for
a negative time. In this case, the field $\Phi_{s}$ can be
considered as a simple reflection of the one for $s>0$. We discuss
an example of $a(t)\simeq \vert t\vert$ in Section~\ref{sec10}.

We return to de Sitter expanding metric (57). From Equation (69),
\begin{displaymath}
\Gamma=a^{3}\frac{d}{dt}\ln\Big(a^{-\frac{3}{2}}Z_{\nu}\big(\frac{k}{H}\exp(-Ht)\big)\Big).
\end{displaymath}
Hence,
\begin{displaymath}\begin{array}{l}
\Phi_{s}=\exp(-\int_{t_{0}}^{s}(a^{-3}\Gamma)(t-\tau)d\tau)\Phi+\sqrt{i\hbar}
\int_{t_{0}}^{s}\exp(-\int_{\tau}^{s}(a^{-3}\Gamma)(t-\tau^{\prime})d\tau^{\prime})
a(t-\tau)^{-\frac{3}{2}}dW_{\tau}\end{array}\end{displaymath}or
\begin{equation}\begin{array}{l}
\Phi_{s}=a(t-s)^{-\frac{3}{2}}Z_{\nu}(\frac{k}{H}\exp(-Ht+Hs))a(t-t_{0})^{\frac{3}{2}
}\cr\Big(Z_{\nu}(\frac{k}{H}\exp(-Ht+Ht_{0}))\Big)^{-1}\Phi +
\sqrt{i\hbar}a(t-s)^{-\frac{3}{2}}Z_{\nu}(\frac{k}{H}\exp(-Ht+Hs))\cr\int_{t_{0}}^{s}
\Big(Z_{\nu}(\frac{k}{H}\exp(-Ht+H\tau))\Big)^{-1}dW_{\tau}.\end{array}\end{equation}

The solution of the stochastic equation determines the field
correlation functions (from Equations (47) and (85), $\hat{\Phi}$
denotes the quantum field, we set $t_{0}=0$)
 \begin{equation}\begin{array}{l}
(\psi_{0}^{g},\hat{\Phi}_{t}({\bf k})\hat{\Phi}({\bf
k}^{\prime})\psi_{0}^{g}) =\int d\Phi
\vert\psi_{t}^{g}(\Phi)\vert^{2}\Phi({\bf k})
E\Big[\Phi_{t}(\Phi,{\bf k}^{\prime})\Big] \cr
=i(\Gamma(t)-\Gamma(t)^{*})^{-1}Z_{\nu}(\frac{k}{H})a(t)^{\frac{3}{2}
}\Big(Z_{\nu}(\frac{k}{H}\exp(-Ht))\Big)^{-1}\delta({\bf k}+{\bf
k}^{\prime}),
\end{array}\end{equation}
where we have used the covariance (from Equation (40))
\begin{equation}
\int d\Phi \vert\psi_{t}^{g}(\Phi)\vert^{2}\Phi({\bf k})\Phi({\bf
k}^{\prime})=i\hbar(\Gamma-\Gamma^{*})^{-1}\delta({\bf k}+{\bf
k}^{\prime}).
\end{equation}
In general,
\begin{displaymath}
Z_{\nu}=\beta H_{\nu}^{(1)}+\alpha H_{\nu}^{(2)},
\end{displaymath}
where $H_{\nu}$ are the Hankel functions (in order to satisfy
canonical commutation relations we must have $\vert
\alpha\vert^{2}-\vert\beta\vert^{2}=1$).

In order to calculate the rhs of Equation (87) we apply (with
$z=\frac{k}{H}\exp(-Ht)$)
\begin{equation}\begin{array}{l}
\Gamma(t)-\Gamma(t)^{*}=a^{3}Hz(Z_{\nu}^{*}Z_{\nu})^{-1}\Big(\frac{d}{dz}Z_{\nu}^{*}Z_{\nu}
-\frac{d}{dz}Z_{\nu}Z_{\nu}^{*}\Big).
\end{array}\end{equation}

The rhs of the expression (87) may have any  sign. For $\beta=0$
and $\alpha=1$ (the Bunch--Davies 
 vacuum \cite{bunch}), we have
\begin{equation}
\Gamma-\Gamma^{*}=i\Big(H_{\nu}^{(2)*}H_{\nu}^{(2)}\Big)^{-1}
\end{equation}
Then, inserting Equation (89) in Equation (86), we obtain
\begin{equation}\begin{array}{l} (\psi_{0}^{g},\hat{\Phi}_{t}({\bf
k})\hat{\Phi}({\bf k}^{\prime})\psi_{0}^{g}) =\hbar
H^{(2)}_{\nu}(\frac{k}{H})a(t)^{\frac{3}{2}
}\Big(H^{(2)}_{\nu}(\frac{k}{H}\exp(-Ht)\Big)^{*} \delta({\bf
k}+{\bf k}^{\prime}).\end{array}\end{equation} From Equation (47),
using Equation (90), we can also calculate (for the massless
field) \vspace{-6pt}
\begin{displaymath}\begin{array}{l}
(\psi_{0}^{g},\hat{\Phi}_{t}({\bf k})\hat{\Phi}_{t}({\bf
k}^{\prime})\psi_{0}^{g})=(\psi_{t}^{g},\Phi({\bf k})\Phi({\bf
k}^{\prime})\psi_{t}^{g})\cr=\hbar \delta({\bf k}+{\bf
k}^{\prime})H_{\frac{3}{2}}^{(2)*}H_{\frac{3}{2}}^{(2)}(\frac{k}{H}\exp(-Ht))
\end{array}\end{displaymath}
and
\begin{displaymath}\begin{array}{l}
(\psi_{0}^{g},\Phi({\bf k})\Phi({\bf
k}^{\prime})\psi_{0}^{g})-(\psi_{0}^{g},\hat{\Phi}_{t}({\bf
k})\hat{\Phi}_{t}({\bf k}^{\prime})\psi_{0}^{g})\cr= \hbar
\delta({\bf k}+{\bf k}^{\prime})\frac{1}{2k}(1-\exp(-2Ht))
\end{array}\end{displaymath}
a result obtained in \cite{ford,starlect,star,linde}.

 In
a similar way, using Equation (47), we calculate higher order
correlation functions
\begin{displaymath}\begin{array}{l}
(\psi_{0}^{g},\hat{\Phi}_{t}({\bf k}_{1})\hat{\Phi}_{t}({\bf
k}_{2})\hat{\Phi}({\bf k}^{\prime}_{3})\hat{\Phi}({\bf
k}_{4}^{\prime})\psi_{0}^{g})\cr= \int
d\Phi\vert\psi_{t}^{g}(\Phi)\vert^{2}\Phi({\bf k}_{1})\Phi({\bf
k}_{2})E\Big[\Phi_{t}(\Phi,{\bf k}^{\prime}_{3})\Phi_{t}(\Phi,{\bf
k}_{4}^{\prime})\Big]\end{array}\end{displaymath} In this
equation, we insert the solution (82). The integral over $\Phi$ is
the Gaussian integral with the covariance (87). The expectation
value in the formula for the correlations is
\begin{equation}
\begin{array}{l} E\Big[\Phi_{t}(\Phi,{\bf
k}^{\prime}_{3})\Phi_{t}(\Phi,{\bf k}_{4}^{\prime})\Big]
\cr=(u_{0}\Phi)({\bf k}^{\prime}_{3})((u_{t}\Phi)({\bf
k}^{\prime}_{3}))^{-1}(u_{0}\Phi)({\bf
k}^{\prime}_{4})((u_{t}\Phi)({\bf k}^{\prime}_{4}))^{-1}
\cr+i\hbar u_{0}^{2}\delta({\bf k}^{\prime}_{3}+{\bf
k}^{\prime}_{4})\int_{0}^{t}u_{\tau}^{-2}({\bf
k}_{3}^{\prime})a_{\tau}^{-3}d\tau\equiv E_{cl}+G_{t},
\end{array}\end{equation}
where the $G_{t}$ term (it still will be discussed in subsequent
sections) being the quantum fluctuation of $\Phi_{t}$ is
proportional to $\hbar$. The integral (91) can explicitly be
calculated for $M=0$. Then, inserting $H_{\frac{3}{2}}^{(2)}$
\vspace{-6pt}
\begin{displaymath}
a_{\tau}^{\frac{3}{2}}u_{\tau}(k)=Cz^{-\frac{3}{2}}(z-i)\exp(-i z)
\end{displaymath}
with a certain constant $C$ and $z(\tau)=\frac{k}{H}\exp(-H\tau)$,
we obtain the integral in Equation (91). The integral reads
\begin{displaymath}\begin{array}{l}
G_{t}=-i\hbar(\frac{k}{H}-i)^{2}\frac{H^{2}}{k^{3}}\exp(-2i\frac{k}{H})
\cr\int_{\frac{k}{H}}^{\frac{k}{H}\exp(-Ht)}
y^{2}(y-i)^{-2}\exp(2iy)dy.
\end{array}\end{displaymath}
For a small $k$, the leading  infrared behavior of the real part
of $G_{t}$ is $\frac{Ht}{k}$ confirming the diffusive behavior
 of $\Phi_{t}$ discovered in
\cite{ford,starlect,star,linde}.

 In the case of an inverted metric (58) describing the Euclidean
 anti-de Sitter  space the correlation function is an
analytic continuation of Equation (90)
\begin{equation}\begin{array}{l}
H^{(2)}_{\nu}(i\frac{k}{H})a(t)^{-\frac{3}{2}
}\Big(H^{(2)}_{\nu}(i\frac{k}{H}\exp(-Ht)\Big)^{*} \delta({\bf
k}+{\bf k}^{\prime})\cr\simeq
K_{\nu}(\frac{k}{H})a(t)^{-\frac{3}{2}
}K_{\nu}(\frac{k}{H}\exp(-Ht)\Big)\delta({\bf k}+{\bf k}^{\prime})
\end{array}\end{equation}
with $\nu=\frac{3}{2}\sqrt{1+\frac{4\mu^{2}}{9H^{2}}}$. The
correlation function (90) coincides with the one derived
\mbox{in~\cite{tagirov,bunch,schom}} describing the quantum free
field in de Sitter space, with the ground state invariant under
the de Sitter group. The two-point function (92) describes the
Euclidean version of the quantum field on the anti-de Sitter space
\cite{witten,polyakov}.

 With the solution (85)--(90), we shall
have the same problem with a perurbative construction of an
interaction (ultraviolet divergencies and renormalization) as in
the QFT in the Minkowski space-time. In Sections~\ref{sec11}
and~\ref{sec12}, we discuss a method to construct interactions
when the solution of the wave equation $u(t)$ is a real function,
and we do not insist on the existence of the ground state. In the
derivation of Equation (90), we have chosen as $u$ the Hankel
function $H_{\nu}^{(2)}(\frac{k}{H}\exp(-Ht))$. In such a case,
$\Gamma$ is complex. Hence, $\psi_{t}^{g}$  is not a pure phase
WKB solution. If $M=0$, then $\nu=\frac{3}{2}$. We could chose as
$u$ the Bessel functions $J_{\frac{3}{2}}$ or $Y_{\frac{3}{2}}$,
which are expressed by trigonometric functions. In such a case,
$\Gamma$ is real and $\psi_{t}^{g}$ is a pure phase. We obtain
another representation of de Sitter field with caustic
singularities, which will be discussed in another model in
Section~\ref{sec10}. The Bessel functions of an imaginary argument
$K_{\nu}$ in Equation (92) give a solution in Euclidean AdS. In
such a case,  $u_{\nu}$ is a real function without caustic poles.
We can construct interaction without an ultraviolet cutoff by
means of the Feynman--Kac formula, as discussed in
Section~\ref{sec12}.

 Let us still determine the stochastic field  (76)
resulting from the Hamiltonian (60). It is the solution of the
stochastic equation
\begin{equation}
d\Phi_{lm}(s)=-\partial_{\tau}\ln(u_{lm}(\tau-s))\Phi_{lm}(s)ds+\sqrt{i\hbar}H^{\frac{3}{2}}\cos(\tau-s)dw_{lm}(s),
\end{equation}
where $u_{lm}$ is the solution of Equation (63) and $w_{lm}$ are
Gaussian processes with mean zero and the covariance
\begin{equation}
E[w_{lm}(s)w_{l^{\prime}m^{\prime}}(s^{\prime})]=\delta_{ll^{\prime}}\delta(m+m^{\prime})min(s,s^{\prime}),
\end{equation}
where $m=(j,\sigma)$ (as explained at Equation (59)) and
$\delta(m+m^{\prime})=\delta(j-j^{\prime})\delta(\sigma+\sigma^{\prime})$.

The Euclidean Hamiltonian (62) generates the stochastic equation
\begin{equation}\begin{array}{l}
d\Phi_{lm}(s)^{E}=-\partial_{\tau}\ln(u^{E}_{lm}(\tau-s))\Phi^{E}_{lm}(s)d(-s)
+\sqrt{\hbar}H^{\frac{3}{2}}\cosh(\tau-s)dw_{lm}(-s),
\end{array}\end{equation}
where $u^{E}_{lm}$ is the solution of Equation (65).

The solution defines a real diffusion process solving the
diffusion Equation (26) for $\tau\leq 0$.

The solution of Equation (93) is 
\begin{equation}\begin{array}{l}
\Phi_{lm}(s)=u_{lm}(\tau-s)u_{lm}(\tau)^{-1}\Phi_{lm}+\sqrt{i\hbar}H^{\frac{3}{2}}u_{lm}(\tau-s)\int_{0}^{s}(u_{lm}(\tau-t))^{-1}
\cos(\tau-t)dw_{lm}(t).\end{array}
\end{equation}

Equation (95) has the solution \vspace{-12pt}

\begin{equation}\begin{array}{l}
\Phi_{lm}^{E}(s)=u_{lm}^{E}(\tau-s)u^{E}_{lm}(\tau)^{-1}\Phi_{lm}
+\sqrt{\hbar}u_{lm}^{E}(\tau-s)H^{\frac{3}{2}}\int_{0}^{s}(u_{lm}^{E}(\tau-t))^{-1}\cosh(\tau-t)dw_{lm}(t),
\end{array}\end{equation}

where $\Phi_{lm}$ is the initial condition at $s=0$. We need to
calculate
\begin{displaymath}\begin{array}{l}
E[(\Phi_{lm}(s)-E[\Phi_{lm}(s)])(\Phi_{lm}(s^{\prime})-E[\Phi_{l^{\prime}m^{\prime}}(s^{\prime})])]
=G_{lm}(s,s^{\prime})\delta_{ll^{\prime}}\delta(m+m^{\prime}).
\end{array}\end{displaymath}
We have
\begin{equation}\begin{array}{l}
G_{lm}(s,s^{\prime})=i\hbar
u_{lm}(\tau-s)u_{lm}(\tau-s^{\prime})H^{3}\int_{0}^{m(s,s^{\prime})}
dt(u_{lm}(\tau-t))^{-2}\cos^{2}(\tau-t)
\end{array}\end{equation}
for the field (96) and
\begin{equation}\begin{array}{l}
G^{E}_{lm}(s,s^{\prime})=\hbar u^{E}_{lm}(\tau-s)u^{E}_{lm}(\tau-
s^{\prime})H^{3}\int_{0}^{m(s,s^{\prime})}
dt(u_{lm}^{E}(\tau+t))^{-2}\cosh^{2}(\tau-t)
\end{array}\end{equation}
for the inverted metric of Equation (56).

 For a general $l$ and $M$, the solution of
the wave equation is defined by the Legendre functions
\cite{tagirov}. We are unable to calculate the integrals in
$G_{lm}(s,s^{\prime})$ (98) and (99) exactly. For $l=0$ and $M=0$,
we have
\begin{equation}
u_{00}(\tau)=\tau+\frac{1}{4}\sin(2\tau)
\end{equation}
and
\begin{displaymath}
u_{00}^{E}(\tau)=\tau+\frac{1}{4}\sinh(2\tau).
\end{displaymath}
For a large $l$, we can obtain the WKB solution (the odd solution)
of the wave equation
\begin{equation}
u_{lm}(\tau)=(\frac{dS}{d\tau})^{-\frac{1}{2}}\sin(S(\tau))\cos(\tau),
\end{equation}
where
\begin{displaymath}
S(s)=\int_{0}^{s}d\tau\sqrt{l(l+2)+1+(M^{2}-2H^{2})H^{-2}\cos^{-2}(\tau)}.
\end{displaymath}
In the even solution, we replace $\sin$ in Equation
(101) by $\cos$. For the Euclidean Equation~(65), the WKB solution
is of the form (101), but the trigonometric functions are replaced
by hyperbolic functions. So, the odd solution reads
\begin{equation}
u^{E}_{lm}(\tau)=(\frac{dS^{E}}{d\tau})^{-\frac{1}{2}}\sinh(S^{E}(\tau))\cosh(\tau)
\end{equation}
 with\begin{equation}
S^{E}(s)=\int_{0}^{s}d\tau\sqrt{l(l+2)+1+(\mu^{2}-2H^{2})H^{-2}\cosh^{-2}(\tau)}.
\end{equation}
For the Euclidean version, an exponential solution of Equation
(65) will be needed (the sum of even and odd solutions)
\begin{equation}
u_{lm}(\tau)=(\frac{dS^{E}}{d\tau})^{-\frac{1}{2}}\exp(S^{E}(\tau))\cosh(\tau).
\end{equation}
We can  calculate $G_{lm}$ using  the WKB approximation with the
result
\begin{equation}\begin{array}{l}
G_{lm}(s,s^{\prime})=i\hbar H^{3}
u_{lm}(\tau-s)u_{lm}(\tau-s^{\prime})(\cot(S(\tau-m(s,s^{\prime}))
-\cot S(\tau)) \end{array}\end{equation} for $u_{lm}$ defined by
Equation (101) and
\begin{equation}\begin{array}{l}
G_{lm}(s,s^{\prime})=\hbar H^{3}
u^{E}_{lm}(\tau-s)u^{E}_{lm}(\tau-s^{\prime})(\coth(S^{E}(\tau-m(s,s^{\prime}))
-\coth S^{E}(\tau)) \end{array}\end{equation} for the inverted
metric in Equation (102).

 In the case of  the exponential solution (104), we obtain
\vspace{-12pt}

\begin{equation}\begin{array}{l}
G_{lm}(s,s^{\prime})=\hbar H^{3}
u^{E}_{lm}(\tau-s)u^{E}_{lm}(\tau-s^{\prime})(\exp(-2S^{E}(\tau-m(s,s^{\prime}))
-\exp(-2 S^{E}(\tau))).\end{array}
\end{equation}

For the even solutions of the WKB  form  defined by $\cos$ and
$\cosh$ in Equations (101) and (102), the cotangent functions in
Equations (105) and (106) are replaced by the tangents. The
approximate expression for $G_{lm}(s,s^{\prime})$ can be applied
for calculations of the propagators and field correlations in
subsequent sections. In Appendix F, we express fields
in de Sitter space and their correlations in the cosmic time.
There, we also discuss de Sitter fields in two dimensions (see
earlier papers \cite{figari,figari2}), where we express the
correlation functions by elementary functions.

With the trigonometric functions entering $G_{lm}(s,s^{\prime})$,
we have the difficulty with the integrability over time in the
definition of the evolution kernel in Section~\ref{sec7} of
quantum field theory in the Schr\"odinger formulation (because of
the poles of the trigonometric functions in Equation (105)).  We
note that the replacement of the trigonometric functions by
hyperbolic functions (signature inversion) in the approximate WKB
solutions (106) avoids this difficulty. This analytic continuation
allows a definition of the Feynman--Kac formula for a real time
(but with an inverted spatial signature), as will be  discussed in
Sections~\ref{sec11} and~\ref{sec12}.

\section{The Propagator for the Schr\"odinger Evolution of the Scalar Field}\label{sec7}
We can express the time evolution either by the solution of the
stochastic equation as in Equation (45) or by an evolution kernel
defined by

\begin{displaymath}
(U_{t}\psi)(\Phi)=\int d\Phi^{\prime}
\tilde{K}_{t}(\Phi,\Phi^{\prime})\psi(\Phi^{\prime}).
\end{displaymath}
We write $\psi$ in the form (5). Then, the definition of the
kernel is rewritten as \vspace{-6pt}
\begin{equation}
(U_{t}\psi_{0}^{g}\chi)(\Phi)=\psi_{t}^{g}\int d\Phi^{\prime}
K_{t}(\Phi,\Phi^{\prime})\chi(\Phi^{\prime}).
\end{equation}
We represent $\chi$ as a Fourier transform
\begin{displaymath}
\chi(\Phi)=\int
d\Lambda\tilde{\chi}(\Lambda)\exp(i(\Lambda,\Phi)).
\end{displaymath}
Then, from Equation (45), we obtain

\begin{equation}\begin{array}{l}
K_{t}(\Phi,\Phi^{\prime}) =\int d\Lambda
E\Big[\exp\Big(i(\Lambda,\Phi_{t}(\Phi)-\Phi^{\prime})\Big)\Big].
\end{array}\end{equation}
We  calculate the expectation value of $\Phi_{t}$ with the result
\begin{equation}\begin{array}{l}
E\Big[\exp\Big(i(\Lambda,\Phi_{t}(\Phi)-\Phi^{\prime})\Big)\Big]=
\exp\Big(
i(\Lambda,u_{0}u_{t}^{-1}\Phi-\Phi^{\prime})-\frac{1}{2}(\Lambda,
G_{t}\Lambda)\Big).\end{array}
\end{equation}
Then, calculating the $\Lambda$ integral we obtain (up to a
normalization constant)
\begin{equation}\begin{array}{l}
K_{t}(\Phi,\Phi^{\prime}) =\det G_{t}^{-\frac{1}{2}} \exp\Big(
-\frac{1}{2}\Big((u_{0}u_{t}^{-1}\Phi-\Phi^{\prime}),G_{t}^{-1}(u_{0}u_{t}^{-1}\Phi-\Phi^{\prime})\Big)\Big],
\end{array}\end{equation}
where
 \begin{displaymath}\begin{array}{l}
  G_{t}({\bf k},{\bf k}^{\prime})= E[(\Phi_{t}({\bf k})-<\Phi_{t}>({\bf
  k}))
  (\Phi_{t}({\bf k}^{\prime})-<\Phi_{t}>({\bf k}^{\prime}))]\cr\equiv
 G_{t}(k)\delta({\bf k}+{\bf k}^{\prime})
\end{array}\end{displaymath}
with $<\Phi>=E[\Phi]$.

In a homogeneous expanding space-time for positive time (to be
concise, from now on we omit the $\delta({\bf k}+{\bf
k}^{\prime})$ term in $ G_{t}({\bf k},{\bf k}^{\prime})$)
\begin{equation}
G_{t}=i\hbar
u_{0}^{2}\int_{0}^{t}u_{t-\tau}^{-2}a_{t-\tau}^{-3}d\tau.
\end{equation}
Equation (112), (where
$m(s,s^{\prime})\equiv min(s,s^{\prime})$), 
 is a consequence of
\begin{displaymath}
E\Big[\int_{0}^{s}f_{\tau}dW_{\tau}({\bf
k})\int_{0}^{s^{\prime}}f_{\tau^{\prime}}dW_{\tau^{\prime}}({\bf
k}^{\prime})\Big]=\int_{0}^{m(s,s^{\prime})}f^{2}_{\tau}d\tau\delta
({\bf k}+{\bf k}^{\prime})\end{displaymath} following from the
definition of the stochastic integral \cite{freidlin}.

In the presence of an interaction from Equation (46), we obtain a
generalization of the Formula (111) for the evolution kernel as
\begin{displaymath}\begin{array}{l}
K_{t}(\Phi,\Phi^{\prime}) =\int d\Lambda
E\Big[\exp\Big(i(\Lambda,\Phi_{t}(\Phi)-\Phi^{\prime})\Big)
\exp\Big(-\frac{i}{\hbar}\int_{0}^{t}V(\Phi_{s}(\Phi))ds\Big)\Big].
\end{array}\end{displaymath}
Clearly, we cannot explicitly calculate this expectation value and
the $\Lambda$ integral as we did in Equations (110) and (111), but
we can perform such calculations in a perturbation expansion in
$V$, as will be discussed in Sections~\ref{sec11} and~\ref{sec12}.

 For
negative time in Equation (78),
 the correlation function is
\begin{equation}
G_{t}=i\hbar u_{0}^{(-)}u_{0}^{(-)}\int_{0}^{-t}
(u_{t+\tau}^{(-)})^{-2}a_{t+\tau}^{-3}d\tau
\end{equation}
where $u^{(-)}$ is the solution of the wave equation for a
negative time. Let us note that, for positive time, we obtain an
oscillating propagator $K_{t}$, which is a pure phase, whereas for
an inverted metric (for the negative time), we obtain a real
function describing a transition function for a~diffusion.

The result of (111) and (112) gives an explicit formula for the
evolution kernel. Another way to calculate this propagator
involves a solution of the Cauchy problem for the wave equation,
as discussed briefly at the end of Appendix D (such
calculations are performed in more detail in \cite{prok1}). Note
that a real $u$ leads to a purely imaginary $G_{t}$. Then, we
obtain the propagator (111), which is a pure phase in agreement
with the Feynman Formula (36) and Appendix D. If $u$
is complex, then $\Gamma$ is complex. Hence, $G_{t}$ is complex.
Such a modification of the kernel results from its definition
(108) involving $\psi_{t}^{g}$, which is not a pure phase.

 Using the expectation value (110), we can also calculate the
field correlation functions in a state $\psi_{0}^{g}\chi$ with a
Gaussian state $\chi$ of the form
\begin{equation}
\chi=\det B^{\frac{1}{2}}\exp(-\frac{1}{2}(\Phi,B\Phi)).
\end{equation}
Then, from Equation (110) (up to a constant multiplier)
\begin{equation}\begin{array}{l}
\chi_{t}(\Phi)=\det B\int d\Lambda
\exp(-\frac{1}{2}(\Lambda,B^{-1}\Lambda))\cr
 \exp\Big(
i(\Lambda,u_{0}u_{t}^{-1}\Phi)-\frac{1}{2}(\Lambda,
G_{t}\Lambda)\Big) =\det(1+BG_{t})^{-\frac{1}{2}}\cr \det
B^{\frac{1}{2}}\exp\Big(-\frac{1}{2}(u_{0}u_{t}^{-1}\Phi,(B^{-1}+G_{t})^{-1}u_{0}u_{t}^{-1}\Phi)\Big).
\end{array}\end{equation}
Next,
\begin{equation}\begin{array}{l}
(\Phi\chi)_{t}(\Phi)=u_{0}u_{t}^{-1}B^{-1}(B^{-1}+G_{t})^{-1}\Phi
 \det B^{\frac{1}{2}}\det(1+BG_{t})^{-\frac{1}{2}} \cr
\exp\Big(-\frac{1}{2}(u_{0}u_{t}^{-1}\Phi,(B^{-1}+G_{t})^{-1}u_{0}u_{t}^{-1}\Phi)\Big).
\end{array}\end{equation}
Then, the field correlation function according to Equation (48)
can  be calculated (for a general solution $\psi_{t}^{g}$ of the
Schr\"odinger Equation (25)) from the formula
\begin{equation}\begin{array}{l}
(\psi_{0}^{g}\chi,\hat{\Phi}_{t}({\bf k})\hat{\Phi}({\bf
k}^{\prime})\psi_{0}^{g}\chi)=\delta({\bf k}+{\bf k}^{\prime})
\int d\Phi \vert \psi_{t}^{g}\vert^{2}\chi^{*}_{t}(\Phi)\Phi({\bf
k})(\Phi\chi)_{t}(\Phi,{\bf k}^{\prime}),
\end{array}\end{equation}
where $\chi_{t}$ and $(\Phi\chi)_{t}$ have been evaluated in
Equations (115) and (116).

The integral (117) is Gaussian. Hence, we derive an explicit
(although quite complicated) formula for the correlation function.
The formula simplifies if, e.g., $\vert \psi_{t}^{g}\vert^{2}$ is
integrable (when $i(\Gamma -\Gamma^{*})<0$) and $\chi=1$ (as in
Equation (91)). Then,
$(\Phi\chi)_{t}(\Phi)=E[\Phi_{t}(\Phi)]=u_{0}u_{t}^{-1}\Phi$.

 We can calculate the field correlations in
various states $(\psi_{0}^{g}\chi,\hat{ \Phi}(t,{\bf
x})\hat{\Phi}({\bf x}^{\prime})\psi_{0}^{g}\chi)$ using the
propagator (111) as \begin{equation}\begin{array}{l} \int d\Phi
d\Phi^{\prime}d\Phi^{\prime\prime}\Big(U_{t}(\Phi,\Phi^{\prime})
(\psi_{0}^{g}\chi)(\Phi^{\prime})\Big)^{*}\Phi({\bf x})
U_{t}(\Phi,\Phi^{\prime\prime})\Phi^{\prime\prime}({\bf
x}^{\prime})(\psi_{0}^{g}\chi)(\Phi^{\prime\prime})\cr =\int d\Phi
d\Phi^{\prime}d\Phi^{\prime\prime}
\vert\psi_{t}^{g}(\Phi)\vert^{2}K_{t}(\Phi,\Phi^{\prime})^{*}
K_{t}(\Phi,\Phi^{\prime\prime})\chi(\Phi^{\prime})^{*}\chi(\Phi^{\prime\prime})
\Phi({\bf x})\Phi^{\prime\prime}({\bf x}^{\prime})
\end{array}\end{equation}
with
\begin{equation}
U_{t}(\Phi,\Phi^{\prime})=\psi_{t}^{g}(\Phi)K_{t}(\Phi,\Phi^{\prime})
(\psi_{0}^{g}(\Phi^{\prime}))^{-1}.
\end{equation}
We can express all multi-time correlation functions by the
evolution kernel. As an example,
\begin{equation}\begin{array}{l}
(\psi_{0}^{g}\chi,\hat{ \Phi}_{t}({\bf
x})\hat{\Phi}_{t^{\prime}}({\bf x}^{\prime})\psi_{0}^{g}\chi)=
(U_{t}\psi_{0}^{g}\chi,\Phi({\bf x})U_{t,t^{\prime}}\Phi({\bf
x}^{\prime})\psi_{0}^{g}\chi)
\end{array}\end{equation}
where
\begin{equation}
U_{t,t^{\prime}}(\Phi,\Phi^{\prime})=\psi_{t}^{g}(\Phi)
K_{(t,t^{\prime})}(\Phi,\Phi^{\prime})(\psi_{t^{\prime}}^{g}(\Phi^{\prime}))^{-1}.
\end{equation}
Hence, in terms of the kernels Equation (120) is expressed as
\begin{equation}\begin{array}{l} \int d\Phi
d\Phi^{\prime}d\Phi^{\prime\prime}\Big(U_{t}(\Phi,\Phi^{\prime})
(\psi_{0}^{g}\chi)(\Phi^{\prime})\Big)^{*}\Phi({\bf x})
U_{t,t^{\prime}}(\Phi,\Phi^{\prime\prime})\Phi^{\prime\prime}({\bf
x}^{\prime})(\psi_{0}^{g}\chi)(\Phi^{\prime\prime})=\cr \int d\Phi
d\Phi^{\prime}d\Phi^{\prime\prime}
\vert\psi_{t}^{g}(\Phi)\vert^{2}K_{t}(\Phi,\Phi^{\prime})^{*}
K_{(t,t^{\prime})}(\Phi,\Phi^{\prime\prime})\chi(\Phi^{\prime})^{*}\chi(\Phi^{\prime\prime})
\Phi({\bf x})\Phi^{\prime\prime}({\bf x}^{\prime})
\end{array}\end{equation}

The kernel $K_{(t,t^{\prime})}$ is obtained as in Equation (109)

\begin{equation}
K_{(t,t^{\prime})}(\Phi,\Phi^{\prime})=(\det
G(t,t^{\prime}))^{-\frac{1}{2}}
\exp\Big(-\frac{1}{2}(u_{0}(u_{t-{t}^{\prime}})^{-1}\Phi-\Phi^{\prime})
G(t,t^{\prime})^{-1}(u_{0}(u_{t-t^{\prime}})^{-1}\Phi-\Phi^{\prime})\Big),
\end{equation}

where now the process $\Phi_{s}$ is the solution of the stochastic
equation with an initial value at $t^{\prime}$ (instead of zero)

\begin{equation}
\Phi_{s}=u_{t-s}(u_{t-t^{\prime}})^{-1}\Phi+\sqrt{i\hbar}u_{t-s}
\int_{t^{\prime}}^{s}u_{t-\tau}^{-1}a_{t-\tau}^{-\frac{3}{2}}dW_{\tau}
\end{equation}
Hence, from Equations (109) and (110),
\begin{equation}
G(t,t^{\prime})=i\hbar
u_{0}^{2}\int_{t^{\prime}}^{t}u_{t-\tau}^{-2}a_{t-\tau}^{-3}d\tau
\end{equation}
Similarly to Equation (122) in terms of the kernel
$U_{t,t^{\prime}}(\Phi,\Phi^{\prime})$, we can express the
$n$-point~functions
\begin{equation}
(\psi_{0}^{g}\chi,\hat{ \Phi}_{t_{1}}({\bf x}_{1})\ldots \ldots
\hat{\Phi}_{t_{n}}({\bf x}_{n})\psi_{0}^{g}\chi)=
(U_{t_{1}}\psi_{0}^{g}\chi, \Phi({\bf
x}_{1})U_{t_{1},t_{2}}\Phi({\bf x}_{2})\ldots \ldots
U_{t_{n}-t_{n-1}}\Phi({\bf x}_{n})\psi_{0}^{g}\chi)
\end{equation}

In the presence of interaction, there are additional Feynman--Kac
factors in $K_{(t,t^{\prime})}$ as in Equation (48)
\begin{displaymath}
\exp\Big(-\frac{i}{\hbar}\int^{t}_{t^{\prime}}V(\Phi_{s})ds\Big)
\end{displaymath}
At the end of this section, we would like to discuss different
notions of propagators in the literature \cite{parker,calzetta}
(the one for the metric $a\simeq t$ has been discussed in
\cite{prop1,prop2}). In these papers, the propagator has been
defined as the inverse of the operator ${\cal A}$ appearing in the
action (22) when written in the form $\int dx L= \int dx\Phi{\cal
A}\Phi$. We could also represent this propagator by Schwinger's
proper time
\begin{displaymath}{\cal A}^{-1}=i\int_{0}^{\infty}
d\tau\exp(-i\tau{\cal A}).
\end{displaymath}
By formal functional integration, the functional integral average
$<\phi(x)\phi(x^{\prime})> $ is equal to ${\cal
A}^{-1}(x,x^{\prime})$(the kernel satisfying ${\cal A}{\cal
A}^{-1}=1$). In general, by formal differentiation of the lhs of
Equation (117) using Equation (23) in the metric (67), we obtain
for $ t>0$ under the assumption that the quantum field satisfies
the wave equation
\begin{displaymath}\begin{array}{l}
\partial_{t}^{2}(\psi_{0}^{g}\chi,\hat{\Phi}_{t}({\bf k})\hat{\Phi}({\bf
k}^{\prime})\psi_{0}^{g}\chi)=\partial_{t}(\psi_{0}^{g}\chi,a^{-3}\Pi_{t}({\bf
k})\hat{\Phi}({\bf k}^{\prime})\psi_{0}^{g}\chi)\cr=
(-3H\partial_{t}-a^{-2}k^{2}-M^{2})(\psi_{0}^{g}\chi,\Phi_{t}({\bf
k})\Phi({\bf k}^{\prime})\psi_{0}^{g}\chi) \delta({\bf k}+{\bf
k}^{\prime}).
\end{array}\end{displaymath}
Hence, the correlation function also satisfies the wave equation.
If $\chi=1$ and $ \vert \psi_{t}^{g}\vert^{2}$ is integrable, then
\begin{displaymath}\begin{array}{l}
(\psi_{0}^{g},\hat{\Phi}_{t}({\bf k})\hat{\Phi}({\bf
k}^{\prime})\psi_{0}^{g}) =\int d\Phi \vert
\psi_{t}^{g}\vert^{2}\Phi({\bf k})E[\Phi_{t}({\bf
k}^{\prime})]\cr=u_{0}({\bf k})u_{t}^{*}({\bf k})\delta({\bf
k}+{\bf k}^{\prime})
\end{array}\end{displaymath}
because $E[\Phi_{t}({\bf k}^{\prime})]=(u_{0}u_{t}^{-1}\Phi)({\bf
k})$ (from Equation (82)) and $i(\Gamma-\Gamma^{*})\simeq
(u_{t}u_{t}^{*})^{-1}$ from the Wronskian (as has been exploited
in the particular case of the de Sitter metric (57) in
Equation~(90)). We can conclude that, although the correlation
function in any state is a solution of the wave equation, then the
solution of the equation for the correlation functions is not
unique, because it depends on the state under consideration. If
there is a unique ground state (as in  de Sitter space) invariant
under a symmetry group, then we can distinguish the solution
having this invariance \cite{schom}. In other states, the
two-point correlation function must be determined through
calculations, e.g., from Equation (117) by means of the propagator
(111) (as will be discussed at the end of Section~\ref{sec9}).

In QFT in the Minkowski space-time  in the ground state (11), we
obtain ${\cal A}^{-1}$ as the Lorentz invariant Green function of
$\partial_{t}^{2}-\triangle +M^{2}$  equal to $\frac{1}{2}
\nu^{-1}\exp(-i\nu\vert t-t^{\prime}\vert)$, where
$\nu=\sqrt{-\triangle+M^{2}}$. In the case of the de Sitter
space-time, the calculation with $\chi=1$ and $\psi^{g}$ defined
by $\Gamma$ in Equation (90) coincides (as discussed after
Equation (117)) with the result of an expectation value computed
either by a formal functional integration or derived by an
expansion of the field in creation and annihilation operators
defined by de Sitter invariant vacuum
\cite{tagirov,schom,kirsten}. In the time-dependent metric of this
section (as well as in Section~\ref{sec8}), there is no candidate
for a vacuum. Hence, it remains unclear in which state
$\psi_{0}^{g}\chi$ the two-point correlation function could be
equal to ${\cal A}^{-1}$. If there is no unique vacuum for the
quantum field in an expanding metric, then the physical meaning of
${\cal A}^{-1}$ is obscure, whereas the propagator defined in this
section (and the correlation functions defined by the propagator)
has a clear meaning for the canonically quantized field theory of
Section~\ref{sec3}. A relation of the propagator $K_{t}$  to
${\cal A}^{-1}$, defined as the causal propagator, has been
discussed in ref.~\cite{prok1}. In this paper, the definition of
the propagator $K_{t}$  is related to the solution of the Cauchy
problem, as expressed by the causal propagator. We briefly discuss
the method of the calculation of $K_{t}$ using the solution of the
Cauchy problem for ${\cal A}$ at the end of
Appendix D.

\section{Power-Law Expansion }\label{sec8}
In this and subsequent sections, we discuss some soluble wave
equations. We are unable to derive an explicit solution of the
wave Equation (68) in a homogeneous space-time for an arbitrary
expansion $a(t)$. In Appendix B, we solve this
equation for a large $k$ by means of the WKB method (it can be
made exact as in \cite{parker}). Then, in this approximation, we
can calculate the propagator (111) and the field correlation
functions (Appendix B).

In this and subsequent sections, we discuss soluble models. First,
we consider  the model with the power-law expansion
 $a^{2}=b_{0}^{-2}(t+\gamma)^{2\alpha}$  with a general $\alpha\in R$
 and  $t+\gamma>0$ (we shift the initial time by $\gamma$ in order to pose the initial condition at $t=0$,
 even if $\gamma=0$ corresponds to a degenerate metric).  If $t+\gamma<0$, then $a$ is a complex
 function in general (but even powers of $t$ lead to admissible models with a topology change
 between positive and negative time \cite{horowitz}). We discuss the case $\alpha=\frac{1}{2}$ for $t+\gamma<0$ in Section~\ref{sec10}.
The expansion law $a=b_{0}^{-1}\vert t+\gamma\vert ^{\alpha}$ for
both positive $t+\gamma$  and negative $t+\gamma$ appears in
 cosmological models resulting from string theory  \cite{string,string2}.
 Such a metric cannot be a solution of general relativity, because it is not differentiable at
 $t+\gamma=0$.

 The wave equation for $t+\gamma>0$ has the form
\begin{equation}
\partial_{t}^{2}u+3\alpha (t+\gamma)^{-1}\partial_{t}u+
b_{0}^{2}(t+\gamma)^{-2\alpha}k^{2}u+M^{2}u=0.
\end{equation}

The corresponding stochastic equation is
\begin{displaymath}
d\Phi_{s}=-u^{-1}(t-s)\partial_{t}u(t-s)ds+\sqrt{i\hbar}a^{-\frac{3}{2}}dW_{s}.
\end{displaymath}
For $t+\gamma<0$ and $a=b_{0}^{-1}\vert t+\gamma\vert ^{\alpha}$
(in string inspired models \cite{string,string2}), the wave
equation is
\begin{displaymath}
\partial_{t}^{2}u+3\alpha (t+\gamma)^{-1}\partial_{t}u+
b_{0}^{2}\vert t+\gamma\vert^{-2\alpha}k^{2}u+M^{2}u=0.
\end{displaymath}

  Equation (127) is
explicitly soluble if $M=0$ (\cite{kamke}, 2.162 Equation (1a)).
The solution is (for $t+\gamma>0$, we may choose here either
complex-valued or real-valued Bessel functions)
\begin{equation}
u=(t+\gamma)^{\frac{1-3\alpha}{2}}Z_{\nu}(\frac{b_{0}}{1-\alpha}k(t+\gamma)^{1-\alpha}),
\end{equation}
where
\begin{displaymath}
\nu=\frac{1-3\alpha}{2(1-\alpha)}.\end{displaymath} $\alpha$ is
related to $w$ in  the equation of state ($p=w\rho$)
\begin{displaymath}
\alpha=\frac{2}{3(1+w)}.\end{displaymath} Note that
$\nu=-\frac{1}{2}$ if $\alpha=\frac{1}{2}$ (radiation,
$w=\frac{1}{3}$) and $\nu=-\frac{3}{2}$ if $\alpha= \frac{2}{3} $
(dust, $w=0$). We discuss $\alpha=\frac{1}{2}$ in detail in
Section~\ref{sec10}.

Let us briefly consider the interesting  case of $\alpha=
\frac{2}{3} $. Then, a real solution valid for both $t+\gamma>0$,
as well as $ t+\gamma<0$, which is continuous at $t+\gamma=0$, is
expressed by elementary functions
($T=3b_{0}(t+\gamma)^{\frac{1}{3}}$ is the conformal time)
\begin{displaymath}\begin{array}{l}
u=(t+\gamma)^{-\frac{2}{3}}\Big(
\cos(3b_{0}k(t+\gamma)^{\frac{1}{3}})-
 (3b_{0}k(t+\gamma)^{\frac{1}{3}})^{-1}
 \sin(3b_{0}k(t+\gamma)^{\frac{1}{3}})\Big).
\end{array}\end{displaymath}
The real solutions $u$ are distinguished in our approach as they
lead to real $\Gamma$  and purely imaginary $G_{t}$ in Equation
(112). This property is relevant for a construction of
interactions  via the Feynman--Kac formula (see
Section~\ref{sec11}). We also consider a complex solution, which
gives a square integrable $\psi_{t}^{g}$
\begin{displaymath}
u=(
t+\gamma)^{-\frac{1}{2}}H^{(2)}_{\frac{3}{2}}(3b_{0}k(t+\gamma)^{\frac{1}{3}})
\end{displaymath}
The two-point function in this state is discussed at the end of
this section.

For a general $\nu$ from Equation (69), we obtain
\begin{displaymath}
\Gamma=(t+\gamma)^{3\alpha}\frac{d}{dt}\ln\Big((t+\gamma)^{\frac{1}{2}-\frac{3\alpha}{2}}
Z_{\nu}(\frac{b_{0}}{1-\alpha}k(t+\gamma)^{1-\alpha})\Big).
\end{displaymath}

 The stochastic Equation (80) reads

\begin{equation}\begin{array}{l}
d\Phi_{s}=-\frac{d}{dt}\ln\Big((t+\gamma-s)^{\frac{1}{2}
-\frac{3\alpha}{2}}Z_{\nu}(\frac{b_{0}}{1-\alpha}k(t+\gamma-s)^{1-\alpha})\Big)\Phi_{s}ds
 +\sqrt{i\hbar}(t+\gamma
-s)^{-\frac{3\alpha}{2}}dW_{s}.\end{array}
\end{equation}

The solution is
\begin{equation}\begin{array}{l} \Phi_{s}=
\Big((t+\gamma-s)^{\frac{1}{2}-\frac{3\alpha}{2}}Z_{\nu}(\frac{b_{0}}{1-\alpha}k(t+\gamma-s)^{1-\alpha})\Big)
\cr
\Big((t+\gamma-t_{0})^{\frac{1}{2}-\frac{3\alpha}{2}}Z_{\nu}(\frac{1}{1-\alpha}
k(t+\gamma-t_{0})^{1-\alpha})\Big)^{-1}\Phi \cr
+\sqrt{i\hbar}\Big((t+\gamma-s)^{\frac{1}{2}-\frac{3\alpha}{2}}
Z_{\nu}(\frac{b_{0}}{1-\alpha}k(t+\gamma-s)^{1-\alpha})\Big) \cr
\times\int_{t_{0}}^{s}d\tau\Big((t+\gamma-\tau)^{\frac{1}{2}-\frac{3\alpha}{2}}
Z_{\nu}(\frac{b_{0}}{1-\alpha}k(t+\gamma-\tau)^{1-\alpha})\Big)^{-1}
(t+\gamma-\tau)^{-\frac{3}{2}\alpha}dW_{\tau}.\end{array}\end{equation}
 Let
\begin{displaymath}
z(t+\gamma)=\frac{b_{0}k}{1-\alpha}(t+\gamma)^{1-\alpha}.
\end{displaymath}
The general complex solution which can give a normalizable
$\psi_{t}^{g}$ is a superposition of Hankel functions. The
Wronskian for the Hankel functions is (here $H^{(2)}=H^{(1)*}$)
\begin{displaymath}
(\frac{d}{dz}H^{(2)})H^{(1)}-(\frac{d}{dz}H^{(1)})H^{(2)}=-\frac{4i}{\pi
z}
\end{displaymath}
Hence, for  $Z_{\nu}=H^{(1)}_{\nu}$, we obtain
\begin{equation}
\Gamma-\Gamma^{*}=\frac{4i}{\pi}(H^{(2)}_{\nu}H^{(2)*}_{\nu})^{-1}(t+\gamma)^{3\alpha}H
\end{equation}
Now, $\Gamma-\Gamma^{*}$ of Equation (131) leads to a normalizable
Gaussian state (allowing a  computation of correlation functions,
an analog of Equation (90))
\begin{displaymath}\begin{array}{l}
(\psi_{0}^{g},\hat{\Phi}_{t}({\bf k})\hat{\Phi}({\bf
k}^{\prime})\psi_{0}^{g}) =\int d\Phi
\vert\psi_{t}^{g}(\Phi)\vert^{2}\Phi E\Big[\Phi_{t}(\Phi)\Big] \cr
=\frac{\pi\hbar}{4H}H^{(2)}_{\nu}(z(\gamma))H^{(2)*}_{\nu}(z(t+\gamma))(t+\gamma)^{1-\frac{3}{2}\alpha}
\delta({\bf k}+{\bf k}^{\prime}).
\end{array}\end{displaymath}
We can  express the correlations in the case of dust
$\alpha=\frac{2}{3}$, $\nu=-\frac{3}{2}$ by elementary functions.
The two-point function at small $k$ behaves as \vspace{-6pt}
\begin{displaymath}
k^{-3}\exp(iz(t+\gamma)-iz(\gamma))
\end{displaymath}
It has the infrared singularity. $G_{t}$ is complex but the small
$k$ expansion contains a real part
\begin{displaymath} \Re G_{t}\simeq \frac{t}{k}\delta({\bf k}+{\bf
k}^{\prime})
 \end{displaymath}
 similar to the one in de Sitter space.

 The
random fields defined by the Hankel function $H_{\nu}^{(2)}$ do
not allow definition of the interaction by means of the
Feynman--Kac integral in Sections~\ref{sec11} and~\ref{sec12}. We
can establish the Feynman integral with the choice $J_{\nu}$ or
$Y_{\nu}$ as $Z_{\nu}$. In such a case, after solving the
Schr\"odinger equation with an interaction in Section~\ref{sec11},
we must look for states which are square integrable and define the
correlation functions by means of Equation (48) (these could be
the Gaussian states of Equation (116)).

 \section{The Expansion \boldmath{$a^{2}(t)=\frac{\epsilon}{b_{0}^{2}}(t+\gamma)^{2}$}}\label{sec9}
 This is the limiting case ($\alpha\rightarrow 1$) of Equation (127). It describes the $w=-\frac{1}{3}$ fluid corresponding to the coasting
 cosmology \cite{coasting} or Dirac--Milne 
  cosmology \cite{milne}.
 The scale $a(t)$ is invariant under
 $t+\gamma\rightarrow -t-\gamma$)
 (it is contracting for $t+\gamma<0$ and expanding for $t+\gamma>0$; we choose $\gamma>0$ for $t>0$
 and $\gamma<0$ for $t<0$). The scalar field theory of these models has been also
 discussed in \cite{prop1,prop2}. As noticed in Section~\ref{sec5},
 the Einstein equations which can appear in the path integral of
 quantum gravity together with $a^{2}\simeq t^{2} $ give also the
 solution with $a^{2} \simeq -t^{2}$ (with the same
 $w=-\frac{1}{3}$). The model with $a^{2}<0$ is interesting
 for the construction of an interaction via the Feynman--Kac formula, as will be
 discussed in Section~\ref{sec12}.

 The wave equation reads (with
 $M=0$)
 \begin{equation}
 \frac{d^{2}u}{dt^{2}}+3(t+\gamma)^{-1}\frac{du}{dt}+\epsilon (t+\gamma)^{-2}b_{0}^{2}k^{2}u=0.
 \end{equation}
In Equation (132), we add a parameter $\epsilon=\pm 1$ in order to
 describe a model with an inverted spatial metric
resulting from the solution of Friedmann equations
$a^{2}=-\frac{1}{b_{0}^{2}}(t+\gamma)^{2}$ with $w=-\frac{1}{3}$.
Choosing $\sqrt{-g}=i\sqrt{\vert g\vert}$, we obtain a diffusion
equation for positive time instead of the Schr\"odinger equation.
The diffusion equation makes sense only in one direction of time
(either positive or negative). Our choice of the square root of
$\sqrt{-g}$ in this section gives the diffusion equation for a
positive time. The diffusion Equation (26) reads
\begin{equation}\begin{array}{l}
\hbar\partial_{t}\psi_{t}= \frac{1}{2}\int d{\bf x}
\Big(\hbar^{2}b_{0}^{3}\vert
t+\gamma\vert^{-3}\frac{\delta^{2}}{\delta\Phi({\bf x})^{2}}
-\frac{\vert t+\gamma\vert
}{b_{0}}(\nabla\Phi)^{2}-\mu^{2}b_{0}^{-3}\vert
t+\gamma\vert^{3}\Phi^{2}\Big)\psi_{t}\end{array}
\end{equation}
{The equation for $\chi$ is}
\begin{equation}\partial_{t}\chi_{t}=\int d{\bf x}
\Big(\hbar \frac{1}{2} b_{0}^{-3}\vert
t+\gamma\vert^{-3}\frac{\delta^{2}}{\delta\Phi({\bf
x})^{2}}-u^{-1}\partial_{t}u \Phi({\bf
x})\frac{\delta}{\delta\Phi({\bf x})}\Big)\chi_{t},
\end{equation}
where  $u$ is the solution of Equation (132) with $\epsilon=-1$.
We could consider the model of Section~\ref{sec3}, where there is
$d{\bf x}\sqrt{\vert g\vert}$ as the volume element. In such a
case, we  still have the Schr\"odinger Equation (instead of the
diffusion equation). Then, the equation for $\chi$ is
\begin{displaymath}\partial_{t}\chi_{t}=\int d{\bf x}
\Big(\hbar \frac{i}{2} b_{0}^{3}\vert
t+\gamma\vert^{-3}\frac{\delta^{2}}{\delta\Phi({\bf
x})^{2}}-u^{-1}\partial_{t}u \Phi({\bf
x})\frac{\delta}{\delta\Phi({\bf x})}\Big)\chi_{t},
\end{displaymath}
where $u$ satisfies Equation (132) with $\epsilon=-1$ . We could
also treat the introduction of $ \epsilon$ in Equation (132) as a
technical step for a derivation of solutions, which subsequently
are to be continued analytically to $\epsilon=1$. There is no
continuous transition between  $\epsilon=-1$ and $\epsilon=1$.

  The solution of Equation (132) is   \cite{kamke} (true for $t+\gamma>0$ as well as for $
  t+\gamma<0$)
 \begin{equation}
  u= C_{1}\vert t+\gamma\vert ^{-1-\mu}+C_{2}\vert t+\gamma\vert ^{-1+\mu},
  \end{equation}
  where
  \begin{equation}
  \mu=\sqrt{1-\epsilon b_{0}^{2}k^{2}}.
  \end{equation}
The conformal time is
\begin{displaymath}
T=\int dt a^{-1}=b_{0}\ln(t+\gamma).\end{displaymath} Hence, the
classical solution in conformal time \begin{displaymath}u=C_{1}
\exp(-b_{0}^{-1}T(\mu+1))+ C_{2}\exp(-b_{0}^{-1}T(-\mu+1))
\end{displaymath}
for large momenta (with $\epsilon=1$) $\mu\simeq ikb_{0}$ looks
like a free wave (with a decaying amplitude).

We choose $C_{1}=0$. Then,
\begin{displaymath}
i\Gamma=-\vert t\vert^{3}u^{-1}\partial_{t}u=-t^{2}(\mu-1)<0.
\end{displaymath}
Hence, $\psi_{t}^{g}$ (40) is square integrable.

  The solution of the stochastic Equation (80) for the model (134) is
   \begin{equation}
  \Phi_{s}=u_{t-s}u_{t}^{-1}\Phi+\sqrt{\hbar}u_{t-s}b_{0}^{\frac{3}{2}}
\int_{0}^{s}  u_{t-\tau}^{-1}
  (t+\gamma-\tau)^{-\frac{3}{2}}dW_{\tau}.
  \end{equation}
We are interested in an explicit calculation of the evolution
  kernel and correlation functions. In such calculations, we need to evaluate
 \begin{displaymath}\begin{array}{l}
  G_{s,s^{\prime}}({\bf k},{\bf k}^{\prime})= E[(\Phi_{s^{\prime}}({\bf k})-<\Phi_{s^{\prime}}>({\bf
  k}))
  (\Phi_{s}({\bf k}^{\prime})-<\Phi_{s}>({\bf k}^{\prime}))]\equiv
 G_{s,s^{\prime}}(k)\delta({\bf k}+{\bf k}^{\prime}),
\end{array}\end{displaymath}
where
 $<\Phi>\equiv E[\Phi]$.

 The operator $\mu=\sqrt{1+\epsilon b_{0}^{2}\triangle}$ with $\epsilon=1$  cannot be defined
  in the infinite dimensional setting.
  For this reason, we consider the model with the inverted signature  ($\epsilon=-1$)
  $\triangle\rightarrow -\triangle$ when $\mu$ is a self-adjoint
  positive operator in $L^{2}(d{\bf x})$. When $C_{1}=0$ then, with the inverted signature,
  the operator
  $u_{t-s}u_{t}^{-1}=\frac{t+\gamma}{t+\gamma-s}(\frac{t+\gamma-s}{t+\gamma})^{\mu}$
is well-defined as a contractive semigroup acting upon $\Phi$ in
Equation (137).

We have (with $C_{1}=0$ and $t+\gamma>0$) the following for Model
(134):
  \begin{equation}\begin{array}{l}
  G_{s,s^{\prime}}(k)=    \hbar u_{t-s}u_{t-s^{\prime}}
\int_{0}^{m(s,s^{\prime})}d\tau u_{t-\tau}^{-2}
(t+\gamma-\tau)^{-3}\cr=\frac{\hbar}{2\mu}
(t-s+\gamma)^{-1+\mu}(t-s^{\prime}+\gamma)^{-1+\mu}\cr\Big((t+\gamma-m(s,s^{\prime}))^{-2\mu}
-(t+\gamma)^{-2\mu}\Big),
  \end{array}\end{equation}
where $m(s,s^{\prime})=min(s,s^{\prime})$.

  It can be checked that the rhs of Equation (138) is expressed by a well-defined
  operator $\exp(-r\mu)$  ($\epsilon=-1$),  where $r\geq 0$.

 For equal time in Model (138), we  have
 \begin{equation}\begin{array}{l}
  G_{t}=  \frac{\hbar}{2\mu}
\gamma^{-2}\Big(1-(1+\frac{t}{\gamma})^{-2\mu}\Big)
  \end{array}\end{equation}

 The limit $\gamma=0$ is infinite expressing the degeneracy of the metric
 at $t+\gamma=0$. The evolution kernel is
defined by $G_{t}$ in Equation (111). We can express the
correlation function of Equation (138) by the two-point function
$G^{E}_{M}$ of the scalar free field with a mass {M}
 using the formula
($\nu=\sqrt{k^{2}+M^{2}}$)
\begin{equation}\begin{array}{l}
(2\pi)^{-3}\int d{\bf k}\exp(i{\bf kx})(2\nu)^{-1}\exp(-s\nu)
\cr=(2\pi)^{-4}\int dk_{0}d{\bf k}\exp(i{\bf
kx}+ik_{0}s)(k_{0}^{2}+{\bf k}^{2}+M^{2})^{-1}=G^{E}_{M}(s,{\bf
x})\end{array}
\end{equation}
It can be seen from Equations (138) and (140) that the random field $\Phi_{s}({\bf k})$
has correlation function with the same large $k$ behavior (the
same short distance behavior) as the quantum Euclidean free field
(after the signature inversion). This holds true for all
stochastic fields defined in this paper. In the construction of
polynomial interactions (Sections~\ref{sec11} and~\ref{sec12}),
and in the expansion of non-polynomial interactions in powers of
the fields \cite{salam}, we shall have the same ultraviolet
singularity and the same renormalization problem as in the
conventional Fock space approach or in the constructive Euclidean
framework. However, in the standard approach to QFT in four
dimensions, we cannot go beyond the perturbative framework because
after a renormalization the Feynman--Kac factor becomes unbounded.
In non-polynomial interactions the superpropagator \cite{salam}
becomes extremely singluar. This can change in the stochastic
approach (as discussed in Sections~\ref{sec11} and~\ref{sec12}),
because we can work with (bounded)
 Feynman--Kac oscillatory factors (in particular, the superpropagator becomes an oscillatory hence Lebesgue integrable function).

    \section{The Radiation Background \boldmath{$ a^{2}(t)=c_{0}^{-1}(t+\gamma)$}}\label{sec10}
    In this section, we
      consider the massless   scalar field with $\alpha=\frac{1}{2}$ in Equation (132).
           We insert $\gamma\geq 0$ for $t\geq 0$ and $\gamma\leq 0$ for $t\leq
     0$, so that when $\gamma\neq 0$, the metric is not degenerate.
     The metric (for  $t+\gamma>0 $) is the solution of the Friedmann equation for
     radiation with the energy density  $\rho=\rho_{0}a^{-4}$ and
     the pressure $p=\frac{1}{3}\rho$. It is usually rejected at
     $t+\gamma< 0$ because the inverted signature has no classical
     meaning (\cite{landau}, Section 112),
 as it violates the local
     special relativity principles. The inverted metric can appear
     as a stationary point in quantum gravity, defined as an
     average over the metric tensor. The causal structure in
     quantum gravity can  disagree with the classical one at
     the Planck scale.

     We are interested in the behavior of the quantum scalar field
     evolution for the metric  $ a^{2}(t)=c_{0}^{-1}(t+\gamma)$
     for positive as well as negative time in the limit $\gamma=0$.
     The Gaussian solution (40) is determined by  a solution of the equation
     \begin{equation}
     \frac{d^{2}u}{dt^{2}}+\frac{3}{2}(t+\gamma)^{-1}\frac{du}{dt}
      +(t+\gamma)^{-1}c_{0}k^{2}u=0
      \end{equation}
      true for   $t+\gamma>0$, as well as for $t+\gamma<0$.

For $t+\gamma<0$ we choose $\sqrt{-g}=-i\sqrt{\vert g\vert}$. With
such a choice of the square root, the diffusion Equation (26)
reads
 \begin{equation}\begin{array}{l}
\hbar\partial_{t}\psi_{t}= \frac{1}{2}\int d{\bf x}
\Big(-\hbar^{2}c_{0}^{\frac{3}{2}}\vert
t+\gamma\vert^{-\frac{3}{2}}\frac{\delta^{2}}{\delta\Phi({\bf
x})^{2}}\cr+c_{0}^{-\frac{1}{2}}\vert
t+\gamma\vert^{\frac{1}{2}}(\nabla\Phi)^{2}-M^{2}c_{0}^{-\frac{3}{2}}\vert
t+\gamma\vert^{\frac{3}{2}}\Phi^{2}\Big)\psi_{t}\end{array}
\end{equation}
The equation for $\chi$ is
\begin{equation}\partial_{t}\chi_{t}= \int d{\bf x}
\Big(-\hbar\frac{1}{2}c_{0}^{\frac{3}{2}}\vert
t+\gamma\vert^{-\frac{3}{2}}\frac{\delta^{2}}{\delta\Phi({\bf
x})^{2}}-u^{-1}\partial_{t}u \Phi({\bf
x})\frac{\delta}{\delta\Phi({\bf x})}\Big)\chi_{t}.
\end{equation}
Equations (142) and (143) are well-defined for $t+\gamma<0$,
because the generator of the diffusion has a positively definite
second order differential operator.

We could consider the metric $g_{lk}=\delta_{lk}c_{0}^{-1}\vert
t+\gamma\vert$, which does not have the second derivative at
$t+\gamma =0$. For this reason, it cannot be  a solution  of
Einstein equations, but appears as a solution in the effective
field theory resulting from the string theory \cite{string}. For
this metric, the Schr\"odinger Equation (25) reads
\begin{displaymath}\partial_{t}\chi_{t}= \frac{1}{2}\int d{\bf x}
\Big(i\hbar\frac{1}{2}c_{0}^{\frac{3}{2}}\vert
t+\gamma\vert^{-\frac{3}{2}}\frac{\delta^{2}}{\delta\Phi({\bf
x})^{2}}-u^{-1}\partial_{t}u \Phi({\bf
x})\frac{\delta}{\delta\Phi({\bf x})}\Big)\chi_{t},
\end{displaymath}
where  $u$ is the solution of the equation \vspace{-6pt}
\begin{equation}
     \frac{d^{2}u}{dt^{2}}+\frac{3}{2}(t+\gamma)^{-1}\frac{du}{dt}
      +\vert t+\gamma\vert^{-1}c_{0}k^{2}u=0.
      \end{equation}

In order to obtain a solution of Equation (144), it is useful to
change the cosmic time $t$ into the conformal time  $T$ as
$T=2c_{0}^{\frac{1}{2}}\sqrt{t}$ for $t>0$ and
$T=-2c_{0}^{\frac{1}{2}}\sqrt{-t}$
 for $t<0$ (then $a^{2}(T)\simeq T^{2}$ as in \cite{cpt}).
Inserting  $\tilde{u}=Tu$ into Equation (144), we can see that
$\tilde{u}$ satisfies the oscillator equation. Hence, the solution
of Equation (144) for $t+\gamma> 0$ is a superposition of
plane~waves
\begin{displaymath}
u=A_{1}T^{-1}\exp(ikT)+A_{2}T^{-1}\exp(-ikT).
\end{displaymath}
It follows that the solution of Equation (144) for positive as
well as negative time is
\begin{equation}\begin{array}{l}
       u_{t}=C_{1}\vert t+\gamma\vert^{-\frac{1}{2}}\cos(2k\sqrt{c_{0}}\sqrt{\vert t+\gamma}\vert)
       +C_{2}\vert t+\gamma\vert^{-\frac{1}{2}}\sin(2k\sqrt{c_{0}}\sqrt{\vert t+\gamma\vert})
       \end{array}\end{equation}

We obtain a different solution  of Equation (141). We express it
by real functions (for $t+\gamma> 0$)
       \begin{equation}\begin{array}{l}
       u_{t}=C_{1}(t+\gamma)^{-\frac{1}{2}}\cos(2k\sqrt{c_{0}}\sqrt{t+\gamma})+C_{2}(t+\gamma)^{-\frac{1}{2}}\sin(2k\sqrt{c_{0}}\sqrt{t+\gamma})
       \end{array}\end{equation}
       If $t+\gamma<0$, then the solution of Equation (141) is
    
   \begin{equation}\begin{array}{l}
          u_{t}^{(-)}=C_{1}(-t-\gamma)^{-\frac{1}{2}}\cosh(2k\sqrt{c_{0}}\sqrt{-t-\gamma})+C_{2}(-t-\gamma)^{-\frac{1}{2}}\sinh(2k\sqrt{c_{0}}\sqrt{-t-\gamma})
         \end{array} \end{equation}

For $t+\gamma<0$ Equation (141) (with $k^{2}\rightarrow
-\triangle$) is an elliptic equation. It does not describe~waves.

 The limit $\gamma\rightarrow 0$ of $u_{t}$
  exists for all $\vert t\vert\geq 0$ (see a discussion of continuity in \cite{horowitz,visser1,visser2,ellis}) only if $C_{1}=0$.
Then, the limit $t+\gamma \rightarrow 0$ for
 positive time, as well as  for the negative time, is equal to
 $u_{0}=2C_{2}\sqrt{c_{0}}k$. The limit  $t+\gamma\rightarrow 0$ of $
 \partial_{t}u_{t}$ also exists from both sides
 \begin{displaymath}
(\partial_{t}u_{t})_{\vert
t=0}=-\frac{4}{3}C_{2}\sqrt{c_{0}}c_{0}k^{3}.
\end{displaymath}

        The solution of the stochastic Equation (80) for $t\geq
        s\geq 0$ is
        \begin{equation}\begin{array}{l}
        \Phi_{s}(\Phi)=u_{t-s}u_{t}^{-1}\Phi+\sqrt{i\hbar}u_{t-s}c_{0}^{\frac{3}{4}}\int_{0}^{s}u_{t-\tau}^{-1}
        \vert t+\gamma-\tau\vert^{-\frac{3}{4}}dW_{\tau}.
        \end{array}\end{equation}
      For $-t\geq -s\geq 0$ of the (string) metric, $a(t)\simeq
       \vert t\vert^{\frac{1}{2}}$ the field $\Phi_{s}$ determined by Equations~(77) and (84) is
\begin{equation}\begin{array}{l}
        \Phi_{s}(\Phi)=u_{t-s}u_{t}^{-1}\Phi+\sqrt{-i\hbar}u_{t-s}c_{0}^{\frac{3}{4}}\int_{0}^{-s}u_{t-\tau}^{-1}
        \vert t+\gamma+\tau\vert^{-\frac{3}{4}}dW_{\tau}.
        \end{array}\end{equation}
      The quantum field theory depends  (for a
        positive time) on the correlation
        function\begin{equation}\begin{array}{l}
            G_{ss^{\prime}}=E[(\Phi_{s}({\bf k})-E[\Phi_{s}]({\bf k}))(\Phi_{s^{\prime}}({\bf k}^{\prime})
            -E[\Phi_{s^{\prime}}]({\bf k}^{\prime}))]
            =\delta({\bf k}+{\bf k}^{\prime})i\hbar c_{0}^{\frac{3}{2}}u_{t-s}u_{t-s^{\prime}}\cr\times\int_{0}^{m(s,s^{\prime})}d
            \tau
              u_{t-\tau}^{-2}   \vert
              t+\gamma-\tau\vert^{-\frac{3}{2}}\equiv \delta({\bf k}+{\bf k}^{\prime})   G_{ss^{\prime}}(k).\end{array}
             \end{equation}
             where we denote $m(s,s^{\prime})\equiv min(s,s^{\prime})$ if $t\geq s\geq 0$
             and $t\geq s^{\prime}\geq 0$.
             The $\delta({\bf k}+{\bf
             k}^{\prime})$ term in Equation (150) will be omitted  in the formulas below.

For a negative time  of $a^{2}\simeq t$ in the model (141)--(143),
there is no $\sqrt{i}$ in Equation (82), which is canceled by the
$\sqrt{i}$ factor in $a^{-\frac{3}{2}}$. Hence, the counterpart of
Equation (149) for the negative time leads to the diffusion
process $\Phi_{s}$ solving the diffusion Equation (143)
\begin{displaymath}\begin{array}{l}
        \Phi_{s}(\Phi)=u^{(-)}_{t-s}(u^{(-)}_{t})^{-1}\Phi
        +\sqrt{\hbar}u^{(-)}_{t-s}c_{0}^{\frac{3}{4}}\int_{0}^{-s}(u_{t-\tau}^{(-)})^{-1}
        \vert t+\gamma+\tau\vert^{-\frac{3}{4}}dW_{\tau},
        \end{array}\end{displaymath}
where, by $u^{(-)}_{t}$, we denote the solution (147) of the
``wave equation'' (141) for the negative time. The correlation
functions of the fields for a negative time are determined by
the~formula \vspace{-6pt}
             \begin{equation}\begin{array}{l}
            G_{ss^{\prime}}=E[(\Phi_{s}({\bf k})-E[\Phi_{s}]({\bf k}))(\Phi_{s^{\prime}}({\bf k}^{\prime})
            -E[\Phi_{s^{\prime}}]({\bf k}^{\prime}))]\cr
            =\delta({\bf k}+{\bf k}^{\prime})\hbar c_{0}^{\frac{3}{2}}u^{(-)}_{t-s}u^{(-)}_{t-s^{\prime}}
            \cr\times\int_{0}^{m(s,s^{\prime})}d
            \tau
              (u^{(-)}_{t-\tau})^{-2}   \vert
              t+\gamma+\tau\vert^{-\frac{3}{2}}\equiv \delta({\bf k}+{\bf k}^{\prime})
                G_{ss^{\prime}}(k),\end{array}
             \end{equation}
             where, for the negative time $m(s,s^{\prime})\equiv min(-s,-s^{\prime})$, if $-t\geq -s\geq 0$ and $-t\geq -s^{\prime}\geq 0$.

As can be seen  from Equation (143), for a negative time,
$\Phi_{s}$ becomes a real diffusion process and $K_{t}$ in
Equation (111) is a real transition function. The Schr\"odinger
evolution equation (together with the Feynman--Kac formula)  takes
the form of an evolution of the diffusion process. The reason for
this is the purely imaginary value of $\sqrt{-g}$. If $a(t)\simeq
\vert t+\gamma\vert^{\frac{1}{2}}$ then the expression for
$\Phi_{s}$ and $G_{ss^{\prime}}$ for negative time resembles the
ones for the positive time (for a negative time it is a reflection
of the one for a positive time).

         We can express the integrals (150) and (151) by elementary functions
                  if   $C_{2}=0$, $C_{1}=0$, $C_{1}=\pm C_{2}$ and $C_{1}=\pm iC_{2}$.

         If $C_{2}=0$, then we
         have, for $t+\gamma >0$,
         \begin{equation}
         u_{t}=(t+\gamma)^{-\frac{1}{2}}\cos(2\sqrt{c_{0}}k\sqrt{t+\gamma}).
         \end{equation}
        If $C_{1}=0$, then
         \begin{equation}
         u_{t}=(t+\gamma)^{-\frac{1}{2}}\sin(2\sqrt{c_{0}}k\sqrt{t+\gamma}).
         \end{equation}
          If  $ -t-\gamma>0$, then

            \begin{equation}    u^{(-)}_{t}=(-t-\gamma)^{-\frac{1}{2}}\cosh(2\sqrt{c_{0}}k\sqrt{-t-\gamma})
            \end{equation}
        and
              \begin{equation}
              u^{(-)}_{t}=(-t-\gamma)^{-\frac{1}{2}}\sinh(2\sqrt{c_{0}}k\sqrt{-t-\gamma}).
            \end{equation}

          When $C_{1}=\pm C_{2}$, then
             \begin{equation}
               u_{t}^{(-)}=(-t-\gamma)^{-\frac{1}{2}}\exp(\pm
               2\sqrt{c_{0}}k\sqrt{-t-\gamma}).
              \end{equation}
           For the solution (152), we obtain
              \begin{equation}\begin{array}{l}
              G_{ss^{\prime}}=i\hbar u_{t-s}u_{t-s^{\prime}} c_{0}^{\frac{3}{2}} \int_{0}^{m(s,s^{\prime})} d\tau\vert t+\gamma-\tau\vert^{-\frac{1}{2}}
                (\cos(2\sqrt{c_{0}}k\sqrt{t+\gamma-\tau})
                )^{-2} \cr
                =i\hbar u_{t-s}u_{t-s^{\prime}}c_{0}\frac{1}{k}(\tan(2\sqrt{c_{0}}k\sqrt{t-m(s,s^{\prime})+\gamma})
                -\tan(2\sqrt{c_{0}}k\sqrt{t+\gamma}))
               \end{array}\end{equation}
            where
               \begin{equation}
               u_{0}=\frac{1}{\sqrt{\gamma}}\cos(2\sqrt{c_{0}}k\sqrt{\gamma}).\end{equation}
              The limit $\gamma\rightarrow 0$ of $u_{t}$ does not
               exist at $ t=0$. Then, the evolution kernel (111)
               is not defined. For the solution (153),
              \begin{equation}\begin{array}{l}
              G_{ss^{\prime}}=i\hbar u_{t-s}u_{t-s^{\prime}} c_{0}^{\frac{3}{2}} \int_{0}^{m(s,s^{\prime})} d\tau\vert t+\gamma-\tau\vert^{-\frac{1}{2}}
                (\sin(2\sqrt{c_{0}}k\sqrt{t+\gamma-\tau}) )^{-2}   \cr
               =i\hbar u_{t-s}u_{t-s^{\prime}}c_{0}\frac{1}{k}(\cot(2\sqrt{c_{0}}k\sqrt{t+\gamma})
               -\cot(2\sqrt{c_{0}}k\sqrt{t-m(s,s^{\prime})+\gamma}))\end{array}
                \end{equation}
      The limit $\gamma\rightarrow 0$ of $u_{0}$ in Equation (153)
            is $2\sqrt{c_{0}}k$. When $t\rightarrow 0$ and $\gamma\rightarrow 0$
then $m(s,s^{\prime})\rightarrow 0$ and $
G_{ss^{\prime}}\rightarrow 0$ in Equation (159).

                 For the solution (154), we obtain at
                 $t+\gamma<0$ in Equation (151)
               \begin{equation}\begin{array}{l}
               G_{ss^{\prime}}=\hbar u^{(-)}_{t-s}u^{(-)}_{t-s^{\prime}}  c_{0}^{\frac{3}{2}}
               \int_{0}^{m(s,s^{\prime})} d\tau\vert t+\gamma+\tau\vert^{-\frac{1}{2}}
                 (\cosh(2\sqrt{c_{0}}k\sqrt{-t-\gamma-\tau}) )^{-2}\cr
                 =\hbar u^{(-)}_{t-s}u^{(-)}_{t-s^{\prime}} c_{0}\frac{1}{k}
                 (\tanh(2\sqrt{c_{0}}k\sqrt{-t-m(s,s^{\prime})
                 -\gamma})\cr-\tanh(2\sqrt{c_{0}}k\sqrt{-t-\gamma})).\end{array}
                \end{equation}
                $u_{0}$ has no limit when $\gamma\rightarrow 0$.
                Hence, the propagator (111) cannot be defined in this limit.

                For $u^{(-)}_{t}$ of Equation (155), we have
                \begin{equation}\begin{array}{l}
                G_{ss^{\prime}}=\hbar u^{(-)}_{t-s}u^{(-)}_{t-s^{\prime}} c_{0}^{\frac{3}{2}}
                \int_{0}^{m(s,s^{\prime})} d\tau\vert t+\gamma+\tau\vert^{-\frac{1}{2}}
                  (\sinh(2\sqrt{c_{0}}k\sqrt{-t-\gamma-\tau}) )^{-2}\cr
                  =\hbar u^{(-)}_{t-s}u^{(-)}_{t-s^{\prime}} c_{0}\frac{1}{k}(\coth(2\sqrt{c_{0}}k\sqrt{-t-\gamma})
                  \cr-\coth(2\sqrt{c_{0}}k\sqrt{-t-m(s,s^{\prime})-\gamma}))
                 \end{array}\end{equation}
             The limit $\gamma\rightarrow 0$ of $u_{0}$ in Equation (161)
together with Equation (111)
                  also defines the evolution
                 kernel  in the limit $\gamma\rightarrow
                 0$.
        There is an apparent singularity as $k\rightarrow 0$ in the correlation $G_{ss^{\prime}}(k)$,
                         but this singularity is canceled by the volume element $d{\bf k}$
                 in the definition of the evolution kernel (111).

$G_{ss^{\prime}}$ (161) is decaying for a large $k$ as
 \begin{displaymath}\begin{array}{l}
          G_{ss^{\prime}}\simeq \hbar\frac{1}{2} c_{0}(-t+s-\gamma)^{-\frac{1}{2}}(-t+s^{\prime}-\gamma)^{-\frac{1}{2}}
           k^{-1} \cr\Big(
     \exp(- 4\sqrt{c_{0}}k\sqrt{-t-m(s,s^{\prime})-\gamma}+2\sqrt{c_{0}}k\sqrt{-t-\gamma+s}
    \cr +2\sqrt{c_{0}}k\sqrt{-t-\gamma+s^{\prime}})\cr-\exp(-4\sqrt{c_{0}}k\sqrt{-t-\gamma}
     +2\sqrt{c_{0}}k\sqrt{-t-\gamma+s}
     \cr+2\sqrt{c_{0}}k\sqrt{-t-\gamma+s^{\prime}})\Big)\end{array}
       \end{displaymath}
       The correlation functions of the field $\Phi_{s}({\bf x})$
       can be expressed (according to Equation (150) as
the  Fourier transform of $ G_{ss^{\prime}}$. From the large $k$
behavior of (161) and Equation (140), we can conclude that the
short distance behavior of the correlations of $\Phi_{s}({\bf x})$
is the same as in the Euclidean free field theory of the scalar
field.

The exponential solutions define the correlation functions
          \begin{equation}\begin{array}{l}
          G_{ss^{\prime}}=\hbar u^{(-)}_{t-s}u^{(-)}_{t-s^{\prime}} c_{0}^{\frac{3}{2}}
          \int_{0}^{m(s,s^{\prime})} d\tau\vert t+\gamma+\tau\vert^{-\frac{1}{2}}
            (\exp(-2\sqrt{c_{0}}k\sqrt{-t-\gamma-\tau}) )^{-2}\cr
     =\hbar u^{(-)}_{t-s}u^{(-)}_{t-s^{\prime}}c_{0}\frac{1}{2k}
     (\exp(4\sqrt{c_{0}}k\sqrt{-t-m(s,s^{\prime})-\gamma})\cr-\exp(4\sqrt{c_{0}}k\sqrt{-t-\gamma}))\end{array}
       \end{equation}
 for the minus sign in (156) and
             \begin{equation}\begin{array}{l}
             G_{ss^{\prime}}=i\hbar u^{(-)}_{t-s}u^{(-)}_{t-s^{\prime}} c_{0}^{\frac{3}{2}}
             \cr\times \int_{0}^{m(s,s^{\prime})} d\tau\vert t+\gamma+\tau\vert^{-\frac{1}{2}}
               (\exp(2k\sqrt{-t-\gamma-\tau}) )^{-2}\cr
        =i\hbar u^{(-)}_{t-s}u^{(-)}_{t-s^{\prime}}c_{0}\frac{1}{2k}( \exp(-4\sqrt{c_{0}}k\sqrt{-t-\gamma})
       \cr -\exp(-4\sqrt{c_{0}}k\sqrt{-t-m(s,s^{\prime})-\gamma}))\end{array}
          \end{equation} for the plus sign. There is no limit $\gamma\rightarrow
                 0$ of $u_{0}$.

At the end of this section, let us consider some  superpositions
of the solutions (152) and (153) with complex coefficients. So for
positive time, let us consider (for a negative time the solution
of the ``wave equation'' (141) is given by Equation (147))
\begin{equation}
u_{t}=(t+\gamma)^{-\frac{1}{2}}\exp(2i\sqrt{c_{0}}k\sqrt{t+\gamma}).
         \end{equation}
Calculating the field correlations, we obtain
\begin{equation}\begin{array}{l}
 G_{ss^{\prime}}=\frac{\hbar}{2}\frac{1}{\sqrt{c_{0}}k}\frac{1}{\sqrt{t+\gamma-s}}\frac{1}{\sqrt{t+\gamma-s^{\prime}}}
\exp(2i\sqrt{c_{0}}k(\sqrt{t+\gamma-s}+\sqrt{t+\gamma-s^{\prime}}))\cr
\Big(\exp(-4i\sqrt{c_{0}}k\sqrt{t+\gamma-m(s,s^{\prime}})-\exp(-4i\sqrt{c_{0}}k\sqrt{t+\gamma})\Big)
\end{array}\end{equation}
For $G_{t}$ in the propagator (111), we have \vspace{-6pt}
\begin{equation}\begin{array}{l}
G_{t}=-\frac{1}{2\gamma\sqrt{c_{0}}k}\Big(\exp(-4i\sqrt{c_{0}}k
\sqrt{t+\gamma})+4i\sqrt{c_{0}}k\sqrt{\gamma})-1\Big)
\cr=-\frac{1}{2\gamma\sqrt{c_{0}}k}\Big(\cos(-4\sqrt{c_{0}}k\sqrt{t+\gamma})+4\sqrt{c_{0}}k\sqrt{\gamma})-1
\cr-i\sin(-4\sqrt{c_{0}}k\sqrt{t+\gamma})+4\sqrt{c_{0}}k\sqrt{\gamma})\Big)
\end{array}\end{equation}
Note that $\Re G_{t}>0$. Hence, the function defining
$K_{t}(\Phi,\Phi^{\prime})$ in Equation (111) is integrable (this
would not be so if, instead of $u_{t}$ in Equation (164), we have
considered $u_{t}^{*}$).

With a complex $u_{t}$, the solution $\psi_{t}^{g}$ (40) of the
Schr\"odinger equation is not a phase factor, and it may grow to
infinity for a large $\Phi$ (this would be so for $u_{t}^{*}$).
Let us calculate from Equation (69) for $u_{t}$ of Equation (164)
\begin{equation}
i\Gamma=a^{3}u^{-1}\partial_{t}u=-\sqrt{c_{0}}k(t+\gamma)
-\frac{i}{2}\sqrt{t+\gamma}.
\end{equation}
Hence, $\Re( i\Gamma)<0$, showing that $\psi_{t}^{g}$ is square
integrable for the complex solution (164) for positive time. The
correlation function is
\begin{displaymath}\begin{array}{l} (\psi_{0}^{g}\hat{\Phi}_{t}({\bf
k})\hat{\Phi}({\bf k}^{\prime})\psi_{0}^{g})=
\gamma^{-\frac{1}{2}}\exp(2i\sqrt{c_{0}}k\sqrt{\gamma})(t+\gamma)^{\frac{1}{2}}\exp(-2i\sqrt{c_{0}}k\sqrt{t+\gamma})
\cr \times(2\sqrt{c_{0}}k(t+\gamma))^{-1}\delta({\bf k}+{\bf
k}^{\prime})
\end{array}\end{displaymath}
Then, for a small time,
\begin{displaymath}\begin{array}{l}
(\psi_{0}^{g},\Phi({\bf k})\Phi({\bf
k}^{\prime})\psi_{0}^{g})-(\psi_{0}^{g}\hat{\Phi}_{t}({\bf
k})\hat{\Phi}({\bf k}^{\prime})\psi_{0}^{g})\simeq
\Big(t(4\sqrt{c_{0}}k\gamma^{2})^{-1}+it\gamma^{-\frac{3}{2}}\Big)\delta({\bf
k}+{\bf k}^{\prime})
\end{array}\end{displaymath}
This real and imaginary parts of the diffusive (linear in $t$)
behavior of $\Phi^{2}$ can also be seen from Equation (165).

 For
the negative time with the solution
\begin{equation}
u^{(-)}_{t}=(-t-\gamma)^{-\frac{1}{2}}\exp(2\sqrt{c_{0}}k\sqrt{-t-\gamma})
         \end{equation}
         $\psi_{t}^{g}$
is also square integrable, as
$i\Gamma=(a^{2})^{\frac{3}{2}}u^{(-)}\partial_{t}u^{(-)}<0$.
 For the
solutions (164) and (168) in the free field theory in the
radiation background, we can calculate correlation functions using
the Formula (47), as we did in the case of the de Sitter
background in Equations (90) and (91). However, with these
solutions, $u_{t}$ it is difficult to define the interaction via
the Feynman--Kac formula, because we are unable to prove that the
Feynman--Kac factor is a bounded function (as discussed in
\cite{habarxiv} and in Sections~\ref{sec11} and~\ref{sec12}).

We may consider more general superpositions of solutions of wave
Equation (141) by an addition of a piece with negative frequency
to Equation (164) for a positive time
\begin{equation}
u_{t}=(t+\gamma)^{-\frac{1}{2}}(\exp(2i\sqrt{c_{0}}k\sqrt{t+\gamma})+(\alpha+i\beta)
\exp(-2i\sqrt{c_{0}}k\sqrt{t+\gamma})
         \end{equation}
 and
\begin{equation}
u^{(-)}_{t}=(-t-\gamma)^{-\frac{1}{2}}(\exp(2\sqrt{c_{0}}k\sqrt{-t-\gamma})+(\alpha+i\beta)
\exp(-2\sqrt{c_{0}}k\sqrt{-t-\gamma}).
         \end{equation}
        for a negative time.
We can calculate $\Gamma=a^{3}u^{-1}\partial_{t}u$ for a positive
time and $\Gamma=-i (\vert
a^{2}\vert)^{\frac{3}{2}}(u^{(-)})^{-1}\partial_{t}u^{(-)}$ for a
negative time. We did not find square integrable   solutions
($\Re(i\Gamma)<0$), except for the cases (164) and (168), leading
to a square integrable wave function for a quantum scalar field in
a radiation background. The limit $\gamma\rightarrow 0$ of the
degenerate metric does not exist from both sides of time except of
the solutions (153) and (155).

 Let us summarize the results of this section.
When we choose real solutions $u$, then, for a positive time (when
$-g>0$), $\Gamma$ is real. Hence, $\psi_{t}^{g}$ is a pure phase.
$G_{t}$ (112) (as well as $G_{ss^{\prime}}$) is purely imaginary.
For negative time, $a$ is purely imaginary
$a=ic_{0}^{-\frac{1}{2}}\sqrt{\vert t+\gamma\vert}$. Then,
\begin{equation}
i\Gamma=c_{0}^{-\frac{3}{2}}\vert
t+\gamma\vert^{\frac{3}{2}}u_{t}^{(-)}\partial_{t}u_{t}^{(-)}
\end{equation}
is a real function. We have checked, using Equation (155), that
$i\Gamma$ is negative (hence $\psi_{t}^{g}$ is square integrable)
 for the solution (155) when we obtain
\begin{displaymath}
ic_{0}^{\frac{3}{2}}\vert
t+\gamma\vert^{-\frac{3}{2}}\Gamma=\frac{1}{2}\vert
t+\gamma\vert^{-1} -\sqrt{c_{0}}k\vert
t+\gamma\vert^{-\frac{1}{2}}\coth(2\sqrt{c_{0}}k\sqrt{-t-\gamma})<0
\end{displaymath}
For small $\vert t+\gamma\vert$, we have $i\Gamma\simeq
-\frac{5}{6}\vert
t+\gamma\vert^{\frac{3}{2}}c_{0}^{-\frac{1}{2}}k^{2}$.
 There can be a smooth limit $\gamma\rightarrow 0$ of
quantum scalar field theory when the metric passes from positive
to negative signature. This happens if  $\Gamma$ in the WKB state
(40) is determined by the classical wave function solution (153)
(for $t+\gamma>0$) and (155) (for $t+\gamma<0$). $G_{t}$ is purely
imaginary for a positive time, $G_{0}=0$ and $G_{t}$ becomes a
real positively definite function for a negative time. For the
exponential solutions, $\Re i\Gamma $ is also negative, as
discussed at Equation (168). We have obtained the dynamics of the
fields $\Phi_{s}({\bf x})$ in all cases (152)--(157). However, the
limit $\gamma\rightarrow 0$ of the generate metric exists only for
solutions (153) and (155).
 The solution (153) (for positive time) and
its continuation to (155) (for negative time) are distinguished
from all solutions of the wave equation as the corresponding
Gaussian wave function being the WKB solution for positive time
becomes a Gaussian normalizable wave function for the negative
time. This behavior resembles the one in the elementary WKB
approach when the wave function  $\exp(\frac{i}{\hbar}\int
dx\sqrt{2(E-V)})$, before a potential barrier takes the form
$\exp(-\frac{1}{\hbar}\int dx\sqrt{2(V-E}))$ inside the barrier,
suggesting a tunneling process for the scalar field after crossing
the classical barrier of an inverted signature.

 If $a(t)\simeq \vert t\vert^{\frac{1}{2}}$, then the quantum scalar
field evolution is unitary, and remains oscillatory, whereas for
$a^{2}(t)\simeq t$, the oscillatory behavior for positive time
becomes a diffusion at negative time as if the system encountered
a barrier. In such a case, the evolution fails to be unitary. If
we stopped the evolution at certain space-like surface according
to the Hawking--Penrose singularity theorem \cite{ellis-hawking},
then unitarity would be violated as well. If, in the Lagrangian
(22) and in the Hamiltonian (24), we replace $\sqrt{-g}$ by
$\sqrt{\vert g\vert}$, then the resulting time evolution would
resemble the one of an upside-down oscillator (it would be still
unitary after an inversion of the signature).

Finally, we note that the correlation functions (157) are infinite
 if $2k\sqrt{c_{0}}\sqrt{t+\gamma}=(n+\frac{1}{2})\pi$. In Equation (159), we obtain a  pole at
$2\sqrt{c_{0}}\sqrt{t+\gamma}k=n\pi$, where $n$ is a natural
number. So the fields $\Phi_{s}(\Phi)$ are well-defined for small
$k\sqrt{t+\gamma}$. This difficulty already appears for a quantum
mechanical oscillator of frequency $k$ (see
Appendix~\ref{secappA}). It means that the time evolution on the
WKB states should be carefully extended from small values of time.
The problem is connected with caustic singularities in
semiclassical expansion \cite{maslov}.

\section{Interaction with an Ultraviolet Cutoff}\label{sec11}
In this section, we discuss the Feynman--Kac formula for  states
of the WKB form $\exp(\frac{i}{\hbar}\Phi\Gamma\Phi)\chi$, where
$\Gamma $ is a real bilinear form defined by a real solution of
the wave equation. In models of Section~\ref{sec6}, the real
solution $u$ (leading to a real $\Gamma$) is a real Bessel
function or the mode function in the expansion in spherical
functions in Section~\ref{sec5}. In \cite{habarxiv}, we have shown
in quantum mechanics and in QFT in the Minkowski space-time that
if the first (WKB) term on the rhs of Equation (44) is real, then
$G_{ss^{\prime}}$ is purely imaginary (as it is for the inverted
oscillator). In such a case for trigonometric or exponential
interactions,  the perturbation series is absolutely convergent.
 In this section, we consider the free fields  of Sections~\ref{sec3}--\ref{sec10} and  potentials  of the form
 \vspace{-6pt}
\begin{equation}
V(\Phi)=\lambda\int_{B} d{\bf x}\sqrt{\vert
g\vert}d\mu(\alpha):\exp(i\alpha\Phi_{s}({\bf x})):,
\end{equation}
where $\mu$ is a complex measure, $B\subset R^{3}$ is a bounded
domain, $:- :$
 denotes the normal ordering, $\alpha$ can be real or
purely imaginary, and $\chi$ is of the form
\begin{equation}
\chi(\Phi)=\int
d\nu(\alpha_{0})\exp(i\alpha_{0}(f,\Phi))\end{equation}
 with $f\in
L^{2}(d{\bf x})$. The exponential model ($\alpha$ imaginary)
appears as Starobinski model for inflation \cite{starobinsky} and
the trigonometric interaction as the model of natural inflation
\cite{natural}.

 We discuss also polynomial interactions of the
form
\begin{equation}
V(\Phi)=\lambda\int_{B} d{\bf x}\sqrt{\vert g\vert}:\Phi_{s}^{N}:
({\bf x}),
\end{equation}
where  $N=4n$ and $n$ is a natural number. In the latter case, we
consider the Feynman--Kac solution for the holomorphically
extended initial wave function $\psi(\sqrt{i}\Phi)$ (such states
in the Feynman integral have been discussed first in
\cite{hababook,am}). Extensions of the wave functions (a complex
scaling) are studied in the theory of resonances
\cite{simon,reinhardt}. If in the expanding flat metric (67), we
put the scalar field in a spatial box of length $L$, then ${\bf
k}$ is discrete ${\bf k}=\frac{2\pi {\bf n}}{L}$, with ${\bf
n}=(n_{1},n_{2},n_{3})$, where $ n_{j}$ are integers. For a small
time, the Feynman formula will be well-defined if we restrict the
range of ${\bf n}$. First, we consider general formulas without
specifying the number  of ${\bf k}$ modes or  $\Phi_{lm}$ modes
(59). Then, we explain why a restriction to a finite number of
modes (or an ultraviolet cutoff) is necessary in the case of the
pseudoRiemannian metric.

The solution  of the Schr\"odinger equation with the potential
(172) (positive time, a  bounded region $B$) reads
\begin{equation}\begin{array}{l}
\psi_{t}(\sqrt{i}\Phi)=\psi_{t}^{g}(\sqrt{i}\Phi)E\Big[\exp\Big(-\lambda\frac{i}{\hbar}\int_{B}
d{\bf x}\int_{0}^{t}ds a^{3}
\exp(iN\frac{\pi}{4})\cr:\Big(u_{t-s}u_{t}^{-1}\Phi+\sqrt{\hbar}
u_{t-s}\int_{0}^{s}(a^{\frac{3}{2}}u)^{-1}({t-\tau})dW_{\tau}\Big)^{N}:\Big)
\chi(\sqrt{i}\Phi_{t}(\Phi))\Big]
\end{array}\end{equation}
In Equation (175), a necessary and sufficient condition for the
stochastic integral (82) to be well-defined is that the integral
(112) for $G_{t}$ exists (this is equivalent to $G_{ss^{\prime}}$
being finite). With the processes of Section~\ref{sec10}, this
requirement can be achieved (because of the caustic poles
\cite{maslov}) only for a small time if we have a finite number of
modes or an ultraviolet cutoff $\kappa$ ($k<\kappa$), so that
$t\kappa$ is sufficiently small. For a free field, the ultraviolet
cutoff can be introduced by a restriction of states $\chi$ in
Equation (108) as Fourier transforms  to $\Lambda$, which have
their support on $\vert{\bf k}\vert <\kappa$. Another way is to
introduce in the stochastic Equation (80), and in Equation (175),
the ultraviolet regularized Brownian motion
$W_{s}^{\kappa}\rightarrow W_{s}$ with $\kappa\rightarrow \infty
$. The regularized Brownian motion is defined by the covariance
\begin{equation}
E[W^{\kappa}_{s}({\bf k})W^{\kappa}_{t}({\bf
k}^{\prime})]=min(s,t)\delta({\bf k}+{\bf
k}^{\prime})\rho_{\kappa}(k),
\end{equation}
where $\rho$ is an ultraviolet cutoff restricting $k$ to
$k\leq\kappa$ ($\rho_{\kappa}(k)\rightarrow 1$ when $\kappa
\rightarrow\infty$).
 If $N=4n$, then the exponential in Equation (175)
is bounded by 1. In such a case,
\begin{equation}\begin{array}{l}
\vert\psi_{t}^{\kappa}(\sqrt{i}\Phi)\vert\leq
E\Big[\vert\psi(\sqrt{i}\Phi_{t}(\Phi))\vert\Big]
\end{array}\end{equation}
The rhs of Equation (177) is finite for a large class
of functions (e.g., the ones of Equation~(173)). A renormalization
of the polynomial interaction (175) is required if the
perturbation expansion in the coupling constant $\lambda$ is to be
finite.

  For the trigonometric interaction (172) an expansion of the exponential in Equation~(46)
(inside $E[..]$) leads to integrals
$d\mu(\alpha_{j})d\nu(\alpha_{0})$ of functions of the form
\begin{equation}\begin{array}{l}
\frac{\lambda^{n}}{n!}\int ds_{1}\ldots ds_{n}d{\bf x}_{1}\ldots
d{\bf x}_{n}  E\Big[:\exp(i\alpha_{1}\Phi_{s_{1}}({\bf
x}_{1})):\ldots.:\exp(i\alpha_{n}\Phi_{s_{n}}({\bf x}_{n})): \cr
\times\exp(i\alpha_{0}\int d{\bf x}f({\bf x})\Phi_{t}({\bf
x}))\Big].
\end{array}\end{equation}
The expectation value (178) is ($j\neq r$ because of the normal
ordering)
\begin{equation}\begin{array}{l}
\exp\Big(-\frac{1}{2}\sum_{j\neq r}\alpha_{j}\alpha_{r}\int d{\bf
k}\exp(i{\bf k}({\bf x}_{j}-{\bf x}_{r}))G_{s_{j}s_{r}}(k)\Big)
\end{array}\end{equation} times  functions depending on the
initial value $\Phi$
\begin{equation}
\exp\Big(i\sum_{j}\alpha_{j}(u_{t-s_{j}}u_{t}^{-1}\Phi)({\bf
x}_{j})+i\alpha_{0}\int d{\bf x}(u_{0}u_{t}^{-1}\Phi)({\bf
x})f({\bf x})\Big)
\end{equation}
Most explicitly, the problem with the interaction in the
Feynman--Kac Formula (46) appears in the model of
Section~\ref{sec10}, where $G_{ss^{\prime}}$ has been calculated
exactly. For a finite number of modes and small $s_{j}<t$, the
covariance $G_{s_{j},s_{r}}(k)$ is well-defined (as seen in
Equations (157) and (180)) if the modes satisfy
$2k\sqrt{c_{0}}\sqrt{t+\gamma}<\frac{\pi}{2}$. The Formulas (175)
and (178) do not extend to arbitrary time and arbitrarily large
$k$, because the Formula (157) gives an infinite
$G_{ss^{\prime}}(k)$
 if $2k\sqrt{c_{0}}\sqrt{t+\gamma}=(n+\frac{1}{2})\pi$ (a pole at
this value of $k$). In Equation (159), we obtain a similar pole at
$2\sqrt{c_{0}}\sqrt{t+\gamma}k=n\pi$,  where $n$ is a natural
number. However, the Lebesgue integral over $s_{j}$ and
$\alpha_{j}$ in Equation (178) exists for a finite number of modes
if $G_{t}$ is purely imaginary, even if $k$ is large so that the
trigonometric functions have poles. This is a consequence of the
Lebesgue theorem saying that the Lebesgue integral of a bounded
function with singularities on a set of Lebesgue measure zero
exists. We may hope that, after an integration over $s_{j}$ and
$\alpha_{j}$, we can go to the limit of an infinite number of
modes.

 In de Sitter space in the massless case
($M=0$), we can obtain real solutions $J_{\nu}$ and $Y_{\nu}$ of
the wave equation. With  a finite number of modes for a small
time, we can define the interaction (172) in the Feynman--Kac
formula in de Sitter space-time. The expression (179)  will again
be a pure phase factor (bounded by 1).
 We are unable
to calculate $G_{ss^{\prime}}$ explicitly in this case. However,
we expect that an integral of oscillating functions defining
$G_{ss^{\prime}}$ will again show caustic singularities. Such
caustic poles appear already in the evolution kernel (the Mehler
formula for an oscillator) of free field theory.  For a free
field, even though the Mehler kernel (see
 Appendix A) has the caustic singularities at $t=\frac{\pi
}{\omega}(n+\frac{1}{2})$, the calculation of expectation values
in the ground state gives correlation functions, which have an
extension to an arbitrary time (this may be a consequence of the
fact that the ground state is time-independent). We do not know
whether such an extension of correlation functions is possible in
the models of Sections~\ref{sec8}--\ref{sec10}. In
Sections~\ref{sec6} and~\ref{sec8}, we have discussed complex $u$,
leading to an integrable $\vert\psi_{t}\vert^{2}$. Such a wave
function $u$ defines the free quantum field in de Sitter
space-time and in the universe expanding as $\vert
t\vert^{\alpha}$. The correlation functions in such states show no
caustic poles. In particular, we can construct de Sitter invariant
correlation functions taking the Hankel function $H_{\nu}$ as the
solution of the wave Equation (see Equation (90)).

 In the next section, we show that the inversion of the signature
which removes the caustics allows to define the Feynman integral
without the ultraviolet cutoff (or the restriction on the number
of modes).

\section{Inverted Metric without an Ultraviolet Cutoff}\label{sec12}

In~\cite{habarxiv}, we have discussed the models (172)--(174) (see
Appendixes C and  E in this paper)  in the
Minkowski space-time. We have obtained a convergent perturbation
series by an inversion of the signature of $(\nabla\Phi)^{2}$ in
the Lagrangian (3) (but in contradistinction to Euclidean quantum
field theory the time remains real). In the models of
Sections~\ref{sec8}--\ref{sec10}, the inversion of the signature
means $k^{2}\rightarrow -k^{2}$ in Equation (68) or $k\rightarrow
ik$.

In Section~\ref{sec11}, we have shown that if $\Gamma$ is real,
then we have a well-defined Feynman formula for a finite number of
modes (or an ultraviolet cutoff) and a small time $t$. In this
section, we discuss the Feynman integral for fields on a manifold
with an inverted (Euclidean) metric in the interpretation (27).
Then, we have the Schr\"odinger equation for $t>0$, as well as
$t<0$ and $G_{ss^{\prime}}$ which is purely imaginary and without
the caustic poles, hence the model similar to that discussed in
\cite{habarxiv}. We obtain such a Feynman integral if, in the
integral over metrics (with the interpretation of $\sqrt{-g}$ as
$\sqrt{\vert g\vert}$ for an inverted metric),  there appear
stationary points (as solutions of Einstein equations) with an
inverted signature. The stationary point may appear as the
four-sphere (in addition to the de Sitter space of
Section~\ref{sec6}), as the metric $a^{2}\simeq -t^{2}
  $ in the coasting cosmology of Section~\ref{sec9}, and  in the
   background $a^{2}\simeq t$ of Section~\ref{sec10}, if the Feynman integral is considered for
  negative time. With the inverted signature,
there are no caustics in $G_{ss^{\prime}}$. If the term
$\sqrt{-g}$ in the Feynman integral (or in the Hamiltonian (24))
is replaced by $\sqrt{\vert g\vert}$ (or if the inversion of
signature is considered as a technical tool on an intermediate
stage of the construction of quantum fields), then
$G_{ss^{\prime}}$  will be purely imaginary.
 In such a case, we can apply the Feynman formula of Section~\ref{sec11}
 to trigonometric and $\Phi^{4n}$ interactions with an arbitrarily large $t$
 and without an ultraviolet cutoff (owing to the bound (177)).
 The ultraviolet limit exists, although the terms in the
 perturbation expansion (in $\lambda$ in Equation (175) and in $\alpha$ in Equation (172)) are divergent.
Note that the Formula (175) is well-defined for an arbitrarily
large time with an ultraviolet cutoff, ensuring that the functions
in the exponential are bounded. Because of the bound (177), we may
apply the Lebesgue dominated convergence theorem to claim that the
limit $\kappa\rightarrow \infty$  with
$\rho_{\kappa}(k)\rightarrow 1$ exists. The interaction $\Phi^{4}$
deserves a  detailed studies because of its role in the standard
model. A renormalization of the  interaction  is necessary if the
perturbation series in $\lambda$ is to be finite. The counterterms
for  $\Phi^{4}$ are the same as in the conventional perturbation
expansion because the ultraviolet singularity of the stochastic
field is the same as the one of the quantum field.

We discuss the removal of regularization in more detail in the
case of the trigonometric interactions (172). Using the
representation (172) and (173), we calculate the expectation value
over the Brownian motion $W_{s}^{\kappa}$ of the n-th order term
in the perturbation expansion. We obtain a perturbation series of
the form
\begin{equation}\begin{array}{l}
\lambda^{n}\frac{1}{n!}\int \prod_{r}ds_{r}d{\bf x}_{r}A(\Phi)\exp
\Big(-\sum_{j\neq r}\frac{1}{2}\alpha_{j}\alpha_{r}
G^{\kappa}_{s_{j},s_{r}}({\bf x}_{j},{\bf x}_{r})\Big),
\end{array}
\end{equation}
where $A(\Phi)$ is a bounded function depending  on the initial
value $\Phi$ of the field $\Phi_{s}$. The exponential factor in
Equation (181) is a pure phase as the covariance
$G^{\kappa}_{s,s^{\prime}}$ is purely imaginary. As a consequence,
the perturbation series is absolutely convergent if $\int
d\vert\mu\vert <\infty$. In the model of Section~\ref{sec10}  with
the interpretation (27) of the Hamiltonian for the negative time
(when $a^{2}<0$), there will be no caustics poles so that
$G_{s_{j},s_{r}}({\bf x}_{j},{\bf x}_{r})$ (after the removal of
the ultraviolet cutoff) is well-defined outside the coinciding
points, which have the zero Lebesgue measure $ds_{1} d{\bf
x}_{1}\ldots \ldots ds_{n}d{\bf x}_{n}$. As a consequence of the
Lebesgue theorem, the integral~(181) exists also after the removal
of the ultraviolet cutoff. In the model of Section~\ref{sec9} with
$\epsilon=-1$,
 the function $G_{ss^{\prime}}(k)$ is
non-singular as a function of k. The Fourier transform of
$G_{ss^{\prime}}(k)$ in Equation (179) is singular when
$(s_{j},{\bf x}_{j})\rightarrow (s_{r},{\bf x}_{r})$, because
$G_{ss^{\prime}}(k)$ does not fall fast enough for large $k$.
 So, $G_{ss^{\prime}}({\bf x}_{j}-{\bf x}_{r})$  is divergent at the
coinciding points $(s,{\bf x}_{j})\rightarrow (s^{\prime},{\bf
x}_{r})$. This is the reason (as discussed at the end of
Section~\ref{sec2}) that, in the Euclidean space,  the
non-polynomial interactions cannot be defined. In the Minkowski
space-time, the two-point function is a complex distribution
singular on the light-cone, so that the Formula (181) would be
untractable \cite{salam}. However, with the real time and the
inverted metric, the exponential in Equation (181) is a pure phase
bounded by 1. The integral over the singular points can be treated
by means of the Lebesgue lemma. According to the Lebesgue lemma,
an integral of a bounded function with singularities on a set of
measure zero exists.  When we consider an ultraviolet regularized
model (as in Section~\ref{sec11}) then, applying the Lebesgue
dominated convergence theorem, we can conclude that, in the limit
 $\kappa\rightarrow \infty$, the expression (178) is integrable
and the integrals are bounded by $\vert\lambda\vert^{n}\vert
B\vert^{n}t^{n}\frac{1}{n!}$, because the exponential (181) is a
pure phase.

An analogous formula can be derived in the (Euclidean) de Sitter
 space in the metric (56) and the interpretation (27) of the Hamiltonian.
In such a case, the process is generated by the Hamiltonian (61).
Then, instead of the Formula~(181), we obtain (where $\omega_{j}$
are coordinates on $s^{3}$) \vspace{-6pt}
\begin{equation}\begin{array}{l}\int \prod_{j}ds_{j}d\omega_{j}
\exp\Big(-\frac{1}{2}\sum_{j\neq r}\alpha_{j}\alpha_{r}\int
\sum_{lm}Y_{lm}(\omega_{r})Y_{lm}(\omega_{j})^{*}G^{lm}_{s_{j}s_{r}}\Big)
\end{array}\end{equation}
$G^{lm}_{s_{j}s_{r}}$ is  purely imaginary and without caustic
singularities. The infinite sum in the exponential (182) is
convergent to a purely imaginary function singular at the
coinciding points $(s_{j},\omega_{j})$ and $(s_{r},\omega_{r})$.
It has the same singularity as the Green function on $S^{4}$.
Hence, similarly as in Equation (181), we can prove the existence
of the integrals and their non-triviality using Lebesgue lemma on
dominated convergence.

In the metric (57), the analytic continuation of the metric to the
Euclidean (58) leads to a field theory on AdS. As discussed at the
end of Section~\ref{sec5} by means of the analytic continuation
$k\rightarrow ik$, we obtain a real solution $u(k)$ of the ``wave
equation'' on the Euclidean anti-de Sitter
 space. With the real $u$, the corresponding stochastic field has
 the $G_{ss^{\prime}}$ correlation function which is purely
 imaginary. In such a case, the expression (179) is integrable,
 defining the Euclidean quantum field theory on AdS.

\section{Summary}\label{sec13}

We have developed a functional Schr\"odinger description of the
time evolution of the scalar field in an external metric, which is
a solution of Einstein equations.  We do not restrict ourselves to
the solutions with the Lorentzian signature, but also discuss
solutions of Einstein equations with an Euclidean signature. Such
a metric can be relevant when averaging the scalar field theory
over all metrics in quantum gravity. In such a case, all saddle
points contributing to the path integral over the scalar field
should be taken into account. We work in the functional
formulation of quantum field theory. Solutions of the
Schr\"odinger equation define a random field whose correlation
functions determine correlation functions of the quantum field. We
consider Gaussian solutions of the Schr\"odinger equation for free
field theory. The interaction is introduced by the Feynman--Kac
formula. A proper choice of the Gaussian solution facilitates a
definition of an integrable Feynman--Kac factor.
  We have shown that Gaussian
solutions of the Schr\"odinger equation for a free field theory
are determined by the classical solutions of the wave equation in
an external metric. We studied in detail fields on de Sitter
space-time and on the flat expanding space-time. The stochastic
field correlation functions in a flat expanding space-time can be
expressed by well-known cylinder functions, whereas the ones in de
Sitter space require infinite series of Legendre functions.
 We have found a particular solution in a
radiation background, which is of the WKB form  (a Gaussian phase
factor) for a positive time and a Gaussian square integrable
function for a negative time (one may consider it as composed from
two solutions for positive and negative time). This is an analog
of the Gibbons--Hartle--Hawking solution in de Sitter space-time,
which is glued from a solution for a positive time and another one
for an imaginary time (positive and negative signature). We
construct a general solution of the Schr\"odinger equation as a
perturbation of the Gaussian (WKB) solution.  We show how to
calculate correlation functions of quantum fields in terms of
correlation functions of stochastic fields. Then, the Feynman--Kac
formula can be applied to construct solutions of the Schr\"odinger
equation with an interaction. We discuss polynomial and
trigonometric interactions. It is shown that the WKB solutions
determined by real solutions of the classical wave equation (with
an inverted signature) are distinguished from the point-of-view of
the Feynman--Kac formula. The WKB solutions define stochastic
fields,  which allow for the definition of the Feynman--Kac factor
as a bounded function. We briefly discussed the $\Phi^{4}$
interaction with a positive and negative coupling, important for
the standard model and models of inflation. We pointed out that,
with an inverted signature, an analytic continuation and the
standard renormalization, the Feynman--Kac factor can be bounded
and defined with finite renormalized perturbation series. This
model requires further investigation. We studied, in detail, the
trigonometric interaction showing that the perturbation series of
the Schr\"odinger wave function evolution can be expressed by pure
phase factors. The phase factors are well-defined for a small time
and a finite number of modes in the Lorentzian metric, and for an
arbitrary time with an infinite number of modes (no ultraviolet
cutoff) in the case of the inverted metric with the Hamiltonian
(27). Then, a further Lebesgue integration of these phase factors,
which are singular on a set of zero Lebesgue measure, gives a
finite result. This is in contradistinction to the standard
Minkowski  QFT of the trigonometric interaction, when the vacuum
correlation functions are exponentials of unbounded complex
functions (or distributions) which have infinite Lebesgue
integrals. In this paper, a functional Schr\"odinger evolution has
been derived for a scalar field. We shall extend this formulation
to gauge fields when random self-duality equations will play the
role of stochastic equations for the scalar field.

\section{Appendices A-H}

{\bf A: Expansion
 around a Time-Dependent
Solution in Euclidean and Minkowski QFT}

\setcounter{equation}{0}

We consider a solution of Equation (17)

\begin{equation}{1}
u_{t}=\cosh(\nu t),
\end{equation}where $\nu=\sqrt{-\triangle +M^{2}} $.

Then, $\Gamma_{t}=\nu\tanh(\nu t)$. The stochastic equation (9)
reads
\begin{equation}
d\Phi_{s}=-\nu\tanh(\nu(t-s))\Phi_{s} ds+\sqrt{\hbar}dW_{s}.
\end{equation}
With $u_{t}$ of Equation (14.1) solution of Equation (14.2) is
\begin{equation}\begin{array}{l}
\Phi_{s}=\cosh(\nu(t-s))(\cosh(\nu t))^{-1}\Phi\cr+ \sqrt{\hbar}
\cosh(\nu(t-s)) \int_{0}^{s} \Big(\cosh(\nu
(t-\tau))\Big)^{-1}dW_{\tau}.
\end{array}\end{equation}
We can express the solution of Equation (6) in the form
 \begin{displaymath}\begin{array}{l}
 \chi_{t}(\Phi)=E[\chi(\Phi_{t}(\Phi))]
 =\int d\Phi^{\prime}K_{t}(\Phi,\Phi^{\prime})\chi(\Phi^{\prime}),
  \end{array}\end{displaymath}
   where
   \begin{equation}\begin{array}{l}
  K_{t}(\Phi,\Phi^{\prime})=E[\delta(\Phi^{\prime}-\Phi_{t}(\Phi))]
  =\int d\Omega
 E[ \exp(i(\Omega,\Phi^{\prime}-\Phi_{t}(\Phi))].
 \end{array}\end{equation}
We  calculate the expectation value (14.4). For this purpose, we
 need
 \begin{displaymath}\begin{array}{l}
 E\Big[ \Big(\int_{0}^{t} \Big(\Omega, \Big(\cosh(\nu
 (t-\tau))\Big)^{-1}dW_{\tau}\Big)^{2}\Big]
 \cr=  \int_{0}^{t} \Big(\Omega, \Big(\cosh(\nu
    (t-\tau))\Big)^{-2}\Omega\Big)d\tau
 =\Big(\Omega,\frac{1}{\nu}\tanh(\nu t)\Omega\Big)
 \end{array}\end{displaymath}
In order to calculate the evolution kernel, we need to perform the
 $\Omega$ integral in Equation~(14.4) with the result (the Mehler formula
 for an imaginary time)
 \vspace{-6pt}
 \begin{equation}
 \begin{array}{l}
 K_{t}(\Phi,\Phi^{\prime})=\det\Big(\nu\coth(\nu t)\Big)^{\frac{1}{2}}
 \exp\Big( -\frac{1}{2\hbar} \Big(\Phi^{\prime}-(\cosh(\nu
 t))^{-1}\Phi\Big)\cr\times
\nu\coth(\nu t) \Big(\Phi^{\prime}-(\cosh(\nu t )^{-1}\Phi\Big)
\Big)
\end{array}
\end{equation}
The formula is well-defined in an infinite number of dimensions.
 It also follows from the Mehler formula \cite{merzbacher}.

 We can perform the whole procedure in the real-time 
   $t\rightarrow
 it$. Then, the stochastic equation (44) reads
\begin{equation}
d\Phi_{s}=\nu\tan(\nu(t-s))\Phi_{s} ds+\sqrt{i\hbar}dW_{s}
\end{equation}

The formula for the evolution kernel (14.5) takes the form
   \begin{equation}
   \begin{array}{l}
   K_{t}(\Phi,\Phi^{\prime})=\det\Big(i\nu\cot(\nu t)\Big)^{\frac{1}{2}}
   \cr
   \exp\Big( \frac{i}{2\hbar} \Big(\Phi^{\prime}-(\cos(\nu
   t))^{-1}\Phi\Big)
 \nu\cot(\nu t) \Big(\Phi^{\prime}-(\cos(\nu t) )^{-1}\Phi\Big) \Big)
 \end{array}
 \end{equation}
In Equation (14.7), we encounter the difficulty that
    $\cot(\nu t)^{-1} $ is infinite when $\nu(k)t =(n+\frac{1}{2})\pi$.
    Equation (14.7) can make sense for a small time if we introduce an ultraviolet cutoff restricting the range of $k$.
       Let us note that the ultraviolet cutoff is
unnecessary if
 we flip the signature (but the time remains real)
 \begin{displaymath}
 \nabla^{2}\rightarrow -\nabla^{2}
 \end{displaymath}
 together with $M^{2}\rightarrow -\mu^{2}$. In such a case in Equation (14.7), $
 \nu
 \rightarrow i\nu $, $\cos( \omega t) \rightarrow \cosh(\nu t)
 $ and
 $\nu\cot( \nu t) \rightarrow \nu \coth(\nu t)$, where
 $\nu=\sqrt{-\triangle +\mu^{2}}$.
Now, the stochastic Equation (14.6) (real time but an inverted
signature) reads

\begin{equation}
d\Phi_{s}=\nu\tanh(\nu(t-s))\Phi_{s} ds+\sqrt{i\hbar}dW_{s}
\end{equation}
with the solution
\begin{displaymath}
\Phi_{s}=\cosh((t-s)\nu)(\cosh(t\nu))^{-1}\Phi+
\sqrt{i\hbar}\cosh((t-s)\nu)\int_{0}^{s}\Big(\cosh((t-\tau)\nu)\Big)^{-1}dW_{\tau}
\end{displaymath}
We calculate $G_{ss^{\prime}}$ with the result
\begin{displaymath}
G_{ss^{\prime}}=i\hbar\cosh((t-s)\nu)\cosh((t-s^{\prime})\nu)\nu^{-1}
(\tanh(\nu(t-m(s,s^{\prime}))-\tanh(\nu t))
\end{displaymath}

 The propagator  resulting from
 Equation (48)
  (this  is the propagator for an upside-down oscillator discussed in more detail in Appendixes C and D) is
 \begin{equation}
   \begin{array}{l}
   K_{t}(\Phi,\Phi^{\prime})=\det\Big(i\nu\coth(\nu t)\Big)^{\frac{1}{2}}
   \exp\Big( \frac{i}{2\hbar} \Big(\Phi^{\prime}-(\cosh(\nu
   t))^{-1}\Phi\Big)
 \nu\coth(\nu t) \cr\Big(\Phi^{\prime}-(\cosh(\nu t) )^{-1}\Phi\Big) \Big)
 \end{array}
 \end{equation}
This model could be applied for a construction of a time evolution
of trigonometric and $\Phi^{4n}$ interactions, as in
Section~\ref{sec12} and \cite{habarxiv} (where in \cite{habarxiv}
we used $\psi^{g}=\exp(\frac{i}{2\hbar}\Phi\nu\Phi)$ instead of
$\psi^{g}=\exp(\frac{i}{2\hbar}\Phi\nu\tanh(\nu t)\Phi)$
considered in this appendix).

{\bf B:Field Correlations and the Propagator at
Large Momenta}

We consider the wave equation
\begin{equation}
\frac{d^{2}u}{dt^{2}}+3H\frac{du}{dt}+a^{-2}k^{2}u+M^{2}u=0
\end{equation}
at large $k$. Let $u=a^{-\frac{3}{2}}v$, then $v$  satisfies the
equation
\begin{equation}
\frac{d^{2}v}{dt^{2}}+\frac{dS}{dt}v=0
\end{equation}
where
\begin{equation}
S(s)=\int_{0}^{s}dt \sqrt{M^{2}+
a^{-2}k^{2}-\frac{3}{4}a^{-2}(\frac{da}{dt})^{2}-\frac{3}{2}a^{-1}\frac{d^{2}}{dt^{2}}a}
\end{equation}
The field correlation function
\begin{displaymath}\begin{array}{l}
  G_{s,s^{\prime}}({\bf k},{\bf k}^{\prime})= E[(\Phi_{s^{\prime}}({\bf k})-<\Phi_{s^{\prime}}>({\bf
  k}))
  (\Phi_{s}({\bf k}^{\prime})-<\Phi_{s}>({\bf k}^{\prime}))]\cr\equiv
  G_{s,s^{\prime}}(k) \delta({\bf k}+{\bf k}^{\prime})
\end{array}\end{displaymath} is
\begin{equation} G_{s,s^{\prime}}(k)=i\hbar
u_{t-s}u_{t-s^{\prime}}\int_{0}^{m}d\tau
u_{t-\tau}^{-2}a_{t-\tau}^{-3}d\tau,
\end{equation}where $m=min(s,s^{\prime})$.

In the large $k$  (the WKB method), we obtain the approximate
solution of Equation (14.11)
\begin{equation}
v(t)=(\frac{dS(t)}{dt})^{-\frac{1}{2}}\exp(\pm iS(t)).
\end{equation}
We can form the even solution
\begin{equation}
v(t)=(\frac{dS(t)}{dt})^{-\frac{1}{2}}\cos(S(t)).
\end{equation}
In the odd solution, we replace $\cos$ by $\sin$.

With the inverted metric (and $M^{2}\rightarrow -\mu^{2}$)
\begin{equation}
S^{E}(s)=\int_{0}^{s}dt \sqrt{\mu^{2}+
a^{-2}k^{2}+\frac{3}{4}a^{-2}(\frac{da}{dt})^{2}+\frac{3}{2}a^{-1}\frac{d^{2}}{dt^{2}}a}
\end{equation}
Then, the even solution is
\begin{equation}
v^{E}(t)=(\frac{dS^{E}(t)}{dt})^{-\frac{1}{2}}\cosh(S^{E}(t))
\end{equation}and the
exponentially growing solution
\begin{equation}
v^{E}(t)=(\frac{dS^{E}(t)}{dt})^{-\frac{1}{2}}\exp(S^{E}(t))
\end{equation}
With $u=a^{-\frac{3}{2}}v$, we can calculate $
G_{s,s^{\prime}}(k)$ from Equation (14.13)  for the even
solution, for the Euclidean even solution, and for the Euclidean
exponentially growing solutions. The results are expressed by
Equations (105)--(107), where we skip the indices $lm$. So for the
Euclidean exponentially growing solution, we obtain
($u^{E}=a^{-\frac{3}{2}}v^{E}$)
\begin{equation}\begin{array}{l}
G_{s,s^{\prime}}=\hbar
u^{E}(\tau-s)u^{E}(\tau-s^{\prime})(\exp(-2S^{E}(\tau-m(s,s^{\prime}))
-\exp(-2 S^{E}(\tau)) \end{array}\end{equation} For a large $k$,
the correlation $G_{s,s^{\prime}}$ behaves as
$A(s,s^{\prime})k^{-1}\exp(-B(s,s^{\prime})k)$ with certain
functions $A$ and $B$. Hence, it has the same short distance
behavior as the two-point function of the Euclidean free scalar
field (see Equation (140)).

 {\bf C:The Oscillator with an Inverted Signature in Quantum Mechanics}

For the convenience of the reader, we briefly discuss the results
of \cite{habarxiv} in this appendix and Appendix E,
which constitute  simplified version of the models considered in
this paper. We consider the Schr\"odinger equation in quantum
mechanics
\begin{equation}
i\hbar\partial_{t}\psi_{t}=(-\frac{\hbar^{2}}{2}\nabla_{x}^{2}-\frac{\nu^{2}
x^{2}}{2}+\tilde{V}(x))\psi_{t}\equiv
\hat{H}_{0}+\tilde{V}\psi_{t}.
\end{equation}
We write the solution of the Schr\"odinger Equation (14.20) in
the form
\begin{equation} \psi_{t}(x)=\psi_{t}^{g}\chi=\exp(-\frac{\nu}{2}t) \exp(i\frac{\nu
x^{2}}{2\hbar})\chi_{t}(x).
\end{equation}
We express the solution by the Brownian motion.

  For the wave function $\psi_{t}^{g}=\exp(-\frac{\nu}{2}t)\exp (\frac{i\nu}{\hbar}x^{2})$,
    the stochastic Equation (44)
reads
\begin{equation}
dq_{s}=-\nu q ds+\sqrt{i\hbar} dw_{s}.
\end{equation}

We assume that the initial wave function $\chi$ and the potential
$\tilde{V}$ are holomorphic functions. Then, the solution of
Equation (14.20) is given by the Feynman--Kac formula
\begin{equation}
\chi_{t}(x)=E\Big[\exp\Big(-\frac{i}{\hbar}\int_{t_{0}}^{t}ds
\tilde{V}(q_{s}(x))\Big)\chi(q_{t}( x))\Big],\end{equation} here,
$q_{s}(x)$ is the solution of the Langevin Equation (14.22)
with the initial condition $q_{t_{0}}(x)=x$. The solution
(14.23) has been discussed earlier in
\cite{doss1,doss2,habajp,hababook,habaepj}. It is a real-time
version of the Feynman--Kac formula \cite{freidlin,simonbook}.

We define the evolution kernel (for $\tilde{V}=0$) as
\begin{equation}\begin{array}{l}
\exp(-\frac{\nu t}{2}+\frac{i\nu x^{2}}{2})E[\chi(q_{t}(x))] =\int
K(t;x,y)\exp(\frac{i\nu y^{2}}{2})\chi(y) dy
=(U_{t}\psi)(x)\end{array}\end{equation} leading to the result
(the Mehler formula for the evolution kernel of an oscillator with
$\omega\rightarrow i\nu$)~\cite{merzbacher})
\begin{equation}\begin{array}{l}
K(t;x,y)=\exp(t\hat{H}_{0})(x,y)=(2\pi\hbar i\nu^{-1}\sinh(\nu
t))^{-\frac{1}{2}}\cr \exp\Big(\frac{i\nu}{2\hbar\sinh (\nu
t)}\Big((x^{2}+y^{2})\cosh(\nu t)-2xy\Big)\Big).\end{array}
\end{equation}
It is related to the kernel (14.9) through a similarity
transformation by means of the WKB factor
$\exp(\frac{i}{2\hbar}x\nu x)$.

 We consider potentials of the form of the Fourier transforms of a
complex \mbox{measure \cite{abh,habajmp}}

\begin{equation}
\tilde{V}(x)=g\int d\mu(a)\exp(i ax),
\end{equation}
and wave functions of the same form
\begin{equation}
\psi(x)=\int d\rho(a_{0})\exp(i a_{0}x),
\end{equation}where $a\in R$.

 We prove that the solution of Equation (14.20) can be expressed
as a convergent perturbation series if $\int
d\vert\mu\vert<\infty$ and $\int d\vert \rho\vert<\infty$

\begin{equation} \begin{array}{l}
\chi_{t}(x)=E\Big[\sum_{n}\frac{1}{n!}
\Big(-\frac{i}{\hbar}\int_{t_{0}}^{t}\tilde{V}(q_{s})ds\Big)^{n}\chi_{0}(q_{t}(x))\Big],
\end{array}\end{equation}
We show that the perturbation series (14.28) in powers of
$\tilde{V}$ is absolutely convergent.
 With the Fourier representation (14.26), we can see that
the $N$-th order term is  of  the form (a simpler version of
Equation (181)) 
\begin{equation}\begin{array}{l}
\int\prod_{r} d\mu(a_{r})\exp(\sum_{j,k}f(a_{j},a_{k}))\cr\exp
\Big(-\frac{1}{2}i\hbar\sum_{j,k}a_{j}a_{k}
\int_{0}^{min(s_{j},s_{k})}\exp(-\nu(s_{j}+s_{k}-2s))ds\Big),
\end{array}
\end{equation}
where
\begin{equation}
f(a_{j},a_{k})=i\alpha a_{j} x\exp(-\nu s_{j})+i\alpha a_{k}
x\exp(-\nu s_{k}).
\end{equation}
It is clear that absolute values of the terms (14.29)
integrated over $s$ are bounded by 1, leading to a convergent
perturbation expansion in which each term (14.28) is bounded by
$\frac{1}{n!}t^{n}C^{n}$ with a certain positive constant $C$.

{\bf D:QFT in a Formal $ \hbar$ Expansion }

We  calculate the generating functional  in a formal expansion in
$\hbar$ (up to 
 the $O(\sqrt{\hbar})$ terms) for the Lagrangian
(22) with an inverted signature of the spatial metric  in the
Minkowski space-time, and an inverted sign of the mass square
$M^{2}\rightarrow -\mu^{2}$\begin{equation}
\begin{array}{l}Z[J]=\int d\phi\exp(\frac{i}{\hbar}\int dx({\cal
L}+J\phi))= \exp\Big(\frac{i}{\hbar}\int dx ({\cal
L}(\phi_{c})+J\phi_{c})\Big)\cr \det\Big(i(
-\partial_{t}^{2}-\nabla^{2}+\mu^{2}-\tilde{V}^{''}(\phi_{c}))\Big)^{-\frac{1}{2}},
\end{array}\end{equation}
where
\begin{displaymath}
{\cal
L}=\frac{1}{2}((\partial_{t}\phi)^{2}+(\nabla\phi)^{2}+\mu^{2}\phi^{2})-\tilde{V}(\phi)
\end{displaymath}
and $\phi_{c}(t,{\bf x})\equiv \phi^{c}_{t}({\bf x})$ is the
solution of the equation
\begin{equation}
(-\partial_{t}^{2}-\nabla^{2}+\mu^{2})\phi_{c}-\tilde{V}^{\prime}(\phi_{c})=-J
.
\end{equation}
For the propagator, we have the expression
\begin{equation}\begin{array}{l}
K(t;\phi,\phi^{\prime})=\int_{\phi_{0}=\phi,\phi_{t}=\phi^{\prime}}
d\phi\exp(\frac{i}{\hbar}\int dx{\cal L})=
\exp\Big(\frac{i}{\hbar}\int dx {\cal L}(\phi_{c})\Big)\cr\times
\det\Big(i(
-\partial_{t}^{2}-\nabla^{2}+\mu^{2}-\tilde{V}^{\prime\prime}(\phi_{c}))\Big)^{-\frac{1}{2}},
\end{array}\end{equation}where
\begin{equation}
(-\partial_{t}^{2}-\nabla^{2}+\mu^{2})\phi_{c}-\tilde{V}^{\prime}(\phi_{c})=0.
\end{equation}
Equation (14.34) is solved with the boundary conditions
$\phi^{c}_{0}=\phi,\phi^{c}_{t}=\phi^{\prime}$. Equations~(14.31)
and (14.33) have a  form similar to the ones in Euclidean field
theory, but the potential enters with the opposite sign.

For $\tilde{V}=0$, we can obtain explicit formulae from Equations
(14.31)--(14.34)
\begin{equation}
Z[J]=\exp(-\frac{1}{2\hbar}JGJ), \end{equation}
 where
\begin{equation}\begin{array}{l}
G(t,{\bf x};t^{\prime},{\bf x}^{\prime})=i\Big(\exp (-\nu\vert
t-t^{\prime}\vert)(2\nu)^{-1}\Big)({\bf x},{\bf x}^{\prime})\equiv
iG^{E}(t,{\bf x};t^{\prime},{\bf x}^{\prime})\end{array}
\end{equation}
where  $G^{E}$ is the two-point function for the Euclidean free
field (with $\nu=\sqrt{-\triangle +\mu^{2}}$).

In the expanding flat metric with no potential, the formula for
the evolution kernel~reads
\begin{equation}\begin{array}{l}
K(t;\phi,\phi^{\prime})=\int_{\phi_{0}=\phi,\phi_{t}=\phi^{\prime}}
d\phi\exp(\frac{i}{\hbar}\int dx\sqrt{-g}{\cal L})=
\exp\Big(\frac{i}{\hbar}\int dx \sqrt{-g}{\cal
L}(\phi_{c})\Big)\cr\times \det\Big(i(
-\partial_{t}^{2}-a^{-2}\nabla^{2}-3a^{-1}\partial_{t}a\partial_{t}+\mu^{2})\Big)^{-\frac{1}{2}},
\end{array}\end{equation} where
\begin{equation}
(-\partial_{t}^{2}-\nabla^{2}-3a^{-1}\partial_{t}a\partial_{t}+\mu^{2})\phi_{c}=0
\end{equation}
is solved with the boundary conditions
$\phi^{c}_{0}=\phi,\phi^{c}_{t}=\phi^{\prime}$. We would obtain
the form of the kernel (111) if we could solve the boundary
problem (14.38) explicitly. The determinant (for a given
$a(t)$) depends only on time. Then, $\exp\Big(\frac{i}{\hbar}\int
dx {\cal L}(\phi_{c})\Big)$ gives the quadratic form in the
exponential of Equation (111). This way of calculating the
evolution kernel is discussed in \cite{prok1} (without an
inversion of the metric).

{\bf E:Feynman Integral in  QFT of Trigonometric
Interactions}

An extension of Equations (14.20)--(14.22) to QFT   takes the form
(after a subtraction of the infinite vacuum energy)
\begin{equation}
\psi_{t}(\phi)=\exp(\frac{i}{2\hbar}\phi\nu\phi)
E[\chi(\phi_{t}(\phi))] \end{equation} where
\begin{equation}
\phi_{t}(t_{0},\phi)=\exp(-\nu
(t-t_{0}))\phi+\sqrt{i\hbar}\int_{t_{0}}^{t}\exp(-\nu(t-s))dW_{s},
\end{equation} where $\nu=\sqrt{-\triangle+\mu^{2}}$.

From Equation (14.40), we obtain the correlation function  in field
theory as
\begin{equation}\begin{array}{l}
E[\phi_{t}(\phi,{\bf y})\phi_{s}(\phi,{\bf x})]\cr=
(\exp(-(t-t_{0})\nu)\phi)({\bf y})(\exp(-(s-t_{0})\nu)\phi)({\bf
x})\cr+\Big(\frac{1}{2\nu}\exp(-\nu(t+s-2t_{0}))\Big)({\bf x},{\bf
y})+G(t,{\bf y};s,{\bf x})\end{array}\end{equation} with
\begin{equation}\begin{array}{l}
G(t,{\bf y};s,{\bf x})=i(-\partial_{0}^{2}-\triangle
+\mu^{2})^{-1}(t,{\bf y};s,{\bf
x})\cr=\frac{i}{2}\Big(\nu^{-1}\exp(-\nu\vert t-s\vert)\Big)({\bf
x},{\bf y})=iG^{E}(t,{\bf y};s,{\bf x}),
\end{array}\end{equation}
where $G^{E}$ is the two-point function of Euclidean free  quantum
field. The Feynman--Kac formula reads
\begin{equation}\begin{array}{l}
\psi_{t}(\phi)= \exp(\frac{i}{2\hbar}\phi\nu\phi) \cr
E\Big[\exp\Big(-\frac{i}{\hbar}\int_{t_{0}}^{t}
:\tilde{V}(\phi_{s}(\phi,{\bf x})):d{\bf
x}ds\Big)\chi(\phi_{t}(\phi))\Big].
\end{array}\end{equation}
For the  exponential interaction (172), the \emph{n}-th order term
has the form
\begin{equation}
\begin{array}{l}
\int_{\Omega}d{\bf x}_{1}\ldots \ldots d{\bf x}_{n}ds_{1}\ldots
ds_{n}d\mu(a_{1})\ldots d\mu(a_{n})\prod_{j\neq k}\cr
\exp\Big(-\frac{1}{2}i\hbar
a_{j}a_{k}\int_{t_{0}}^{\min(s_{j},s_{k})}
\cr\Big(\exp(-(s_{j}+s_{k}))\nu)\exp(2\tau\nu)\Big)({\bf
x}_{j},{\bf x}_{k})d\tau\Big).
\end{array}
\end{equation}
The absolute value of the integrand (14.44) is 1,
as in quantum mechanics in Appendix C (Equation
(14.29)) and in the models with an inversion of the metric in
Section~\ref{sec12} (\mbox{Equation~(181)}). The formulas of this
appendix exhibit a simpler and more explicit version of the
discussion of Sections~\ref{sec11} and~\ref{sec12}.

{\bf F:De Sitter Space in the Cosmic Time}

 We obtain for the metric (51) the Hamiltonian (24)

\begin{equation}\begin{array}{l} {\cal H}= \frac{1}{2}\sum_{lm}
\Big((\frac{1}{H}\cosh(Ht))^{-3}\Pi_{lm}^{2}+\frac{1}{H}\cosh(Ht)l(l+2)\Phi_{lm}^{2}
+M^{2}(\frac{1}{H}\cosh(Ht))^{3}\Phi_{lm}^{2}\Big)\end{array}
\end{equation}
defining for $t\geq 0$ the Schr\"odinger Equation (25). The
Hamiltonian for the Euclidean metric (54) in the interpretation
(27) is  (we change $M^{2}\rightarrow -\mu^{2}$)
\begin{equation}\begin{array}{l} {\cal H}= \frac{1}{2}\sum_{lm}
\Big((\frac{1}{H}\cos(Ht))^{-3}\Pi_{lm}^{2}-\frac{1}{H}\cos(Ht)l(l+2)\Phi_{lm}^{2}
-\mu^{2}(\frac{1}{H}\cos(Ht))^{3}\Phi_{lm}^{2}\Big)\end{array}
\end{equation}
In the interpretation (26), when $\sqrt{-g}$ is imaginary,  then
we have the diffusion equation  $\hbar\partial_{t}\psi={\cal
H}_{E}\psi$ with

\begin{equation}\begin{array}{l} {\cal H}_{E}= \frac{1}{2}\sum_{lm}
\Big((\frac{1}{H}\cos(Ht))^{-3}\Pi_{lm}^{2}+\frac{1}{H}\cos(Ht)l(l+2)\Phi_{lm}^{2}
+\mu^{2}(\frac{1}{H}\cos(Ht))^{3}\Phi_{lm}^{2}\Big)\end{array}
\end{equation}
The operator ${\cal H}_{E}$ is positive. Hence, the diffusion
equation should be considered for the negative time.

We expand the solution of the wave equation in spherical harmonics
as in Section~\ref{sec5}. Then, $u$ satisfies the equation
\begin{equation}\partial_{t}^{2}u_{lm}+3H\tanh(Ht)\partial_{t}u_{lm}
+M^{2}u_{lm}+l(l+2)(\frac{1}{H}\cosh(Ht))^{-2}u_{lm}=0
\end{equation}
and, in the Euclidean case (54),
\begin{equation}\partial_{t}^{2}u_{lm}-3H\tan(Ht)\partial_{t}u_{lm}
-\mu^{2}u_{lm}-l(l+2)(\frac{1}{H}\cos(Ht))^{-2}u_{lm}=0
\end{equation}
The stochastic equation for the fields is
\begin{equation}
d\Phi_{lm}(s)=-u_{lm}^{-1}\partial_{t}u_{lm}(t-s)\Phi_{lm}(s)ds
+\sqrt{i\hbar}(\frac{1}{H}\cos(H(t-s)))^{-\frac{3}{2}}dw_{lm}(s)
\end{equation}
and the one for the diffusion (26) is
\begin{equation}
d\Phi^{E}_{lm}(s)=-(u^{E}_{lm})^{-1}\partial_{t}u^{E}_{lm}(t-s)\Phi^{E}_{lm}(s)ds
+\sqrt{\hbar}(\frac{1}{H}\cosh(H(t-s)))^{-\frac{3}{2}}dw_{lm}(s)
\end{equation}
The two-dimensional de Sitter model is soluble in terms of
elementary functions, by means of the methods developed in this
paper  (for the standard approach, see \cite{figari,figari2}).
There are minor differences in comparison to Equations
(14.45)--(14.47) resulting from the fact that
$\sqrt{-g}=(\frac{1}{H}\cosh(Ht))^{3}\rightarrow
\frac{1}{H}\cosh(Ht)$, because the three spatial dimensions are
replaced by one dimension. So, Equations (14.48) and (14.49)
are changed into
\begin{equation}\partial_{t}^{2}u_{n}+H\tanh(Ht)\partial_{t}u_{n}
+M^{2}u_{n}+n^{2}(\frac{1}{H}\cosh(Ht))^{-2}u_{n}=0
\end{equation}
and, in the Euclidean case,
\begin{equation}\partial_{t}^{2}u_{n}-H\tan(Ht)\partial_{t}u_{n}
-\mu^{2}u_{n}-n^{2}(\frac{1}{H}\cos(Ht))^{-2}u_{n}=0
\end{equation}
where $n$ is a natural number. $n$ is replacing $l(l+2)$ as an
eigenvalue of the Laplacian on $S^{1}$ (instead of the one on
$S^{3}$). If we introduce the variable
\begin{displaymath}
y=\tanh(Ht)\end{displaymath} then Equation (14.52) in the
massless case reads
\begin{equation}
(y^{2}-1)\partial_{y}^{2}u+y\partial_{y}u-n^{2}u=0
\end{equation}
For Equation (14.53), we set
\begin{displaymath}
y=\tan(Ht)
\end{displaymath}
Then, Equation (14.53) is
\begin{equation}
(y^{2}+1)\partial_{y}^{2}u+y\partial_{y}u-n^{2}u=0
\end{equation}
The solution of Equation (14.54) for de Sitter is the
Tchebyshev polynomial $P_{n}$
\begin{equation}
u_{n}=z^{n}+z^{-n}=P_{n}(y)
\end{equation}
where
\begin{equation}
z=y+i\sqrt{1-y^{2}}
\end{equation}
Then, $G_{ss^{\prime}} $ can be expressed by elementary functions
as
\begin{equation}
G_{ss^{\prime}}=i\hbar
u(t-s)u(t-s^{\prime})\int_{0}^{m(s,s^{\prime})}\frac{dy}{y}P_{n}^{-2}
\end{equation}
The solution for the sphere (14.55) for $n=0 $ is $\ln z $,
where
\begin{equation}
z=y+\sqrt{1+y^{2}}
\end{equation}
For $n>0$, we obtain a polynomial in $y$
\begin{equation}
u_{n}=z^{n}+(-1)^{n}z^{-n}=\tilde{P}_{n}(y)
\end{equation}
The diffusion resulting from the imaginary value of $\sqrt{-g}$
has the correlation
\begin{equation}
G_{ss^{\prime}}=\hbar
u(t-s)u(t-s^{\prime})\int_{0}^{m(s,s^{\prime})}\frac{dy}{y}\tilde{P}_{n}^{-2}
\end{equation}
In the interpretation of the Hamiltonian with
$\sqrt{-g}\rightarrow \sqrt{\vert g\vert}$, we would replace
$\hbar \rightarrow i\hbar$.

We solve the homogenous model with $a(t)=\exp(Ht)$ of Equation
(57). The wave equation reads
\begin{equation}
\partial_{t}^{2}u+H\partial_{t}u+k^{2}\exp(-2Ht)u=0
\end{equation}
The  solution, which gives a negative $i\Gamma$, is
\begin{equation}
u(t)=\exp(-i\frac{k}{H}\exp(-Ht))
\end{equation}
Then, 
\centering 
\begin{equation}
G_{ss^{\prime}}=\frac{\hbar}{2k}
u(t-s)u(t-s^{\prime})\Big(\exp\Big(2i\frac{k}{H}\exp(-H(t-m(s,s^{\prime}))\Big)
-\exp\Big(2i\frac{k}{H}\exp(-Ht)\Big)\Big)
\end{equation}and
\begin{equation}
G_{t}=\frac{\hbar}{2k}
\Big(1-\cos\Big(2\frac{k}{H}(\exp(-Ht)-1)\Big)
-i\sin\Big(2\frac{k}{H}(\exp(-Ht)-1) \Big)\Big)
\end{equation}
So that $\Re G_{t}>0$, proving that the propagator  (111), defines
an integrable function.
 The inverted metric (58) describes the Euclidean version of
the AdS field (the scalar free field on the Poincare upper half
plane). The solution is
\begin{equation}
u^{E}=\exp(-\frac{ k}{H}\exp(-Ht))
\end{equation}
Now,
\begin{equation} G_{ss^{\prime}}=\frac{\hbar}{2k}
u^{E}(t-s)u^{E}(t-s^{\prime})\Big(\exp\Big(2\frac{k}{H}\exp(-H(t-m(s,s^{\prime}))\Big)
-\exp(2\frac{k}{H}\exp(-Ht))\Big)
\end{equation}
 and

\begin{equation} G_{t}=\frac{\hbar}{2k}
\Big(1- \exp\Big(2\frac{k}{H}(\exp(-Ht)-1)\Big)\Big)>0
\end{equation}
In summary, as follows from Equations (14.58), (14.61) and
(14.67), we can express the time evolution propagator of
Section~\ref{sec7} in the two-dimensional massless de Sitter model
by elementary functions. Then, with the trigonometric interaction
of Section~\ref{sec12} (Equation~(181)), we obtain a solution of
the Schr\"odinger equation in the form of a convergent series of elementary functions.

{\bf G: Scalar Field in \boldmath{$d=5$} Dimension}

We may ask the question of whether the signature inversion
described in Section~\ref{sec10} on the background of radiation
($w=\frac{1}{3}$) in four dimensions can happen in other
backgrounds. To discuss this question, we consider the Friedmann
equation in  $d$ dimensional space-time. We assume $p=w\rho$
(where $\rho$ is the energy density and $p$ is the pressure).
Then, from the energy-momentum conservation,
\begin{equation}
\rho=\rho_{0}a^{-(d-1)(1+w)}
\end{equation}
As a consequence, the Friedmann equation  \cite{five} in a flat expanding
universe is
\begin{equation}
(\frac{da}{dt})^{2}=Ca^{-(d-1)(1+w)+2},
\end{equation}
where $C$  is a positive constant. If we insert $a^{2}=t$ in this
equation, then we obtain the condition $4=(1+w)(d-1)$. Hence, for
$d>5$, the index $w$ in the equation of state must be negative.
For a radiation background from the condition $T^{\mu}_{\mu}=0$,
we obtain $w=\frac{1}{d-1}$. Hence, the condition $a^{2}=t$ for
radiation is satisfied only in four dimensions. The signature
would change  when time is changing sign if $a^{2}=t^{2n+1}$,
where $n$ is a natural number. Then, Equation (14.70) can be
satisfied for a particular $w>0$, but in the interval $ -1\leq
w<0$. There remains an interesting case of $d=5$ and $w=0$ (dust)
when $a^{2}=t$. In five dimensions, $g$ does not change its sign
when $t$ becomes negative. In such a case, we have a unitary
evolution for positive as well as  for a negative time with an
inverted metric. The equation for $\chi$ takes the form
\begin{equation}\partial_{t}\chi_{t}= \frac{1}{2}\int d{\bf x}
\Big(i\hbar\frac{1}{2}c_{0}^{2}
(t+\gamma)^{-2}\frac{\delta^{2}}{\delta\Phi({\bf
x})^{2}}-u^{-1}\partial_{t}u \Phi({\bf
x})\frac{\delta}{\delta\Phi({\bf x})}\Big)\chi_{t},
\end{equation}
where  $u$ is the solution of the equation \begin{equation}
     \frac{d^{2}u}{dt^{2}}+2(t+\gamma)^{-1}\frac{du}{dt}
      + (t+\gamma)^{-1}c_{0}k^{2}u=0.
      \end{equation}
The solution can be expressed by the cylinder function
\cite{kamke}
\begin{equation}
u=(t+\gamma)^{-\frac{1}{2}}Z_{1}(2\sqrt{c_{0}}k\sqrt{t+\gamma})
\end{equation}
We can choose the Bessel function as $Z_{1}$
\begin{displaymath}
J_{1}(z)=\frac{z}{2}\sum_{n=0}^{\infty}(-1)^{n}\frac{1}{n!(n+1)!}(z^{2})^{n}
\end{displaymath}
Then, $u$ is defined for positive as well as for negative time.
The stochastic Equation (for positive time) reads
\begin{equation}
d\Phi_{s}=-u^{-1}\partial_{t}u(t-s)
\Phi_{s}ds+\sqrt{i\hbar}(t+\gamma-s)^{-1}dW_{s}.
\end{equation}

\begin{equation}
G_{ss^{\prime}}=i\hbar
u(t-s)u(t-s^{\prime})\int_{0}^{m(s,s^{\prime})}u^{-2}(t-\tau)(t+\gamma
-\tau)^{-2}d\tau
\end{equation}
$G_{ss^{\prime}}$ is purely imaginary. We could construct
interactions in 5 dimensions in the way we did in
Sections 11 and  12.

{\bf H:Matter Field Schr\"odinger Evolution from Wheeler--DeWitt
Equation} 

In this appendix, we would like to show how  the Schr\"odinger
equation in an external metric discussed in this paper can appear
in quantum gravity as a consequence of (Wheeler--DeWitt)
constraint on the canonical variables. The Wheeler--DeWitt
\cite{dw,kiefer0} equation is derived as a Hamiltonian constraint
resulting from the diffeomorphism invariance of the Einstein
action
\begin{equation}
\begin{array}{l}
\Big(-c\hbar \nu_{p}^{-2}\int d{\bf
    x}G_{ijkl}\frac{\delta^{2}}{\delta h_{ij}\delta h_{kl}}-c\hbar
\nu_{p}^{2} \int d{\bf x}\sqrt{h}(R-2\Lambda)+\int d{\bf x}{\cal
    H}(h,{\bf x})\Big)\psi=0,
\end{array}\end{equation}
where ${\cal H}(h,{\bf x})$ is the density of the scalar field
Hamiltonian (24), $\Lambda$ is the cosmological constant and
\begin{equation} G_{ijkl}=\frac{1}{2}h^{-\frac{1}{2}}
(h_{ik}h_{jl}+h_{il}h_{jk}-h_{ij}h_{kl})
\end{equation} is a metric on the set of  symmetric tensors.
$h_{ij}$  is the metric on the spatial hyper-surface and ${\cal
    H}$ is the Hamiltonian (24).

\begin{displaymath}
\nu_{p}=(16\pi G\hbar c^{-3})^{-\frac{1}{2}}
\end{displaymath}
is the inverse of the  the Planck length ($G$ is the Newton
constant). The metric $G_{ijkl}$  does not have a definite sign.
For a conformally flat metric $h_{ij}=a^{2}\delta_{ij}$, the
metric $G_{ijkl}$ is negatively definite. In such a case, Equation
(14.76) becomes an equation of the hyperbolic type (a wave
equation), which can be written in the form
\begin{equation}
\int d{\bf x}\Big(\frac{\delta^{2}}{\delta a({\bf x}) \delta
    a({\bf
        x})}-\nu_{p}^{4}\frac{8}{3}a^{4}(R(a)-2\Lambda)+(c\hbar)^{-1}\nu_{p}^{2}\frac{8}{3}a{\cal
    H}(a,{\bf x})\Big)\psi(a,\phi)=0.
\end{equation}
$R(a)$ for a conformally flat metric depends on derivatives of
$a$. We can solve Equation (14.78) if we assume that $R(a)$ can
be expressed as a function of $a({\bf x})$ (without derivatives).
As an example, this is possible if $a({\bf x})$ depends only on
$\vert{\bf x}\vert$. Then, we can express $\vert{\bf x}\vert$ by
$a$ and, subsequently, $R(\vert{\bf
    x}\vert)$ as $R(a)$.

We obtain the WKB solution treating $\nu_{p}^{-2}$ as a small
parameter
\begin{equation}
\psi_{a}=\exp(\pm i\nu_{p}^{2}S(a))\psi_{a_{0}}.
\end{equation}
with
\begin{equation}
\nu_{p}^{2}S(a)=\int d{\bf x}\int_{a_{0}}^{a({\bf x})}d\alpha
\sqrt{-\frac{8}{3}\nu_{p}^{4}\alpha^{4}(R(\alpha)-2\Lambda)+\frac{8}{3}(c\hbar)^{-1}\nu_{p}^{2}\alpha{\cal
        H}(\alpha,{\bf x})}
\end{equation}
Expanding the square root in Equation (14.80) in powers of the
Planck length $\nu_{p}^{-2}$, we obtain
\begin{equation}\begin{array}{l}
\nu_{p}^{2}S(a)=\int d{\bf x}\int_{a_{0}}^{a({\bf
        x})}d\alpha\Big(\nu_{p}^{2}\sqrt{-\frac{8}{3}\alpha^{4}(R(\alpha)-2\Lambda)}
\cr+ \frac{4}{3}\alpha {\cal H}(\alpha,{\bf
    x})(c\hbar)^{-1}\Big(-\frac{8}{3}\alpha^{4}(R(\alpha)-2\Lambda)\Big)^{-\frac{1}{2}}\Big)+O(\nu_{p}^{-2}).
\end{array}\end{equation}
Let us write
\begin{equation}
\psi_{a}=\exp\Big( \pm i\nu_{p}^{2}\int d{\bf
    x}\int_{a_{0}}^{a({\bf
        x})}d\alpha\sqrt{-\frac{8}{3}\alpha^{4}(R(\alpha)-2\Lambda)}\Big)\chi_{a}\equiv\exp(\pm
i\nu_{p}^{2}S_{cl})\chi_{a},
\end{equation}
where
\begin{equation}
\chi_{a}=\exp\Big(\pm i\frac{4}{3}(c\hbar)^{-1}\int d{\bf
    x}\int_{a_{0}}^{a({\bf x})}d\alpha\alpha {\cal H}(\alpha,{\bf
    x})\Big(-\frac{8}{3}\alpha^{4}(R(\alpha)-2\Lambda)\Big)^{-\frac{1}{2}}\Big)\psi_{a_{0}}.
\end{equation}
Then, $S_{cl}$ in Equation (14.82) satisfies the
Hamilton--Jacobi equation

\vspace{-6pt}
\begin{displaymath}
\int d{\bf x}\frac{\delta S_{cl}}{\delta a({\bf x})}\frac{\delta
    S_{cl}}{\delta a({\bf x})}+\frac{8}{3}\int d{\bf
    x}a^{4}(R(a)-2\Lambda)=0
\end{displaymath}
and $\chi_{a}$ is the solution of the  equation
\begin{equation}
\mp ic\hbar\int       d{\bf x}\frac{3}{4 a({\bf
        x})}\sqrt{-\frac{8}{3}a({\bf x})^{4}(R(a({\bf
        x}))-2\Lambda)}\frac{\delta \chi}{\delta a({\bf x})}=\int d{\bf
    x}{\cal H}(a,{\bf x})\chi.
\end{equation}
In this Schr\"odinger-type equation, there is a functional
derivative over $a$ instead of time. However, when we  insert in
the solution $\chi_{a}$ of Equation (14.84) the classical
solution $a(t,{\bf x})$ instead of $a({\bf x})$, then we have
\begin{equation}
\partial_{t}\chi=\int d{\bf x}\frac{da(t,{\bf
        x})}{dt}\frac{\delta\chi}{\delta a(t,{\bf x})}
\end{equation}
We can replace the functional derivative in Equation (14.84) by
a time derivative if the classical solution satisfies the
equation\begin{equation} \frac{c}{4a(t,{\bf
        x})}\sqrt{-\frac{8}{3}a(t,{\bf x})^{4}(R(a(t,{\bf
        x}))-2\Lambda)}=\frac{da(t,{\bf x})}{dt}.
\end{equation}
Equation (14.86) can be derived from  the Hamilton--Jacobi
formulation of general relativity~\cite{berg}. As an example,
consider a solution of Einstein equations without matter   for the
Robertson--Walker space-time metric with negative scalar curvature
\begin{equation}
ds^{2}=dt^{2}-h_{ij}(t,{\bf x})dx^{i}dx^{j}\equiv dt^{2}-a(t,{\bf
    x})^{2}d{\bf x}^{2},
\end{equation}
where $h_{ij}\equiv \delta_{ij}a(t,{\bf
    x})^{2}=\delta_{ij}a(t)^{2}(1-\frac{1}{4}\vert {\bf
    x}\vert^{2})^{-2} $. Then, $R(h)=-6a(t)^{-2}$. The solution
of the Friedmann equation without matter is $a(t)=ct$, which
agrees with Equation (14.86). In another example, the solution
$a(t)= \exp(\sqrt{\frac{\Lambda}{3}}t)$  of the spatially flat
equation (14.86) ($ R=0$) describing de Sitter space without
matter also leads to the replacement of the functional derivative
in Equation (14.84) by the time derivative.

Let us note that if $R(a)-2\Lambda>0$,  then instead of the
Schr\"odinger equation, we obtain a diffusion equation in the WKB
expansion in powers of $\nu_{p}^{-2}$.

\end{document}